\documentclass[fleqn,usenatbib]{mnras}

\usepackage{newtxtext,newtxmath}
\usepackage[T1]{fontenc}
\usepackage{ae,aecompl}

\usepackage{graphicx}	% Including figure files
\usepackage{amsmath}	% Advanced maths commands
\usepackage{natbib}
\usepackage{siunitx}    % for exponents like \num{1e-10} and degree symbol \ang{30}
\usepackage{float}
\usepackage{placeins}   % for keeping tables/figures where you want them
 \usepackage{subfig}    % the \ContinuedFloat command in it is used for making figures over multiple pages.
\defcitealias{Donnarumma2011}{D11}
\newcommand{\qlens}{\texttt{QLens }}

\newcommand{\cmsg}{$\textrm{cm}^2\textrm{/g}\,$}  % cm squared per gram

\title[A Stringent Upper Limit on Self-Interaction Cross Section]{A Stringent Upper Limit on Dark Matter Self-Interaction Cross Section from Cluster Strong Lensing}

\author[K. E. Andrade et al.]{Kevin E. Andrade$^{1}$\thanks{E-mail: kandrad1@uci.edu},
Jackson Fuson$^1$,
Sophia Gad-Nasr$^1$,
Demao Kong$^1$,
\newauthor
Quinn Minor,$^{2,3}$
M. Grant Roberts$^1$
and Manoj Kaplinghat$^{1}$
\\
$^{1}$University of California, Irvine, Irvine, CA 92697, USA\\
$^{2}$Department of Science, Borough of Manhattan Community College, City University of New York, New York, NY 10007, USA\\
$^{3}$Department of Astrophysics, American Museum of Natural History, New York, NY 10024, USA\\
}

\pubyear{2021}

\begin{document}

\label{firstpage}
\pagerange{\pageref{firstpage}--\pageref{lastpage}}
\maketitle

% Abstract of the paper
\begin{abstract}
We analyze strongly lensed images in 8 galaxy clusters to measure their dark matter density profiles in the radial region between 10 kpc and 150 kpc, and use this to constrain the self-interaction cross section of dark matter (DM) particles. 
We infer the mass profiles of the central DM halos, bright central galaxies, key member galaxies, and DM subhalos for the member galaxies for all 8 clusters using the \qlens code. The inferred DM halo surface densities are fit to a self-interacting dark matter (SIDM) model, which allows us to constrain the self-interaction cross section over mass $\sigma/m$. 
When our full method is applied to mock data generated from two clusters in the Illustris-TNG simulation, we find results consistent with no dark matter self-interactions as expected.
For the eight observed clusters with average relative velocities of $1458_{-81}^{+80}$ km/s, we infer 
$\sigma/m = 0.082_{-0.021}^{+0.027} \rm cm^2/g$
and $\sigma/m <  0.13~ \rm cm^2/g$ at the 95\% confidence level. 
\end{abstract}


%%%%%%%%%%%%%%%%%%%%%%%%%%%%%%%%%%%%%%%%%%%%%%%%%%

%%%%%%%%%%%%%%%%% BODY OF PAPER %%%%%%%%%%%%%%%%%%

\section{Introduction}

The currently-favored theory of the formation of galaxies in the Universe, dark energy plus cold dark matter or $\Lambda$CDM, is remarkably successful in explaining many of the observations of large-scale structures \citep{Vogelsberger2014, Schaye2015}. Dark-matter-only simulations show that structures form in a heirarchical manner, and result in dark matter (DM) halos that are well-approximated by a Navarro-Frenk-White (NFW) density profile \citep{Navarro1996, Gao2012}. 
Statistically significant departures from the NFW halo profile have been reported for galaxy clusters \citep{Sand2008, Pontzen2012, Martizzi2013, Newman2013a, Newman2013b, DelPopolo2012, DelPopolo2014, Annunziatella2017}, which provides a fertile ground for the exploration of DM physics. 
For galaxy clusters, there are a number of studies that point to the existence of DM cores in some galaxy clusters \citep{Sand2008, Newman2013a, Newman2013b, DelPopolo2012, DelPopolo2014, Annunziatella2017}. One frequently proposed explanation for galaxy clusters with cores is that of Active Galactic Nuclei (AGN) feedback, where a black hole accreting gas can cause mass blowout and heating via gravitational interactions, leading to cores \citep{Dubois2010,Teyssier2011, Ragone-Figueroa2012, Martizzi2012, Martizzi2013, Peirani_2017}. In \citet{Martizzi2013}, they show that AGN feedback can potentially cause cores in their simulated halo of mass $\num{1.42e13}M_{\sun}$. On the other hand, \cite{Schaller2015} used the hydrodynamical \textsc{Eagle} simulations of six galaxy clusters with $m_{200} > 10^{14} M_{\sun}$, which employ weaker AGN feedback, and found instead that feedback does not produce cores. Numerical cosmological simulations have advanced steadily in resolution and accuracy over the past two decades, but there is still no consensus about the presence of cores in large clusters of galaxies \citep{Schaller2015,Martizzi2012, Martizzi2013}.

Self-interacting dark matter (SIDM), in which DM has a nonzero self-interaction cross section, is another possible explanation for cored halos. SIDM has been discussed extensively in astro/particle physics literature as another possible model that can produce cores \citep[as key examples]{Spergel2000a, Randall2008, Vogelsberger2012, Rocha2013, Kaplinghat2016,Ren2019}. The basic mechanism of core formation in SIDM is that self-interactions between DM particles in regions with high density cause the DM to thermalize, which results in the transfer of energy to the high density inner region of a halo, thus lowering the core density. One frequently employed measure of self-interaction strength is the scattering cross-section per unit mass, $\sigma/m$, although that does not account for the dependence of scattering cross-section on particle velocity. 
Constant cross-section SIDM models cannot simultaneously have appreciable effects in dwarf galaxies, while being consistent with densities measured in clusters of galaxies \citep{Zavala2013, Ren2019, Kaplinghat2016}. 
A more complete model that accounts for the velocity-dependence of cross section is necessary because of the larger range of velocities probed in going from dwarf galaxies to clusters of galaxies (about 50 to 2000 km/s).

\begin{figure*}
    \centering
    \caption{Schematic diagram of the analysis pipeline. Data sources are at the top. The final results are the parameter posteriors shown at the bottom.}
    \label{fig:analysis_pipeline}
    \includegraphics[height=0.96\textheight]{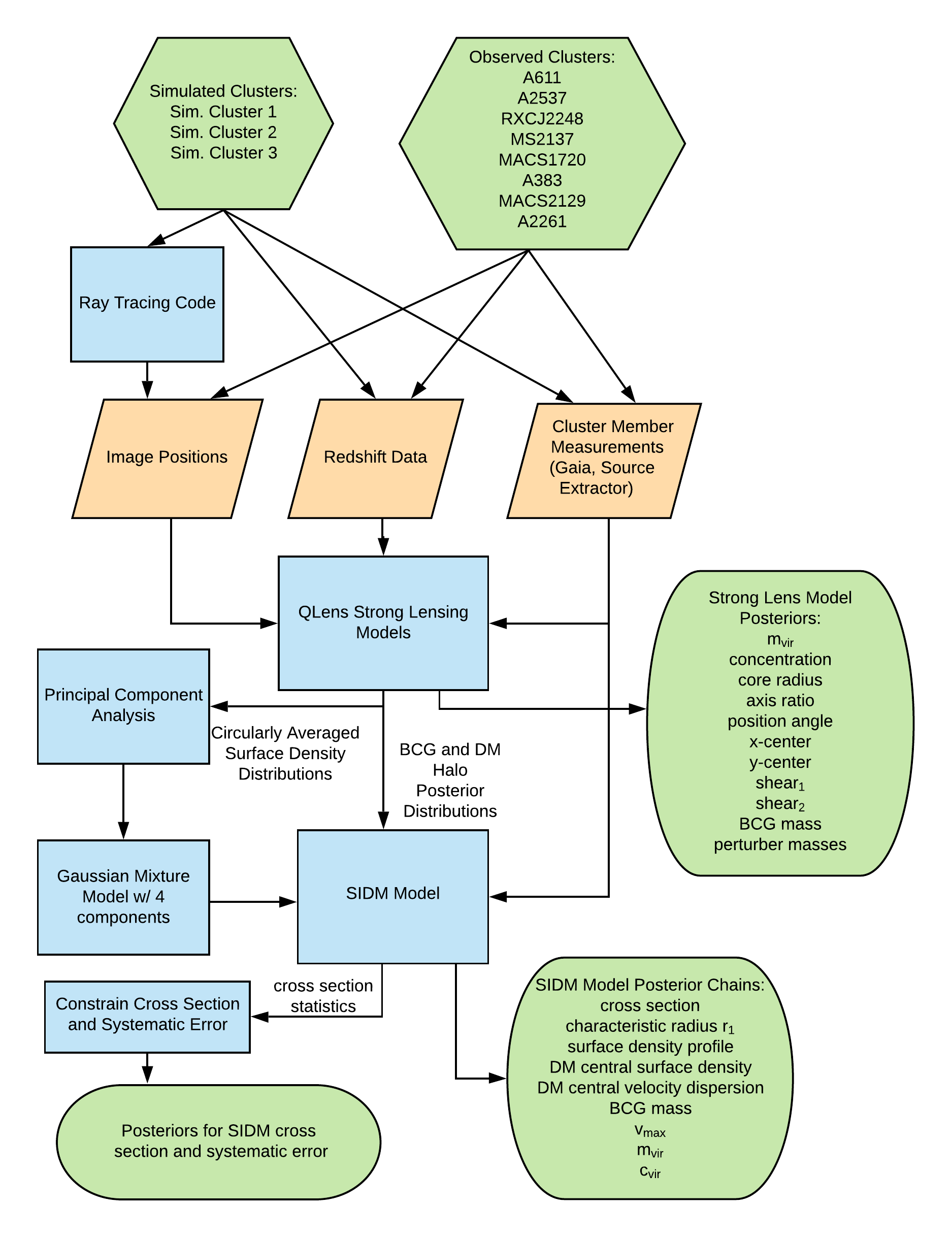}
    \end{figure*}

In this work, we analyze 8 observed galaxy clusters and two simulated clusters in three distinct analysis stages: (1) strong lensing, to determine DM and baryonic density profiles, (2) SIDM halo matching, to determine SIDM cross sections and relative particle speeds for each cluster, and (3) constraints on the cross section, to finally infer the SIDM cross section and systematic error. Our analysis pipeline is shown schematically in Figure~\ref{fig:analysis_pipeline}. In the first stage we extend the strong lensing analysis of \citet{Andrade2019}, in which one cluster (Abell 611) was examined, to include 8 relaxed clusters. The clusters we examine here are in the range of $\num{4e14}M_{\sun}$ to $\num{2e15}M_{\sun}$. These are the among largest bound structures in the universe \citep{Desai2004}. Their centers have dense concentrations of DM, and if such particles self-interact, cores with densities less than that predicted by $\Lambda$CDM could form. Our aim is to put a stronger constraint on the SIDM self-interaction cross-section by inferring the inner distribution of DM in these 8 clusters, using strong lensing alone. In cluster strong lensing, images typically appear near the Einstein radii of the clusters, which usually range from 20 to 100 kpc, well within the scale radii of the clusters \citep{Richard2010}. The DM density profile is constrained with highest accuracy near these image locations. This is also the region where SIDM thermalization would be expected to occur \citep{Kaplinghat2016}. Baryonic effects are also strong in this region, as the brightest cluster galaxy (BCG) is located there, with high baryonic mass. It is therefore important to accurately characterize the BCG's contribution to the total mass profile in the region of interest.
We use photometry to measure cluster member characteristics such as radius, position, ellipticity, orientation angle and luminosity, but leave the stellar mass-to-light ratio as a parameter to be fit for each cluster.

Other techniques that could be employed to determine cluster DM profiles include weak lensing, x-ray temperature analysis and stellar kinematics \citep{Miralda1995, Natarajan1996, Sand2002, Kneib2003, Sand2004, Broadhurst2005,  Mandelbaum2006, Okabe2010, Newman2011, Umetsu2011, Coe2012, Newman2013a, Newman2013b}. Strong lensing directly probes the profiles at the radii at which we are interested, while weak lensing and x-ray analysis probe much larger radii, where the ellipticity and profile may be significantly different than that of the inner region. 
Stellar kinematics of the BCG stars can help significantly in constraining the BCG mass and the slope of the dark matter density profile in the center \citep[as an example]{Newman2013b}. This measurement depends on having the correct priors for radial variation of the stellar velocity dispersion anisotropy \citep{Schaller2015} in spherical or axisymmetric Jeans analysis.  

\citet{He_2020} examined the ability of stellar kinematics, strong and weak lensing data sets to predict dark matter densities in the inner regions of simulated clusters, and found them to be accurate. This validates the approach undertaken by \citet{Newman2013a,Newman2013b} in combining these three data sets to constrain the dark matter density profile of clusters. 
While adding other data sets is well motivated, it is of interest to ask what constraints are possible with just one data set given that biases may creep in because of the assumption of a common density profile with a fixed ellipticity that is inherent in a joint analysis. Somewhat surprisingly, we find that it is possible to infer strong constraints on the SIDM cross section with just strong lensing data. 
In this paper, we develop the method to infer the SIDM cross section using strong lensing data, and leave the analysis including other data sets to future work. \citet{He_2020} also discussed possible ways to reconcile the differences between the inferred dark matter density profile slopes in \citet{Newman2013b} and their simulated clusters in CDM cosmology, and highlighted a possible role for bias resulting from degeneracy with $r_s$, which is constrained in part by the weak lensing data. We will discuss the inference of $r_s$ from strong lensing data, and argue that our results are consistent with expectations from simulations. 

Armed with DM profiles and the aforementioned cluster member data, we employ a fitting process to match an NFW-like outer halo with an SIDM thermalized inner halo, following the process described in \citet{Kaplinghat2016}. The resulting fit yields posterior chains for various parameters, including those for cross section per unit mass and average particle speed, for each cluster. 
In the third analysis stage, we synthesize the results of the the prior stages into an overall distribution for SIDM cross section to obtain an upper limit on self-interaction cross section.

In this work we adopt a flat cosmology with $\Omega_{\Lambda}=0.7$, $\Omega_m=0.3$, and $H_0=70\: km\: s^{-1} Mpc^{-1}$. We define halo mass as $M_{200}$, the mass enclosed by a sphere of radius $R_{200}$, which we define in turn as the radius at which the halo density is 200 times the critical density of the universe at the redshift of the halo.

This paper is organized as follows. In Section~\ref{sec:Strong_Lens_Modeling}, we describe our approach to the strong lens modeling analysis. In Section~\ref{sec:Models_of_Simulated_Clusters} we detail the strong lens models for the simulated clusters, and do the same for the observed clusters in Section~\ref{sec:Models_of_Observed_Clusters}. The results of the strong lensing models for the observed clusters are presented in Section~\ref{sec:Strong_Lensing_Results}. In  Section~\ref{sec:SIDM_Halo_Analysis} we describe our SIDM halo matching model and its results, and also describe the model for putting constraints on the cross section, and its resulting inference of SIDM cross section and systematic error. We summarize our findings in Section~\ref{sec:Conclusions}.

\section{Strong Lens Modeling Methods}
\label{sec:Strong_Lens_Modeling}
Our first step is to use strong lens modeling to determine the surface density profiles for a sample of galaxy clusters. 
For the cluster lensing models, we use previously reduced data in the form of (1) coordinates for multiple images and (2) source redshift data. The availability of multiple image coordinate data was a requirement for all the clusters in our sample. The positional error is assumed to be 0\farcs5 in each dimension (see Section~\ref{sec:SL systematic errors} for a discussion of this choice). We construct mass models of the important features of each cluster: the (potentially cored) DM halo, BCG, important member galaxies ("perturbers"), and the DM subhalos associated with the perturbers. 
Many lensing analyses do not disentangle the DM subhalos from the luminous part, but doing so allows us to model the perturbers in a manner that is more consistent with cosmological simulations. 
The obvious challenge for creating accurate mass models is that the majority of the mass is in the form of invisible dark matter, the distribution of which must be inferred indirectly via lensing or by assuming some correlation with the luminous matter. We have designed a systematic approach that we first test on simulated galaxy clusters from the Illustris TNG simulation \citep{nelson2018, Springel2018, Pillepich_2017, Nelson_2017, Naiman_2018, Marinacci_2018} as described in section \ref{sec:Models_of_Simulated_Clusters} of this work, allowing us to verify the accuracy of the models in reproducing cluster properties. We then apply this approach to actual galaxy clusters as described in section \ref{sec:Models_of_Observed_Clusters}. For each cluster, we obtain a Markov Chain Monte Carlo (MCMC) chain for the model parameters, from which we can estimate parameter posteriors and also infer posterior distribution in other derived parameters. We use the \texttt{Multinest} sampler to create the MCMC chain \citep{Feroz2009}.

In brief, we construct mass models for clusters in the following way.
\begin{itemize}
    \item Starting with a CCD image of the cluster (or a simulated image in the case of Illustris clusters), the coordinates, shapes and fluxes of the BCG and key member galaxies are measured, using one or both of the Starlink Gaia software package \citep{Currie2014} or a proprietary 2D fitting code. A coordinate system is created with its origin at the center of the BCG.
    \item The main DM halo is modeled using a cNFW profile, the parameters of which are mass, scale radius, core radius, position angle, axis ratio, and (x, y) center offset coordinates. All are allowed to vary within ranges defined by wide, uninformative priors. 
    \item The stellar baryonic component of selected significant member galaxies (referred to as ``perturbers") and the BCG are modeled using dPIE profiles. The mass parameter of each is allowed to vary (with perturbers typically tied together via a mass-follows-light scaling relation), while the scale radius, center coordinates, position angle and axis ratio are fixed by photometric measurement.
    \item For the perturbers, each is given an isothermal DM subhalo with a mass related to their baryonic mass via a power law. The scale radii of these DM subhalos are determined by their mass according to another power law, both described in Section~\ref{sec:DM_Subhalo_Masses_and_Radii}. The center coordinates used are the same as those of the underlying baryonic model.
    \item To account for the effects of other masses along the line of sight that are not specifically included in the model, external shear is modeled, requiring two additional varying parameters, $\gamma_1$ and $\gamma_2$ (for a derivation of these, see \citet{Voigt2010}). 
\end{itemize}

\subsection{Lens Profiles}
\label{sec:lens profiles}

Our goal in this analysis is to construct mass models of galaxy clusters that accurately infer the surface density profile in the inner $\sim$200 kpc, and subsequently infer SIDM properties. To model the point images and infer the halo density profile, we use the \qlens software package, as in \citet{Andrade2019}. \footnote{For more information on \qlens contact Quinn Minor: qminor@bmcc.cuny.edu.} 

We employ the cored NFW lens profile (cNFW) as the primary model for the cluster halos. The spherically symmetric form for the cNFW density profile is defined as:
\begin{equation}
\label{eq:cNFW}
    \rho = \frac{\rho_s r_s^3}{\left(r_c + r\right)\left(r_s + r\right)^2}.
\end{equation}
This profile reduces to the canonical NFW form for $r\gg r_c$. 
Other authors have used this profile to fit cored DM halos in both hydrodynamical simulations \citep{penarrubia2013} and actual cluster lenses \citep{Newman2013a,Newman2013b, Andrade2019}.  Analytic formulas for the projected density profile and deflection of the corresponding spherical model are given in Appendix \ref{sec:appendix_lensing_formulae}.

The BCG and member galaxy baryonic components, as well as galaxy member DM subhalos are modeled with a dual pseudo-isothermal ellipsoid (dPIE) profile \citep{Eliasdottir2007}, also known as a "Pseudo-Jaffe" profile. The spherical version of this profile is
\begin{equation}
    \rho = \frac{\rho_{\mathrm{cut}} r_{\mathrm{cut}}^4}{\left(r^2+r_{\mathrm{core}}^2\right)^{2} \left(r^2+r_{\mathrm{cut}}^2\right)^{2}}, 
    \label{eq:dPIE}
\end{equation}
where $\rho_{\mathrm{cut}}$ is the density at the scale radius, $r_{\mathrm{cut}}$ is the scale (or tidal break) radius, and $r_{\mathrm{core}}$ is the core radius.

We model elliptical lens profiles by making the replacement $R^2 \rightarrow qx^2 + y^2/q$ in the projected density profiles. The deflection and Hessian of the lens mapping must be calculated by numerical integration (see \citealt{schramm1990,keeton2001b}), which can be computationally expensive. However, by using Gauss-Patterson quadrature for integration \citep{Davis2007}, \qlens is able to compute such integrals in a reasonable time. This approach avoids using the pseudo-elliptical approximation, in which the lensing \textit{potential} is assumed to be elliptical rather than the mass profile. Using the pseudo-elliptical approximation can lead to inaccurate inferences of the density profile in cases of high ellipticity  (e.g., $q<0.5$, see \citet{Andrade2019}).

\subsection{Photometric Measurement}
We use the Starlink Gaia software package \citep{Currie2014}, and the integrated version of Source Extractor \citep{Bertin1996} to measure the key photometric properties of the BCG and perterbers in each cluster: the coordinates, flux, axis ratio and position angle. We first select the HST image from the available object images that gives a target rest-frame wavelength in center of the visual band, based on the redshift of the lens. For background modeling, we use mesh-based RMS background detection. The projected half-light radii are measured using Source Extractor's "flux radius" feature by setting the flux fraction to 0.5. 
Fluxes are converted to magnitudes, and then corrected for galactic extinction using the tool at https://ned.ipac.caltech.edu. 

\subsection{Central DM Halo}
The main DM halo is modeled using a cNFW profile, the parameters of which are mass, concentration, core radius, position angle, axis ratio, and (x, y) center offset coordinates. Flat log-priors were used on the mass and core radius parameters, while flat priors were used on position angle, axis ratio and center offset.

\citet{Merten2015} studied the mass-concentration relation for cluster halos in 19 X-ray clusters from the CLASH sample.  They observed an average concentration of 3.7, with a standard deviation of 0.65, and a negative correlation of concentration with halo mass. Since several of the clusters in this work are CLASH clusters, we might expect concentration to be in a similar range. However, we note that there are inherent biases in modeling concentration in strong lensing clusters, as outlined in Appendix~\ref{sec:biases_in_triaxial_halos}. Also, \citet{Fielder_2020} found that concentration will be higher for a DM halo when its associated subhalos are excluded from the calculation. Therefore we opt for a weak Gaussian prior on concentration, with a mean of 6.0 and standard deviation of 1.5.  Thus the 2$\sigma$ range is from 3.0 to 9.0, accommodating these anticipated biases if present.

\subsection{BCG and Luminous Contribution of Member Galaxies}
The BCG and perturbers for each cluster were modeled with a dPIE profile, with mass as a free parameter and other parameters set by observed photometry. The core radii of galaxies ($r_\textrm{core}$) in the observed clusters are difficult to measure with high certainty. For our purposes, we are not particularly interested in the core sizes of the member galaxies, as structure at that scale will not have a significant impact on our results. Other authors have assumed a constant core radius for cluster member galaxies, typically in the range of 100 pc to 300 pc \citep{Limousin2007a, Limousin2007}, or alternatively 0.1\arcsec \citep{Limousin2004}, which for our clusters is approximately 300 pc to 600 pc. We adopt a core radius of 300 pc in each member galaxy in our observed clusters. For the Illustris clusters, we fit their two dimensional shapes from the simulated image to a dPIE profile, and use those parameters in our \qlens model. The measured core radii range from 0 kpc to 2.3 kpc. 

We tie together the masses of most of the perturbers in a given cluster and employ a mass-follows-light approach for scaling them. This is a reasonable approach, given that we have separate mass models for the DM subhalos. The perturber masses scale with the observed luminosity of each perturber, resulting in one free mass parameter describing the set of perturbers. One perturber is selected as the anchor, and the others are tied to it. We do this in order to limit the proliferation of model parameters. However, there are some perturbers that are quite close to images and that strongly affect them, in some cases even causing splitting of images into multiple images. For these perturbers, their mass is allowed to vary separately from the group, at the cost of including another degree of freedom in the model. 

\subsection{DM Subhalos}
\label{sec:DM_Subhalo_Masses_and_Radii}

 In many strong lensing analyses, the baryonic and dark components of perturbers are modeled as one object, often based solely upon the photometry of the stellar component. This has potential for bias, since the DM component is likely to be strongly dominant in both mass and size. To guard against this, we model the DM subhalos of each perturbing galaxy.
 
We modeled the dark matter profile of cluster galaxies with a dPIE profile. We set the half-mass radius of the dPIE profile equal to that of a NFW density profile with the median concentration in \citet{Diemer_2019}, as implemented in \texttt{Colossus} package \citep{Diemer2018}. Neglecting the mild redshift dependence in the concentration mass-relation for the relevant lens redshift range (about 0.2 to 0.6), we obtain the half-mass radius $=0.0019 M_{\mathrm{halo}}^{0.37}$. The core radii for the dPIE profiles are set to zero in the case of the observed clusters. For the simulated clusters, we find that the particle size of the Illustris simulation had an impact on the shape of the DM subhalos. In effect, this gives a core to the DM subhalos. The simulation has a DM particle size of \num{5.9e7}$M_{\sun}$, and it has been shown that two-body relaxation effects will give rise to cores at radii fewer than $\sim$ 100 particles and at the densities typical in our model \citep{Power2003}. As the DM subhalos are typically in the range of $\num{2e11}$ to $\num{6e12}\; M_{\sun}$, the 100-particle region has a typical radius 5 kpc to 10 kpc. In addition, we employ a Gaussian filter with a characteristic radius of 2 pixels (2.5 kpc) to help reduce noise in the ray tracing calculations, as noted in Section~\ref{sec:Simulated Image Production}. To model these effects we allow the dPIE DM subhalos to have a core in the simulated cluster models. All subhalos in a given model have the same core, the radius of which is a varied parameter during the MCMC process.   
 
We determine the mass of the subhalos from their fitted stellar mass, as follows. Infalling galaxies and their DM halos will be partially stripped of their DM by tidal forces due to the massive cluster. To quantify this for the simulated clusters, we looked directly at the the stellar-mass to halo-mass (SMHM) relation in Illustris TNG. We selected all subhalos within a 400 kpc radius of the center of the host halos, and with a stellar mass range similar to those of perturbers in the real cluster sample, i.e., between $\num{1e10}M_{\sun}$ and $\num{5e12}M_{\sun}$. These are well fit by a power relation of the form 
 \begin{equation}
 \label{eq:power law}
     M_{\textrm{halo}} = k M_*^{\alpha},
 \end{equation}
 with $k=1.115$ and $\alpha=1.024$. We used this relation to generate DM subhalo masses in the models of the simulated clusters. For the observed clusters, we turned to \cite{Niemiec_2017}, where they examine the stellar mass-halo mass relation for cluster galaxies. We fit a power law to their data (see Figure~\ref{fig:SMHM_Niemiec_fit} in Appendix~\ref{sec:power_law}). The resulting power law (in the same form as Equation~\ref{eq:power law}) gives $k=1.157$ and $\alpha=1.1171$, which is similar to that found for Illustris. As can be seen by the error bars in the figure, this relation has significant uncertainty in the lower bound. We therefore use it as an upper bound only, and explore departures from this upper bound by varying the parameter "k". Ideally, we would fit this parameter with the other parameters through the MCMC sampling process, but unfortunately the \qlens software does not have that capability. Instead, we manually optimize it by making runs with values of k, 0.5k, 0.25k and 0.125k. The value that yields the best model $\chi^2$ is adopted. The value adopted is noted in the individual cluster models.

We find that in Illustris TNG, the position angles of dark matter halos of cluster galaxies are usually quite similar to that of the associated stellar mass, but the dark matter halos are rounder on average. We fix the position angle to be the same as that of the stellar spheroid and choose an axis ratio for the DM subhalos equal to $\sqrt{q_*}$, i.e., the geometric mean between the stellar axis ratio and unity. We ran tests on the effect of this assumption, comparing it to using $q=q_*$ and $q=1$ for three representative clusters, A611, A2537 and MACS2129. The impact on the inferred BCG mass and halo mass was well within one standard deviation for all three clusters.

\begin{figure*}
    \centering
    \caption{Image plane plots for each cluster showing the data image positions (red) and modeling image positions (cyan). The points appear purple where they overlap. Critical curves are shown for a source redshift of 2.0. Redshift values for each family of points are indicated in the box to the right of each plot. North is up, East is to the left. Axes scales are in arc seconds.}
    \includegraphics[width=0.49\textwidth]{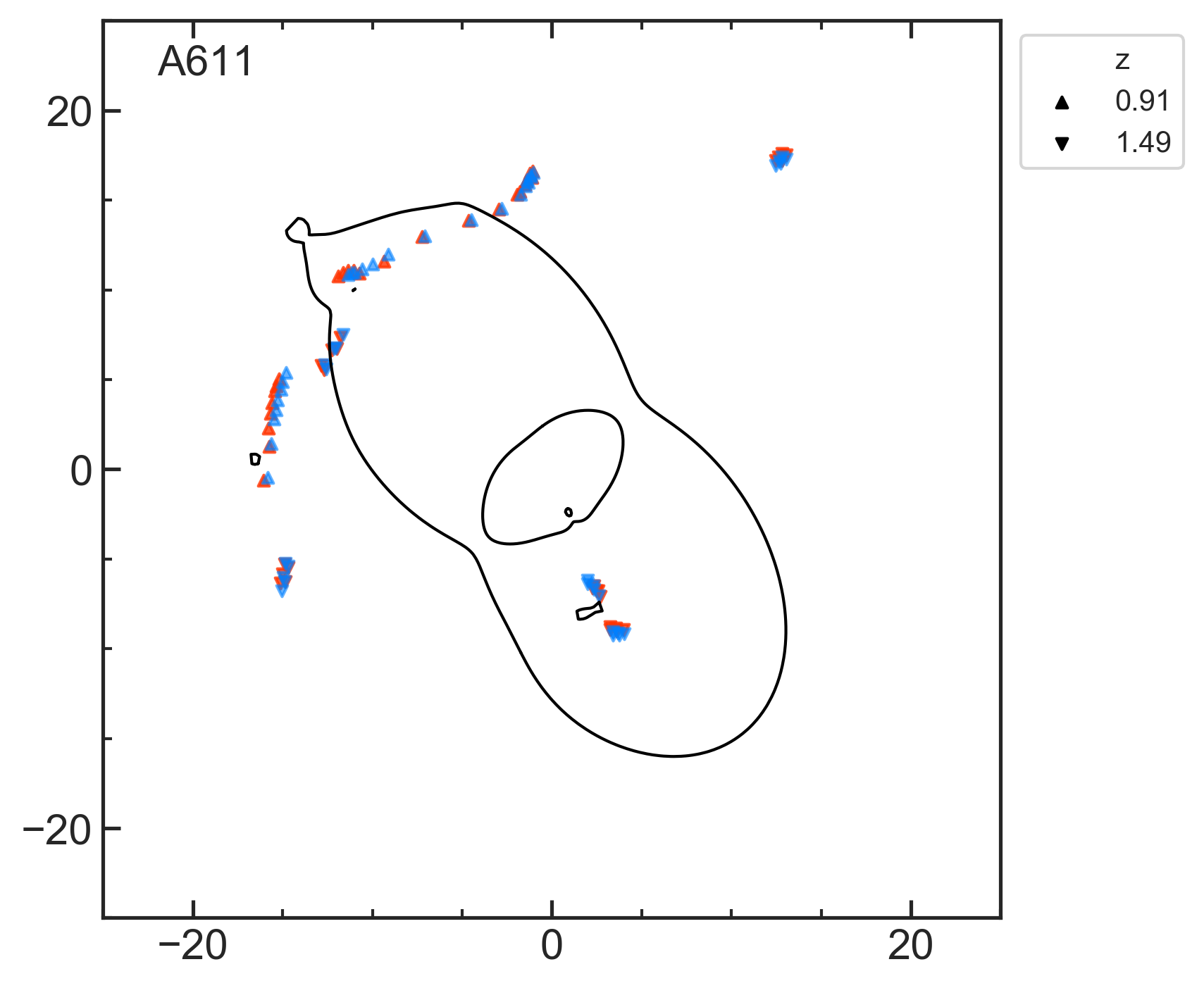}
    \includegraphics[width=0.49\textwidth]{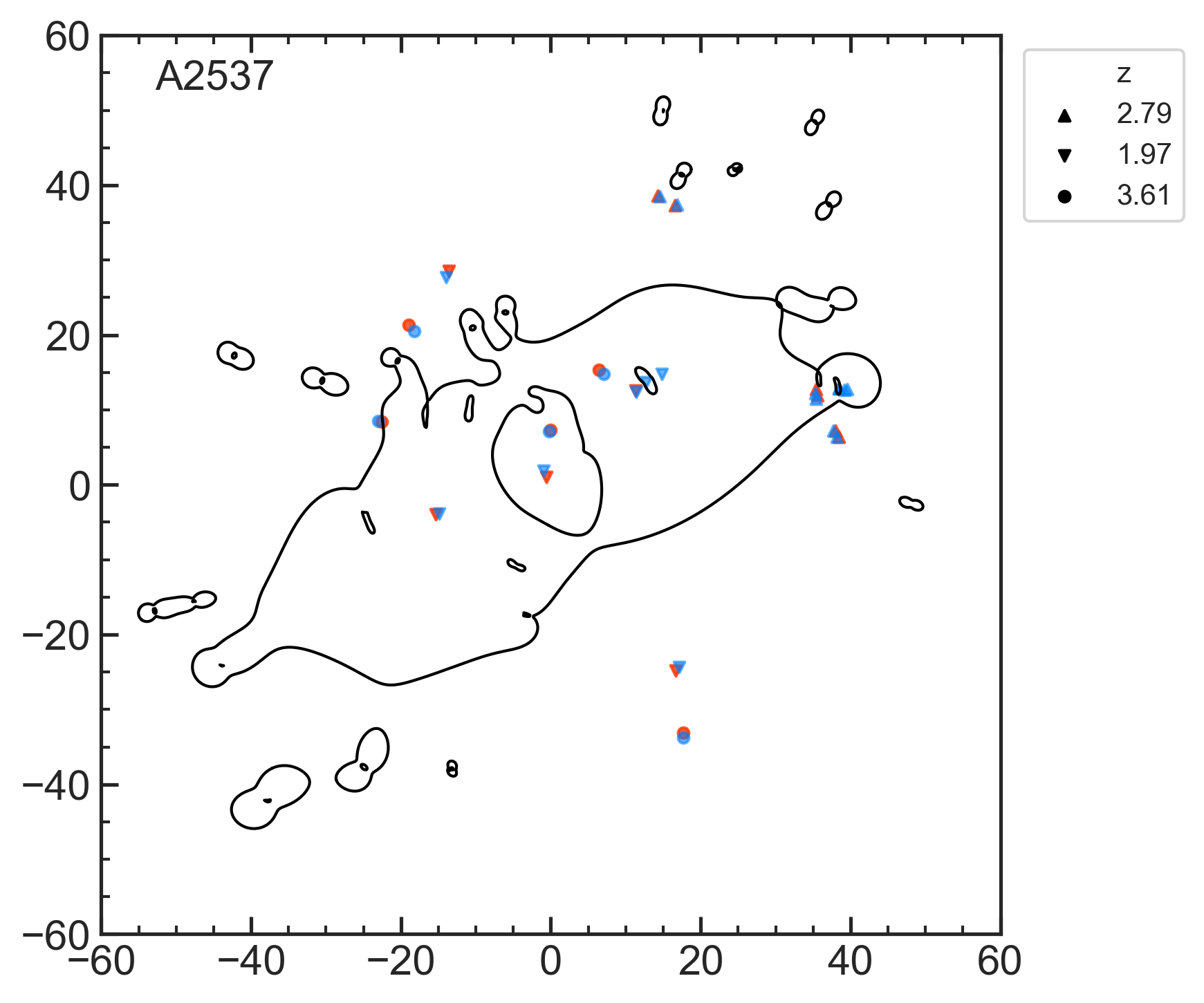}
    \includegraphics[width=0.49\textwidth]{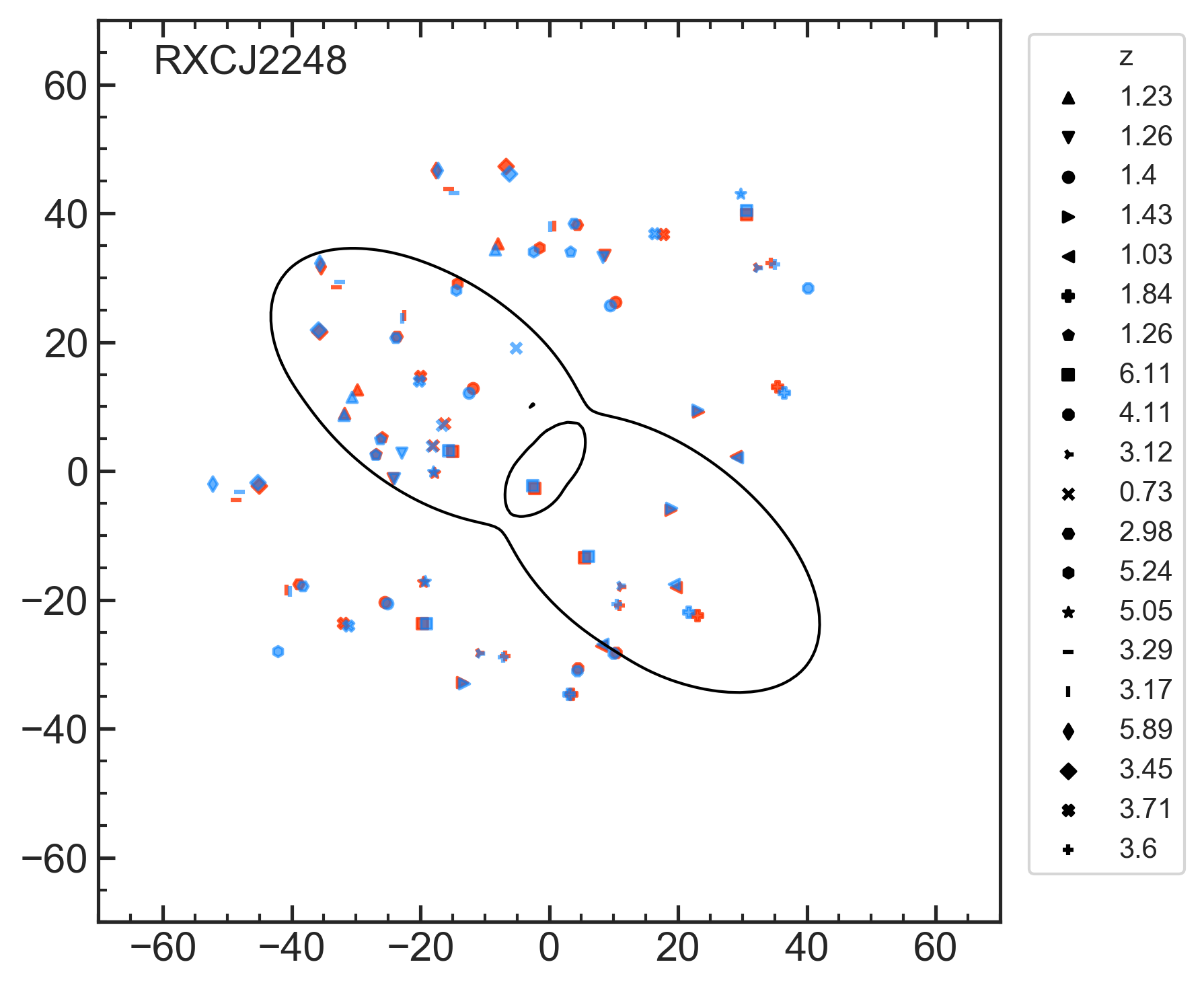}
    \includegraphics[width=0.49\textwidth]{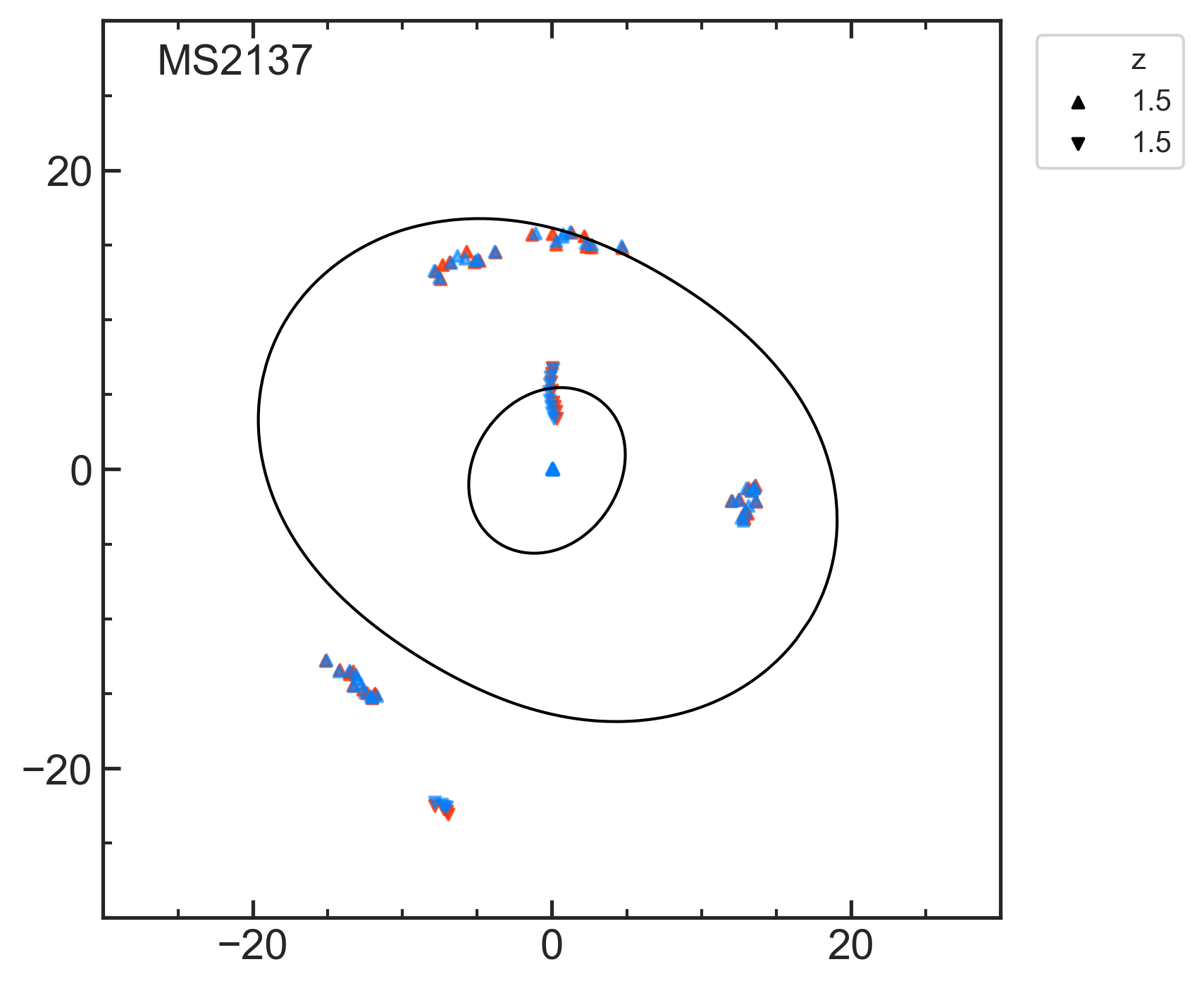}
    \includegraphics[width=0.45\textwidth]{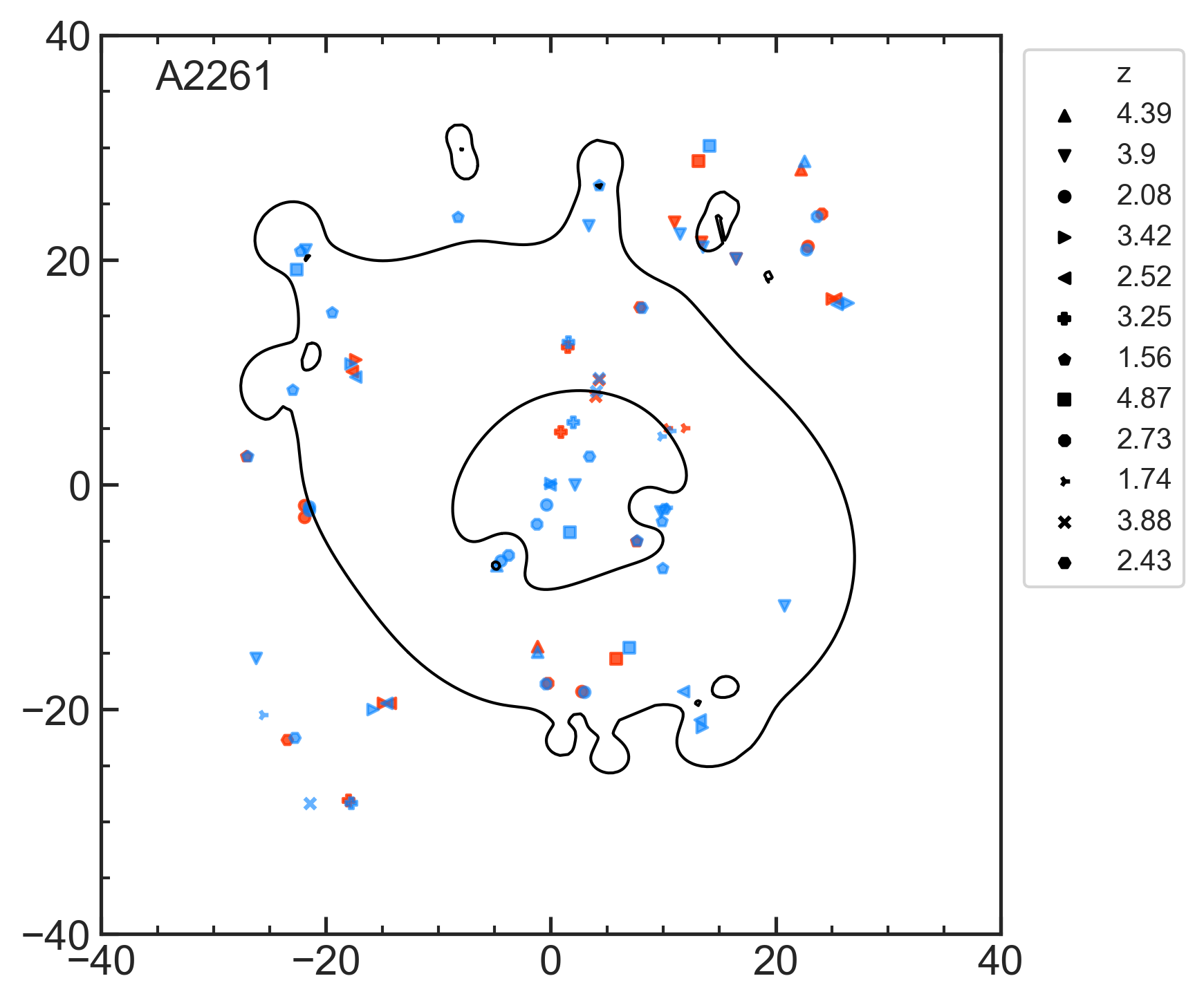}
    \includegraphics[width=0.45\textwidth]{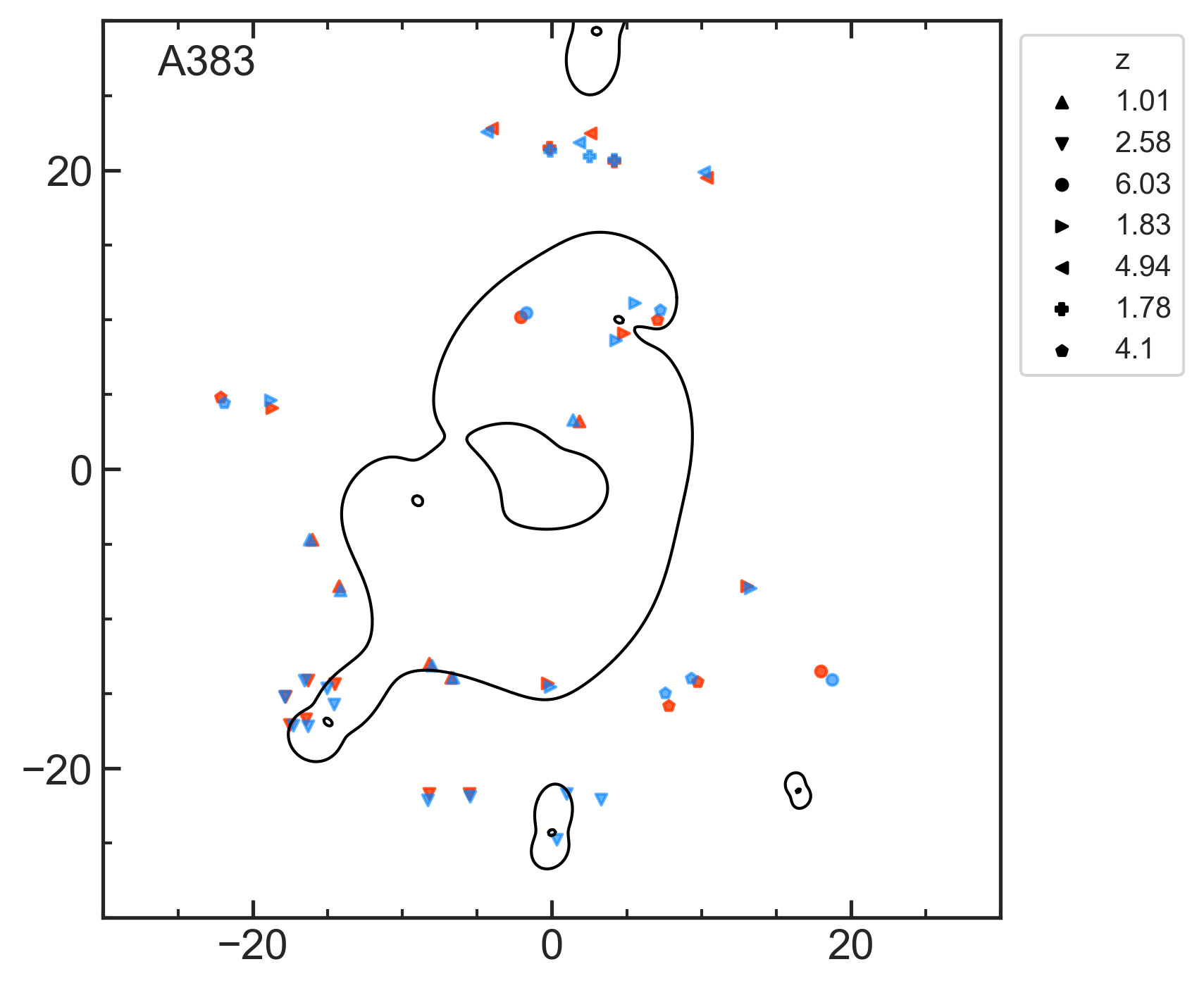}
 
    \label{fig:image_plane_plots}
\end{figure*}

\begin{figure*}\ContinuedFloat
    \centering
    \caption{, continued.}
    \includegraphics[width=0.49\textwidth]{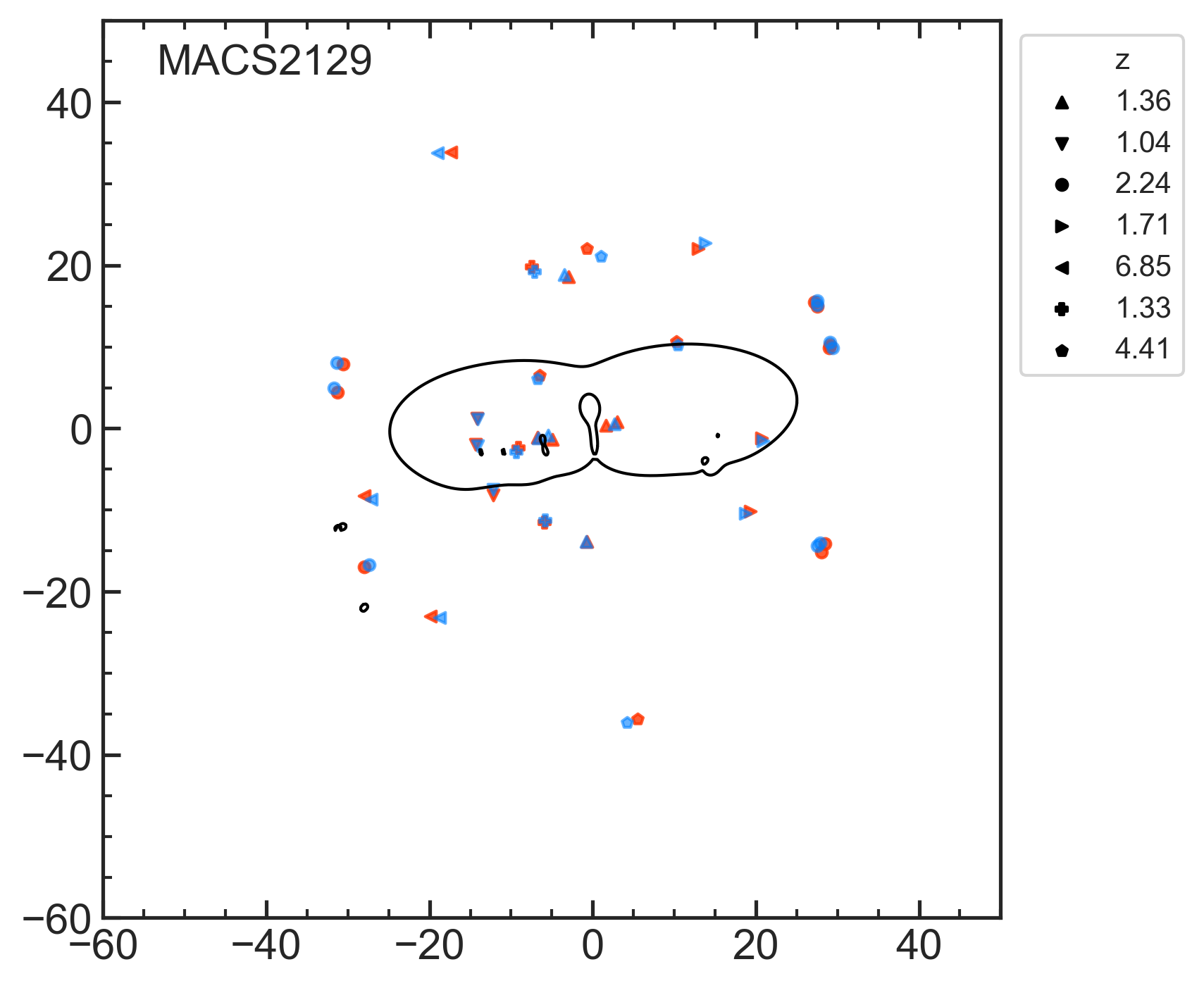}
    \includegraphics[width=0.49\textwidth]{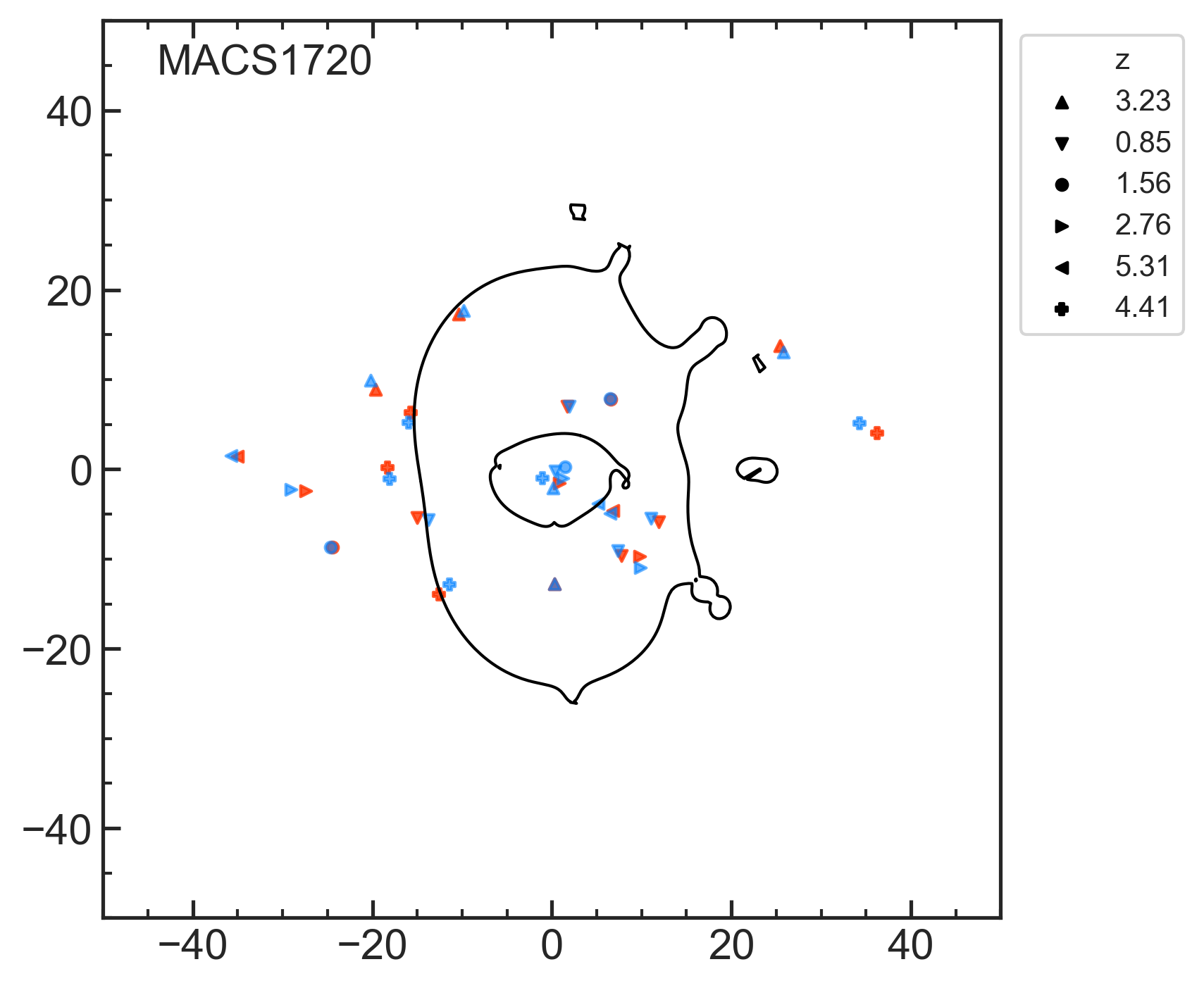}
    \includegraphics[width=0.49\textwidth]{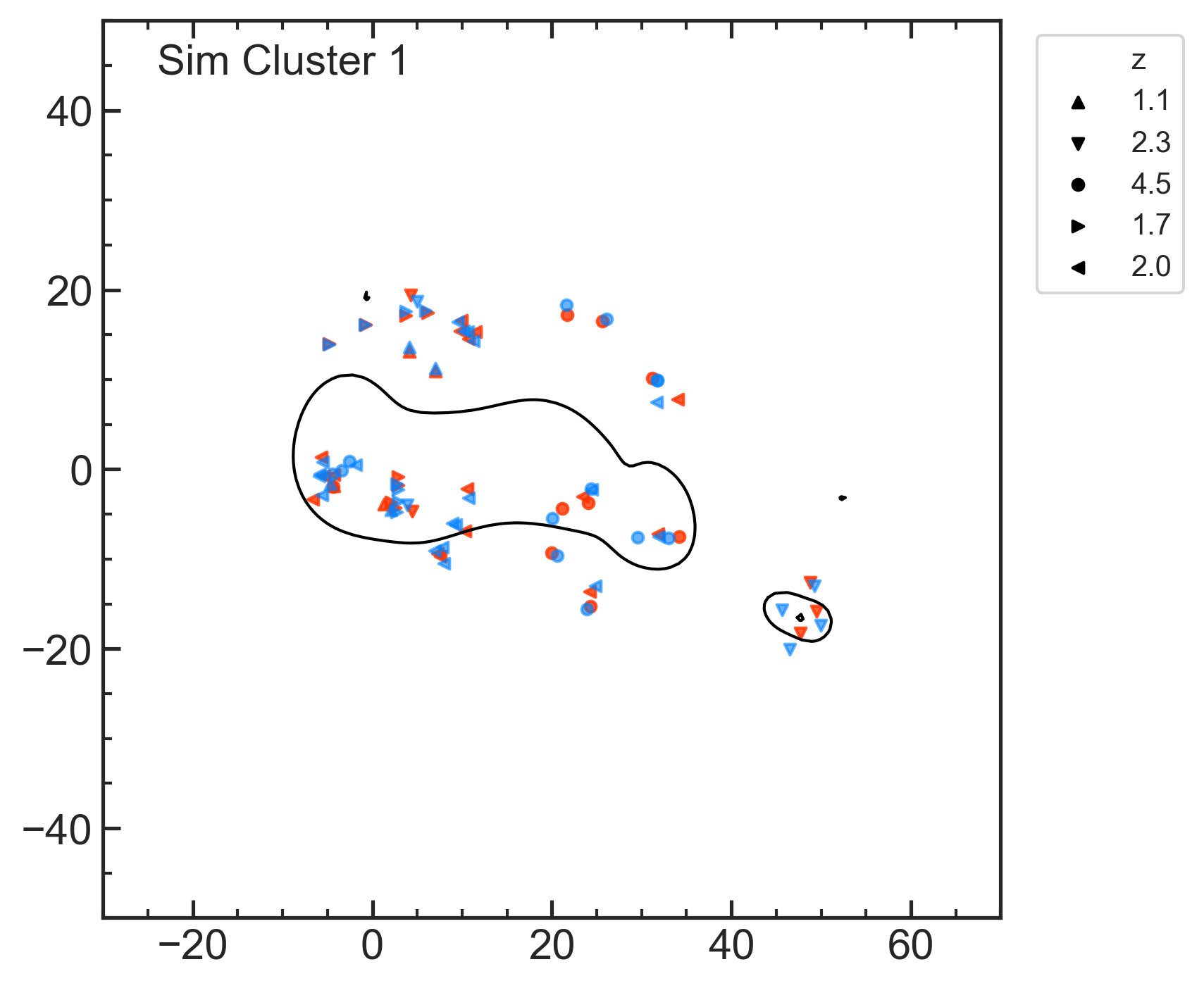}
    \includegraphics[width=0.49\textwidth]{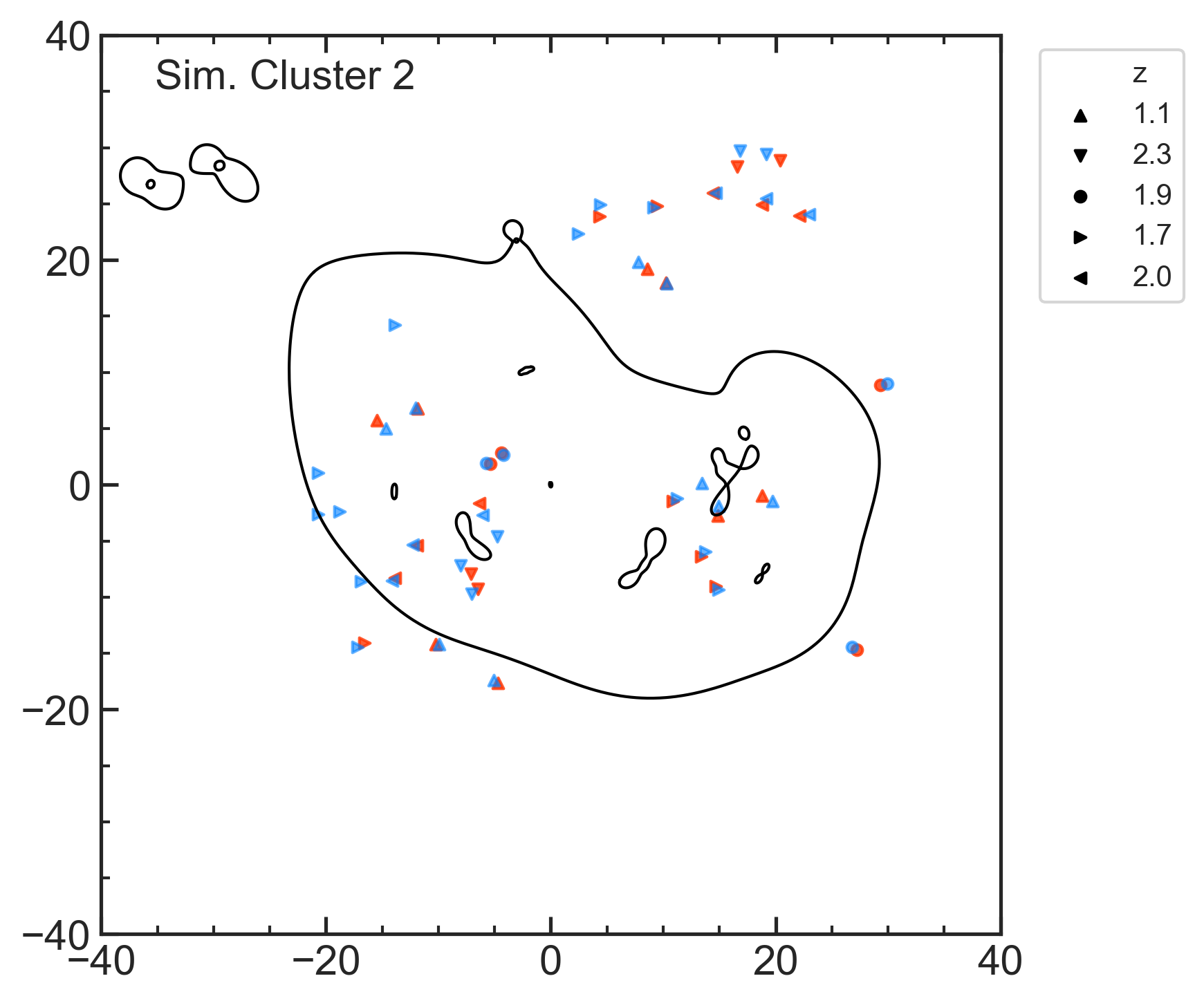}
    \includegraphics[width=0.49\textwidth]{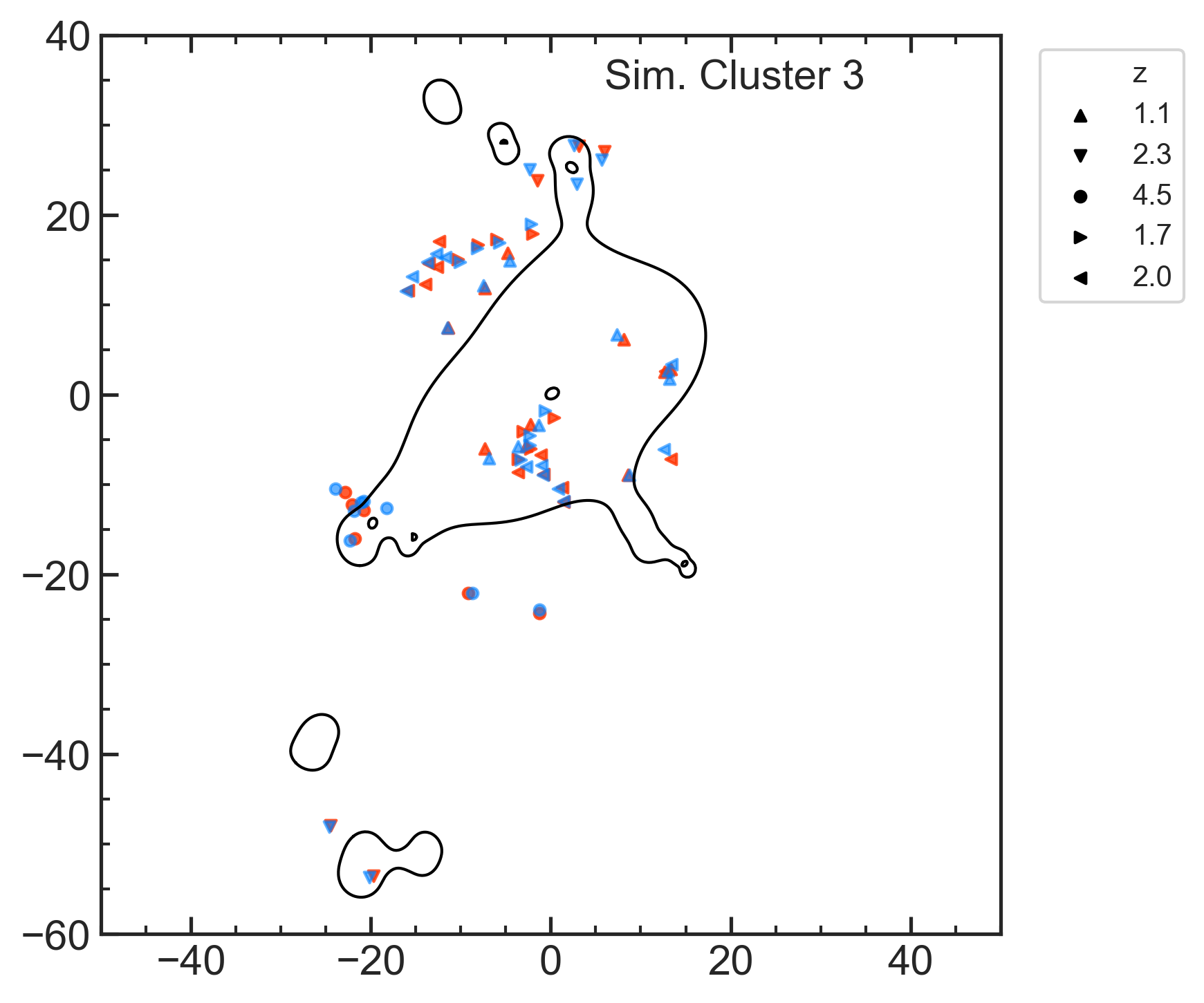}
    \label{fig:image_plane_plots2}
\end{figure*}

\subsection{Use of Spectroscopic Redshift Data}
\label{sec:redshift_data}
The data for each model is in the form of multiple image point locations, together with the redshift of each. Both spectroscopic and photometric redshift data are used, as noted in for each individual cluster. 
Where photometric data with significant confidence intervals was used, we adopted a redshift $z_a$ for each source that minimized  $\chi^2=
\sum_{i=1}^n\frac{(z_i-z_a)^2}{\sigma_i^2}$, 

where n is is the number of images in the image group, 
$z_i$ is the quoted median redshift of the image and $\sigma_i$ is the 68\% uncertainty interval for the measured redshift. In the case that the error in redshift is not given as a single value but in the form $z_{-\sigma\mathrm{low}}^{+\sigma\mathrm{hi}}$, we use ${\sigma\mathrm{hi}}$ as the error when the trial value is above the median and ${\sigma\mathrm{low}}$ when it is below.

\subsection{Use of Central Images}
Gravitationally lensed objects can produce positive-parity "central" images that are typically demagnified and near the center of the lens, and as a result are often obscured by bright objects in the central region of the field. When they are visible, they are helpful in constraining the lens parameters. Central images occur in five of the eight clusters in our sample: A2537, RXCJ2248, MS2137, A2261 and MACS1720. The \qlens software can be configured to look for central images and produce model central images for a given cluster, but the setting is the same for all sources in the cluster, even if only one source has a central image. As a result, for the five clusters with central images, our models produce some images in the central region that do not have a corresponding "observed" match in the data. These predicted images might be present but difficult to observe due to their inherently low magnification and/or proximity to the bright objects usually present in the center of clusters.

\section{Models of Simulated Clusters}
\label{sec:Models_of_Simulated_Clusters}

To test the validity of our methods, we use data from the IllustrisTNG simulation \citep{nelson2018, Springel2018, Pillepich_2017, Nelson_2017, Naiman_2018, Marinacci_2018} to simulate strong lensing in three massive TNG galaxy clusters. Their characteristics are summarized in Table~\ref{tab:simulated_cluster_list}.

\begin{table*}
	\centering
    \caption{Key cluster halo parameters of simulated lensing clusters from Illustris and fitted values from strong lensing models. All simulated clusters were given a redshift of 0.3. The Illustris parameters are from 3D fits using Colossus. We were not able to obtain a meaningful strong lensng fit with Cluster 1 due to its unrelaxed nature. The strong lensing fit $m_{200}$ and concentration values are for the main DM halo only and do not include that in subhalos. RMS position error is for the x and y coordinates, combined in quadrature.}
	\label{tab:simulated_cluster_list}
	\renewcommand{\arraystretch}{1.5}
	\begin{tabular}{lc|ccc|ccccc} 
		\hline
		  &  & \multicolumn{3}{c|}{Illustris 3D Fit} & \multicolumn{5}{c}{Strong Lensing Fit}\\
		   & & & $m_{200}$ & & $m_{200}$ & & & & RMS Position\\
        Cluster Name          & Relaxed? & ID & $(10^{14}M_{\sun})$ & Concentration &   $(10^{14}M_{\sun})$ & Concentration & Image Points & Source Points & Error (arc sec)\\
        \hline
        Cluster 1             &   no & 5941 & 7.08 & 2.96 & N/A & N/A  & 42 & 16 & 1.01   \\
        Cluster 2             &   yes & 10359 & 4.79 & 5.87 & $4.79_{-0.72}^{+0.93}$ & $8.08_{-0.84}^{+0.94}$  & 28   & 12 & 0.69  \\
        Cluster 3             &   yes & 19512 & 3.09 & 5.88 & $4.77_{-0.68}^{+0.83}$ & $7.85_{-0.81}^{+0.89}$ & 41 &  16 & 0.88  \\
		\hline
	\end{tabular}
	\renewcommand{\arraystretch}{1.}
\end{table*}

\subsection{Selection of Simulated Cluster Halos}
\label{sec:Simulated Halos}

To find analogues of our observed cluster sample, we used the largest volume simulation available at a box length of 300 Mpc, in the highest resolution for that suite of simulations, TNG300-1. While the large box size comes at the expense of resolution for DM particle mass ($5.9\times10^7 M_\odot$), the TNG300-1 suite are the best choice among the TNG simulations for studying large (cluster) scales. We used the gravo-magnetohydrodynamical suite of simulations which includes baryons, to account for the effects of baryonic matter on the DM in these halos.

To match the general range of redshifts in our observed cluster sample, we selected three of the most massive host halos as our simulated clusters from the TNG300-1 snapshot corresponding to a redshift of z = 0.3. All three clusters have masses $10^{14} M_\odot < m_{200} < 10^{15} M_\odot$. For each, the bright galaxy at the center of the potential in the central subhalo was identified as the BCG. Of the three simulated clusters, two appear relaxed and have a clearly dominant BCG, while one (Sim. Cluster 1) appears to be merging, as evidenced by a second major galaxy nearly comparable in size to the BCG and located only $\sim$120 kpc away from it. We chose to keep this cluster in our analysis to examine its impact on our inferences. The remaining two hosts have only one BCG.

Cluster strong lensing is subject to selection bias dependent upon the line of sight (LOS), because mass concentrations along specific LOS can increase the surface density and lensing strength for subhalos that depart from spherical symmetry \citep[and see also Appendix~\ref{sec:biases_in_triaxial_halos} for a discussion of concentration bias]{Clowe2004, Hennawi2007, Oguri2012,Sereno_2015}. To simulate this effect and to ensure strong lensing occurs, we shot 10 random lines of sight through each halo and used the LOS with largest central surface density. We then obtained 3D fits of each simulated halo, using the \texttt{Colossus} software package \citep{Diemer2018}. The resulting fitted values for $m_{200}$ and concentration, are listed in Table~\ref{tab:simulated_cluster_list}. Note that the LOS that produced the highest central surface density for a given cluster halo also yielded the highest concentration for that cluster halo, suggesting our simulated cluster analogs are affected by selection bias. 

\subsection{Simulated Image Production}
\label{sec:Simulated Image Production}
After choosing the LOS, we then created a surface density map of the stars, gas, and DM, at a resolution of 1.25 kpc per pixel. We noticed that the surface density map exhibited significant Poisson noise due to the finite particle size of the Illustris simulation. To partially ameliorate the noise, the map was smoothed with a Gaussian kernel, with a 2-pixel characteristic radius. We chose a 2-pixel smoothing radius as a way help reduce the statistical noise arising from the finite particle size but without substantially altering DM structure at scales relevant to our analysis. From the surface density map we calculate the scaled lensing potential:
\begin{equation}
    \psi(\vec{\theta})=\frac{1}{\pi}\int d^2 \theta' \kappa(\vec{\theta}')
\ln|\vec{\theta}-\vec{\theta}'|,
\end{equation}
where $\psi$ is the scaled potential, $\vec{\theta}$ is the deflection angle, and $\kappa$ is the convergence.
The reduced deflection angle $\vec{\alpha}$ can be found by taking the gradient of the scaled potential, i.e.,  $\vec{\alpha}=\vec{\nabla_{\theta}}\:\psi=D_L\vec{\nabla}\:\psi$, where $D_L$ is the distance from the observer to the lens. This, together with the lens equation $\vec{\alpha}=\vec{\beta}-\vec{\theta}$, allows us to solve for image positions for any given angular position $\vec{\beta}$ of a source object.

\begin{figure}
    \centering
    \caption{An example image of a simulated cluster (Sim. Cluster 3).  The BCG is centered, and the red markers indicate 41 simulated point images from 16 source points.}
    \includegraphics[width=0.46\textwidth]{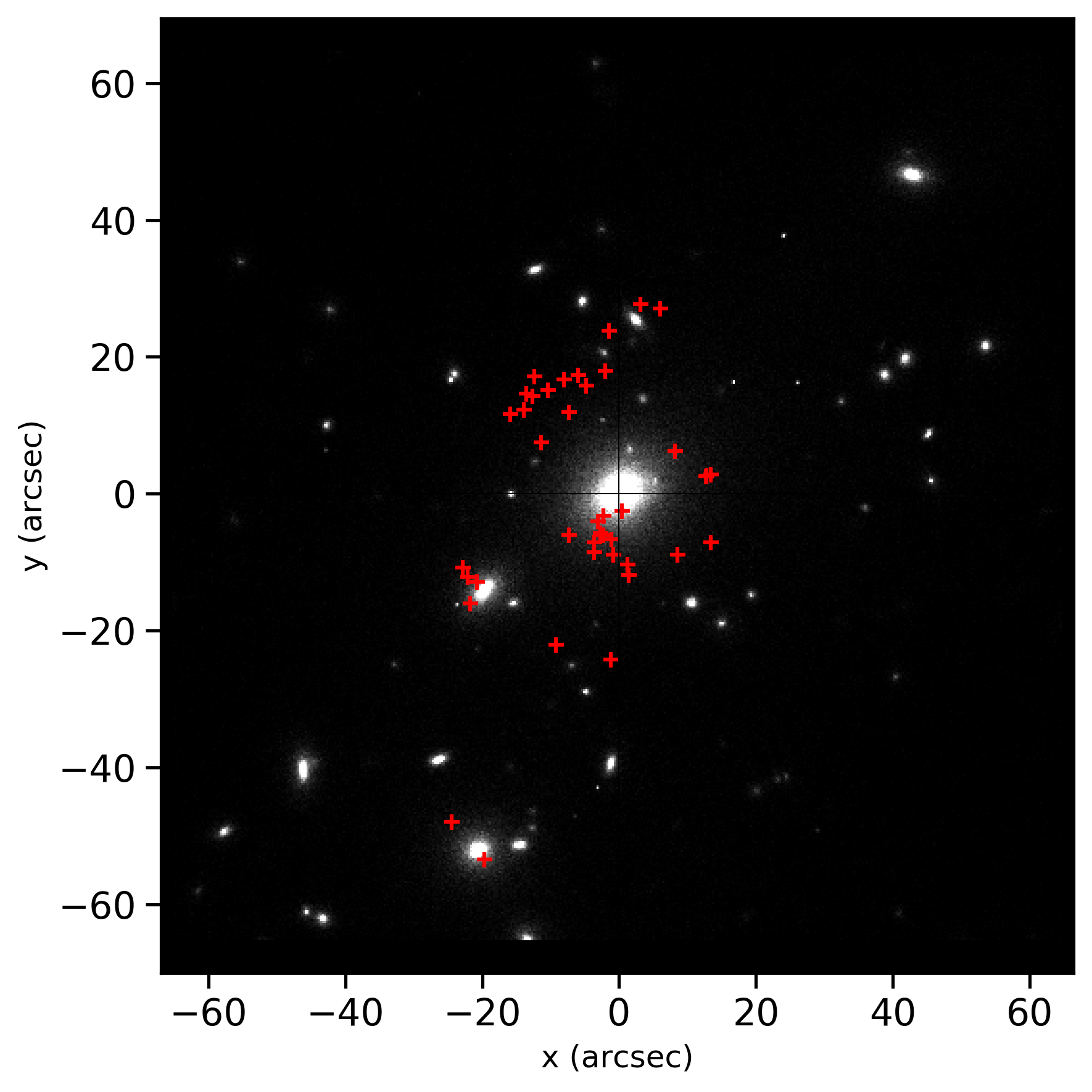}
    \label{fig:mock_halo_4_sim_image}
\end{figure}

 For each cluster, from 12 to 16 point sources were created, each generating 3 or 5 images, resulting in 28 to 41 images. There were 5 unique redshift groups for each cluster, ranging from $z=1.1$ to $z=4.5$, which represent a typical range of source redshifts found in actual clusters. Random Gaussian errors with a standard deviation of 0\farcs5 were added to each x and y coordinate of the images, resulting in a mean position error of 0\farcs71. An example of a mock image for Simulated Cluster 3 is shown in Figure~{\ref{fig:mock_halo_4_sim_image}}.

\subsection{Model Details for Simulated Clusters}
The models for each of the three simulated clusters are discussed below. Key model results are summarized in Table~\ref{tab:simulated_cluster_list}. Image plane plots are shown in Figure~\ref{fig:image_plane_plots}, showing data image locations, modeled image locations and critical curves.
\subsubsection{Cluster 1}
Cluster 1 appears to be a merging cluster, with two large central galaxies separated by approximately 120 kpc. The unrelaxed DM profile proved problematic to model. The data set had 16 source points and 42 images, and we identified 9 perturbers in addition to the two large central galaxies. We attempted several approaches; treating the smaller of the two central galaxies as an ordinary perturber did not work well, nor did using two blended cNFW halos. While we were able to reproduce many of the simulated images, there is not enough freedom in the model to accurately reproduce the DM distribution, which is crucial for accurate inference of SIDM cross section. The modeled images had a root-mean-square (RMS) position error of 1\farcs01 (x and y combined). 

\subsubsection{Cluster 2}
Cluster 2 is a relaxed cluster, and the simulated data set had 12 source points and 28 images. From the created HST-like image, we determined the photometric parameters (luminosity, center coordinates, dPIE radii, position angles and axis ratios) of the BCG and perturbers. We identified 13 perturbing galaxies, choosing those  $\leq 60$ arcsec of the BCG and with a mass $\geq 10^{10}M_{\sun}$. All but one of the perturbers were "anchored" together so that their masses scaled as one group. We individually optimized one perturber that was close to several images. All data images were reproduced by the model, however there were 3 extra images. The modeled images had an RMS position error of 0\farcs69. The surface density was reproduced with an mean precision of 0.048 dex over the radius range where images are located (i.e., from 23 kpc to 157 kpc from the BCG). 

\subsubsection{Cluster 3}\label{halo3}
Cluster 3 is a relaxed cluster with 16 source points and 41 images in the simulated data set. Using the created HST-like image, we reduced photometric measurements of perturbers and selected 10 perturbers, based on their proximity to the center and luminosity. Of these, 9 were anchored together so that their mass was varied as one, and one perturber was optimized individually. To achieve better fitting and image reproduction, we reduced the perturbers' DM halo mass normalization by 50\%. All data images were reproduced by the model, with three extra images. The RMS position error of the modeled images was 0\farcs88. The surface density was reproduced with an mean precision of 0.04 dex over the radius range where images are located (i.e., from 11 kpc to 254 kpc from the BCG).

\subsection{Discussion}
For Sim. Cluster 2, the inferred halo mass from strong lensing closely matches the value obtained from measuring binned data from the Illustris surface density projection, whereas for Sim. Cluster 3, the inferred mass is 35\% lower than the measured value. It should not be expected that the masses inferred from this process would closely match the actual values, because the strong lensing fits only the inner ~200 kpc or so, whereas the $r_{200}$ of these halos is more than 1.5 Mpc.

We measured the size of the BCGs with a 2-D fitting code, assuming an elliptical dPIE model (see Section~\ref{sec:lens profiles}). The measured scale radii values were quite small and compact, with scale radii of 0.5 kpc and 2 kpc respectively for Sim. Clusters 2 and 3. We note that these values are much smaller than the half-light radii quoted in the Illustris catalog, 66 kpc and 76 kpc, respectively. This difference could be due to the extended light distribution in the cluster. As a check, we reran the strong lensing models using the larger BCG radii. In each case, there were modest differences in some parameter posteriors, but in all cases the preferred halo core size was very small, i.e., consistent with no core.

%We plot the DM halo density profiles from strong lensing in Figure~\ref{fig:sim_SD_plot} as the colored bands, which includes DM from the main halo only. The black lines in the figure show the total DM (main halo plus subhalos) from Illustris. Although the comparison is imperfect, it does show that the inferred density profiles from strong lensing at least approximately match the simulation over the range of radii from the position of the innermost image to that of the outermost image. Unfortunately, it is difficult to disentangle the DM of the main halo from the subhalos in the simulation to provide a more direct comparison, but one would expect that excluding the subhalos would make the profiles slightly steeper as the distance from the center increases, making them even more similar to the inferred density profiles. As shown, the root-mean-square (RMS) error of the fit over the radii containing images is 11.7\% (0.048 dex) and 8.9\% (0.037 dex) for Sim. Clusters 2 and 3, respectively, as shown in Figure~\ref{fig:sim_SD_plot}.  

The median inferred concentrations were 30\% to 40\% higher than those found from the 3D fit (although within $2\sigma$). This is consistent with expectation, since the line of sight to the clusters was purposefully chosen to maximize strong lensing, making it likely that the line of sight is preferentially oriented near the major axis of the halo \citep{Clowe2004, Hennawi2007, Oguri2012,Sereno_2015}. The bias of concentration in such cases is explored in more detail in Appendix~\ref{sec:biases_in_triaxial_halos}. Another plausible contributing factor is that the concentration will be higher for a DM halo when mass from its associated subhalos are excluded, as discussed in \citet{Fielder_2020}. This is at least partially true in our models, as we model the subhalos of the largest perturbers individually.

\begin{figure}
    \centering
    \caption{Magnification maps for the simulated clusters. Regions of high magnification appear as yellow bands. The noise in the bands is due to the finite particle size of the simulation. The critical curves found from strong lensing are overlaid as cyan lines. Model image locations are shown as white circles for images matched to data, and red squares for extra images. Axes scales are in arc seconds.}
    \label{fig:mag_maps}
    \includegraphics[width=0.47\textwidth]{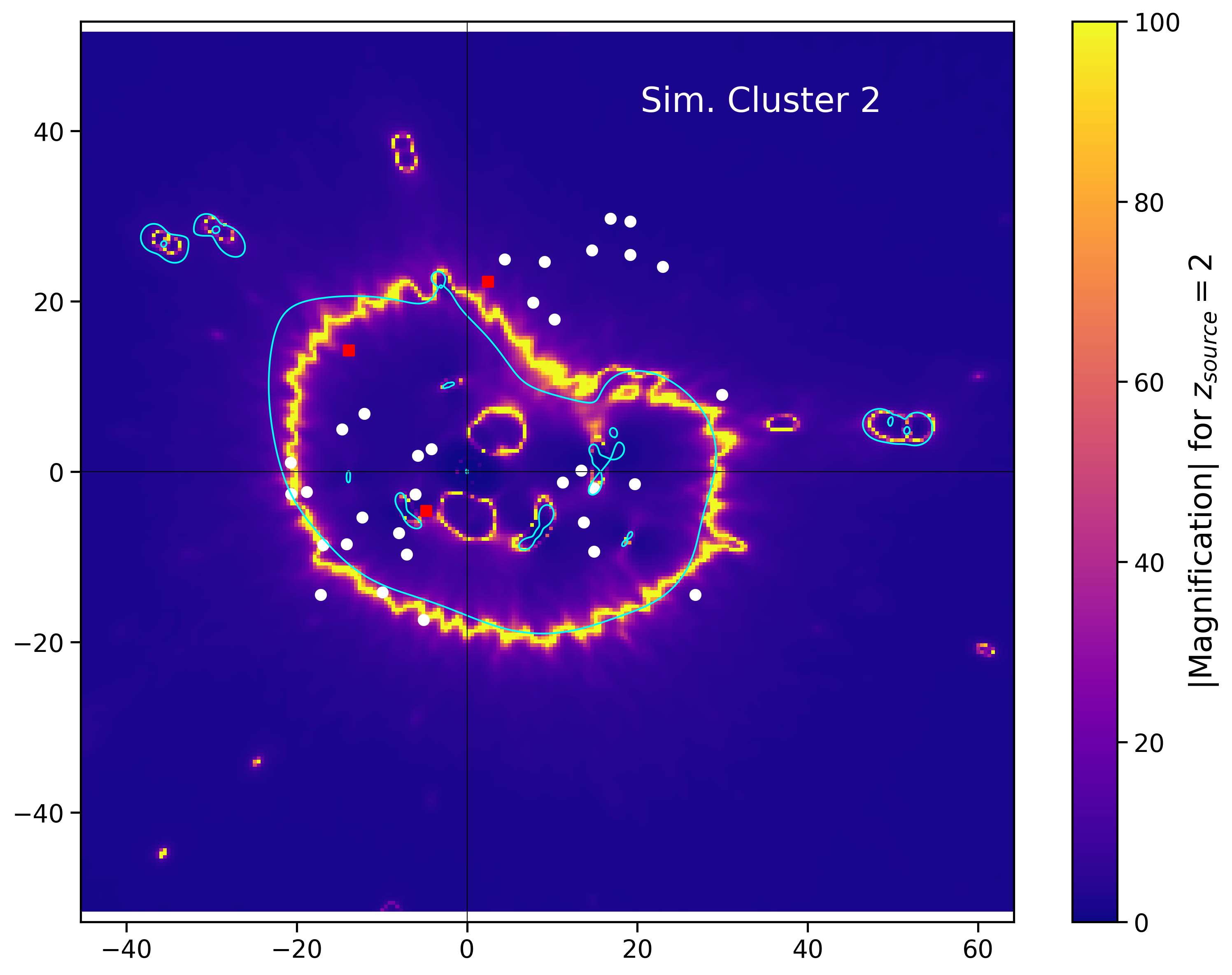}
    \includegraphics[width=0.47\textwidth]{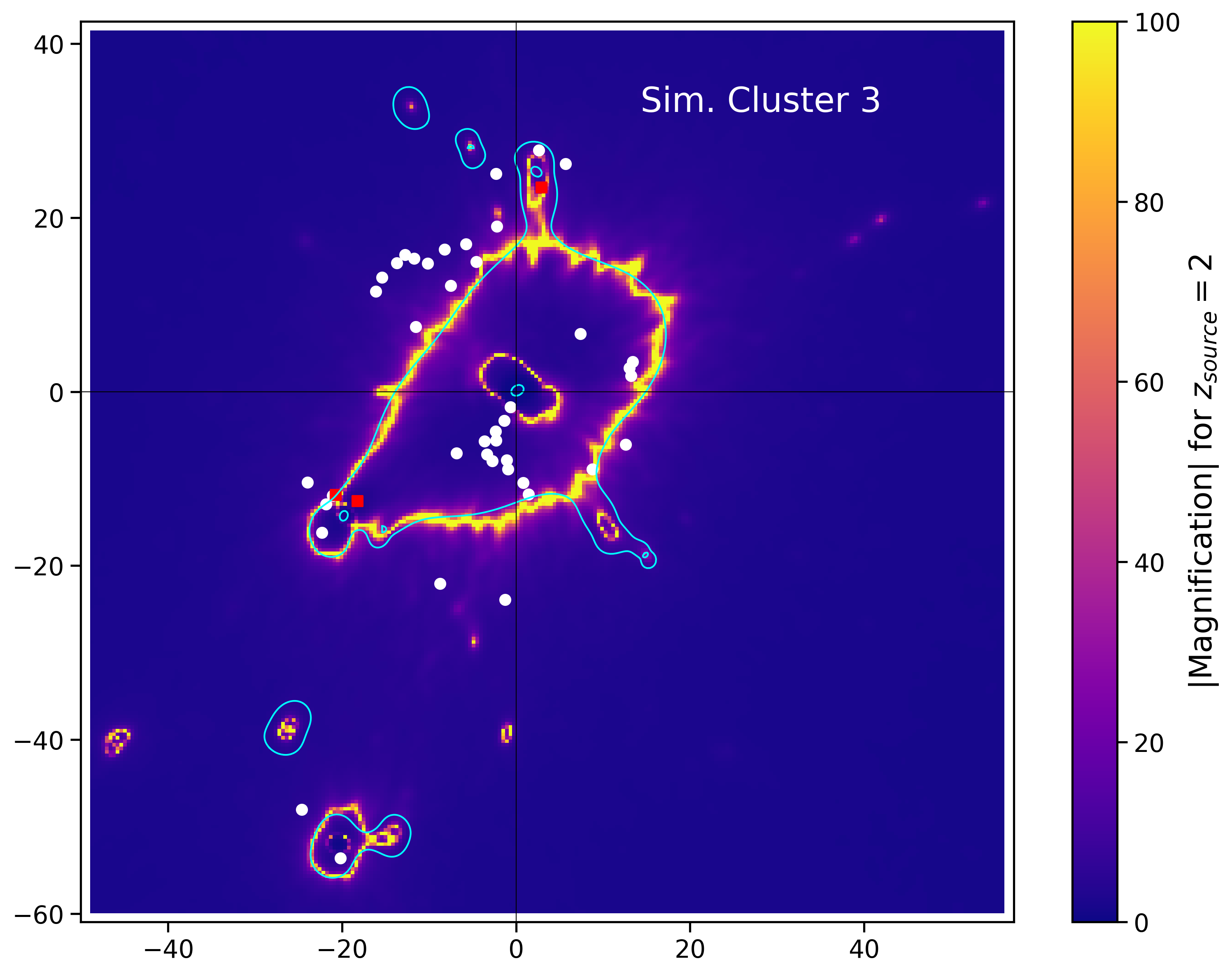}
\end{figure}

Three extra modeled images appeared in each of the relaxed cluster models. Extra images can appear if the source point is close to a caustic curve, or equivalently, if images are close to critical curves in the image plane. To check on this, we compared the critical curves generated by \qlens to the curves of high magnification as generated by the ray tracing code. This is a meaningful comparison because magnification becomes infinite at the critical curves. Figure~\ref{fig:mag_maps} show those comparisons for Sim. Clusters 2 and 3. The extra images are indeed close to critical curves, and the predicted critical curves closely follow the high magnification curves. As can be seen in the figures, the high magnification curves are noisy; this is due to the finite particle size of the Illustris simulation. The ray tracing grid has a pixel size of 1.25 kpc. The critical surface density (where $\kappa=1$, and where the tangential critical curves are located) at a lens redshift of 0.3 and assuming a source redshift of 2.0 is $\Sigma_{\mathrm{crit}} = \num{2.352e9}M_{\sun} / \mathrm{kpc}^2$. Since the mass of a DM particle in the Illustris simulation is $\num{5.9e7}M_{\sun}$, there are an average of only $\sim 60$ particles per pixel, the statistical variation of which can be seen in the magnification curves. By making changes to grid settings in the \qlens software, we could reduce the number of extra images somewhat, but at the cost of higher RMS image position error in the model, which would be undesirable. We checked that the halo and BCG parameters do not change significantly under such changes. We interpret the extra images as due to either artifacts of the modeling process or unmodeled substructure, and do not believe that they significantly impact the inference of the model parameters.

\subsection{Summary of Key Findings from Simulated Clusters}
\label{sec:summary of key findings from simulated clusters}
We used ray tracing to simulate strongly lensed images for three Illustris clusters, and tested our strong lensing analysis pipeline on those clusters. 
Our key findings for those models are summarized here.
\begin{itemize}
    \item  Although the lens models constrain only the inner $\sim 200$ kpc of the DM halo, they were nevertheless able to recover the halo mass of their corresponding simulated clusters to within 20\% to 30\%.
    \item Because of the finite particle size of the simulation, the surface density map exhibits significant Poisson noise. By smoothing the map with a Gaussian kernel with 2-pixel characteristic radius the noise was reduced. The simulation's finite particle size also causes the DM subhalos to suffer from two-body relaxation effects in their centers, effectively giving them cores. We therefore modeled those halos using a cored profile. Doing so moderately improved the accuracy of the image reproduction for Sim. Cluster 2, but yielded only a minor improvement for Sim. Cluster 3.
    \item  The critical curves produced by lens models followed those of the magnification map from the underlying potential, save for noise in the magnification map due to the simulation's finite particle size.
    \item Image positions were recovered with good accuracy, commensurate with the simulated position error added to the mock data. The total simulated position error was 0\farcs71 (0\farcs5 each in x and y), while the RMS position error of the models were 0\farcs69 and 0\farcs88 for Sim. Clusters 2 and 3, respectively.
    \item  Three extra images were produced in each of the simulated cluster models, but did not appear to significantly impact the parameter inference of the models.  
    \item The merging cluster, Sim. Cluster 1, proved difficult to model, as the mass distribution is complex and is not well characterized by a potentially cored NFW halo with additional perturbing bodies. We were not able to recover meaningful parameter values for that cluster. This highlights the importance of choosing clusters that are not major mergers for our analysis. We have been cognizant of this restriction in choosing the sample of observed clusters to model.
\end{itemize}

\section{Models of Observed Clusters}
\label{sec:Models_of_Observed_Clusters}

We began by selecting a subset of galaxy clusters, based on the following criteria:
\begin{enumerate}
    \item a generally relaxed shape, with a dominant main halo and the absence of vigorous current merger activity, based on either X-ray or optical observation;
    \item the presence of strong lensing, resulting in several multiple-image systems
    \item existing multiple image data, including redshifts,
    \item ideally, the availability of galaxy member data, to determine likely lensing perturbers, although in many cases we reduced our own perturber information from HST data.
\end{enumerate}
We targeted a sample size of approximately 8 to 10 halos, so as to enable us to draw cogent general conclusions from the sample. The resulting sample set of 8 clusters is listed in Table~\ref{tab:cluster_list}.
\begin{table*}
	\centering
    \caption{Summary of the observed cluster sample.}
	\label{tab:cluster_list}
	\begin{tabular}{lccccccc} 
		\hline
		                      & Adopted & kpc per & $t_{\text{age}}$ &        & Image and Galaxy \\
        Cluster Name          & Redshift & arcsec & (Gyr) & Relaxed? & Data Sources             & Image Points & Source Points \\
        \hline
        Abell 611          &    0.288 & 4.33 & 10.2  & yes  &  \citet{Donnarumma2011, Newman2011} & 49 & 13 \\
        Abell 2537          &   0.294 & 4.39 & 10.1 & yes* & \citet{Newman2013a, Cerny2018}; & 16 & 4 \\
        
        & & & & & this work & & \\
        
        RXC J2248.7-4431 & 0.348   &  4.92  & 9.7 & yes    & \citet{Bonamigo2018a}; this work & 55 & 20 \\
        MS 2137.3-2353            & 0.314   & 4.60  & 9.9 & yes  & \citet{Donnarumma2009}; & 50 & 14  \\
        
         & & & & & \citet{Newman2013a};this work & & \\
        
        Abell 383                & 0.189   &  3.16  & 11.1 & yes & \citet{Monna2015}; this work & 27 & 9 \\
        Abell 2261               & 0.225   & 3.61   & 10.8 & yes*   & \citet{Coe2012}; this work & 30 & 12 \\
        MACS 2129.4-0741          & 0.589    & 6.63 & 7.8 & yes* & \citet{2017MNRAS.466.4094M}; this work & 31 & 9* \\
        MACS 1720.3+3536          & 0.387   & 5.27  & 9.3 & yes  & \citet{Zitrin2013}; this work & 19 & 6  \\
		\hline
	\end{tabular}
	* but see discussion in the relevant paragraph of Section~ \ref{sec:Models_of_Observed_Clusters}\\
\end{table*}

Figure~\ref{fig:image_plane_plots} shows plots of the image plane for the cluster models, showing the data image positions, their modeled counterparts, and representative critical curves for a redshift of 2.0.

\subsection{Abell 611}
In \citet{Andrade2019} this cluster was modeled in detail, using image position and galaxy member data from \citet{Donnarumma2011}, but with redshift corrections indicated in \citet{Newman2013a, Belli2013}. The redshift data is spectroscopic. In \citet{Andrade2019}, both cNFW and Corecusp halo models were explored, which gave similar results, with the cNFW model being the preferred model. 

In this work, our approach is similar, except that we model DM halos for each of the perturbers, as described in Section~\ref{sec:DM_Subhalo_Masses_and_Radii}. The BCG, the seven perturbing galaxy members and their associated DM halos were modeled with dPIE profiles. The BCG mass parameter was a varied parameter, while the other BCG parameters were fixed based on the photometry from \citet{Newman2013a}. The mass parameters for the perturbing members were varied in three groups, one each for perturbers 1 and 2, and one group for perturbers 3 through 7. The normalization of the stellar-halo mass relation (i.e., parameter "k" in Equation~\ref{eq:power law}) was reduced by 75\%, to allow for the lowest $\chi^2$ while still matching all data and model images. The resulting model reproduced all data images, with no extra images.

\subsection{Abell 2537}
\label{Abell 2537}
Abell 2537 is an efficient gravitational lens that has been studied by several others, including \citet{Newman2013a,Newman2013b, Cerny2018}. It appears relaxed and uniform in X-ray images \citep{Schmidt2007}. \citet{Newman2013a} describes the cluster as likely disturbed, perhaps along the line-of-sight, but we nevertheless were able to construct a satisfactory model that explains the image positions with reasonably good fidelity. The stellar-halo mass normalization "k" was reduced to 12.5\% of its nominal value, to minimize the image position $\chi^2$.

Image positions, spectroscopic redshift and BCG photometry data from \citet{Newman2013a} were used in the model. Perturbing galaxy data from \citet{Cerny2018} was also utilized. Only those perturbers with V-band luminosity greater than $\num{5e10}~L_{\sun}$ and located less than 60\arcsec~from the BCG were included in the model, resulting in 32 perturbers. The BCG mass and one anchor galaxy mass (to which the other 31 perturbers are anchored) were varied parameters. There were three extra non-central images in the model. Those images are in the vicinity of a bright perturbers and could be present but washed out in the perturber light. 

\subsection{RXC J2248.7-4431}
Also known as Abell 1063S, this cluster has been previously modeled in some detail by \citet{Caminha2016} and \citet{Bonamigo2018a}. It appears relaxed and uniform in X-ray images \citep{Schmidt2007}. We used data from \citet{Bonamigo2018a} for the image positions, redshifts and perturbers. We included only those perturbers with V-band luminosity greater than $\num{2e10}~L_{\sun}$ and a distance from the BGC less than 60\arcsec, resulting in 13 perturbers. The mass of the BCG and one anchor galaxy (to which the other 12 perturbers were anchored) were varied. The stellar-halo mass normalization "k" was not reduced from its nominal value.

The best-fit models for this cluster produced 7 extra images. Examination of the HST image \citep{Postman_2012} revealed what may be image candidates at most of the predicted locations, but without spectroscopic data we cannot confirm them. 

\subsection{MS 2137.3-2353}
MS2137 appears very relaxed and uniform in X-ray images \citep{Schmidt2007}. We adopt the image positions and redshift data from \citet{Donnarumma2009}. The point image locations follow two great arcs at nearly identical spectroscopic redshifts. BCG photometry from \citet{Newman2013a} was used. There is one prominent perturber, the position and photometry data of which we reduced using HST data \citep{Postman_2012}. The mass of the BCG and perturber were allowed to vary, while the other parameters were fixed, based on photometric measurements. The stellar-halo mass normalization parameter "k" was reduced by 50\% to achieve better fitting and image reproduction. All data images were reproduced, with no extras.

\subsection{Abell 383}\label{A383 section}
Abell 383 is a relaxed cluster with 9 lensed sources for a total of 27 images. Following \citet{Monna2015}, we also adopt 19 perturbers for this cluster. For the BCG and all other cluster members, the photometric parameters were reduced from the HST F814W filter \citep{Postman_2012} using SourceExtractor provided in the Gaia software package. As in \citet{Monna2015} we use the GR galaxy as our reference galaxy and anchor all other perturber masses to the GR galaxy for optimization. Due to localized radial arcs near the G1 and G2 perturbers, we individually optimize these perturbers in addition to the BCG and GR. The stellar-halo mass normalization "k" was reduced to 25\% of its nominal value, to minimize the image position $\chi^2$. The model reproduced all data images but yielded 6 extra images. Examination of the HST image for potential new images at their predicted locations was inconclusive due to noise in the image.

\subsection{Abell 2261}
Abell 2261 is a borderline relaxed cluster with 12 strongly lensed sources for a total 30 images (see Table 3 of \citet{Coe2012}). Our search of past work on this cluster did not yield data on potential perturber membership in the cluster. We therefore adopted 12 perturbers by choosing the galaxies which distort critical curves in Figure 1 of \citet{Coe2012}. We measured effective radius, axis ratio, luminosity and position angle from the HST F775W filter \citep{Postman_2012} using SourceExtractor provided in the Gaia software package for the BCG and perturbers. We individually optimized 4 perturbers which are located in close proximity to images; the others were grouped and optimized as one. The stellar-halo mass normalization "k" was reduced to 50\% of its nominal value, to minimize the image position $\chi^2$. The model was able to match all data images with an RMS position error of 0\farcs83, however 8 extra non-central images with |magnification|>1 were produced. 

\subsection{MACS 2129.4-0741}\label{sec:MACS2129}
MACS2129 is described by \citet{2012MNRAS.420.2120M} as a recent but well separated merger, although it appears relaxed. It has 8 lensed sources and a total of 31 images according to \citet{2017MNRAS.466.4094M} (see Table 3 there). However, system 4 is much better characterized as a system of two images being lensed rather than one image being lensed, thus we use 9 sources as in \citet{2017MNRAS.466.4094M}, where they also make a distinction between the two images in system 4. We adopt 10 perturbing galaxies, two of which are clear strong lensing sources that impact image positions significantly. The masses of those two were optimized individually. The masses of the other perturbers were anchored to a reference galaxy, the mass of which was varied and optimized. The stellar-halo mass normalization "k" was reduced to 12.5\% of its nominal value, to minimize the image position $\chi^2$. The model reproduced 30 out of 31 data images, with no extra images. The model matched data images with an RMS position error of 0\farcs79. The radii, position angle, and axis ratio of all cluster members were measured from the HST F814W filter \citep{2007ApJ...661L..33E}.

\subsection{MACS 1720.3+3536}
MACS1720 also appears relaxed and uniform in X-ray images \citep{Schmidt2007}. We use the image position and photometric redshift data of \citet{Zitrin2015}. A single redshift value for each image group was selected by the photo-z optimization process described in Section~\ref{sec:redshift_data}. Image set 7 from \citet{Zitrin2015}, a 3-image set in the far southern part of the image, was problematic to model. It contains a pair of images within close proximity to one another, but no apparent perturbing bodies nearby. In addition, the photometric redshift of this source has a high uncertainty. We therefore excluded that source point. Interestingly, we were able to accurately produce the three images in Image set 7 if we included a perturbing subhalo of mass $\sim\num{2e12}M_{\sun}$ at coordinates of (12.5, -25.5) arcsec relative to the BCG. However, inspection of the HST images does not reveal any significant luminous body at that location.

BCG photometry and perturber data were reduced from HST images \citep{Postman_2012}. We included only those perturbers with F814W luminosity greater than $\num{5e9}~L_{\sun}$ and a distance from the BGC less than 30\arcsec, resulting in 13 perturbers. The masses of these 13 were varied as a group. The stellar-halo mass normalization "k" was not reduced from its nominal value. The resulting model had one extra image with |magnification| > 1.0, a magnification 2.7 image located at (4.2, -3.5) arcsec relative to the BCG. Examination of the HST reveals a good candidate object near that location, although we cannot confirm it absent redshift measurement. 

\section{Strong Lensing Results}
\label{sec:Strong_Lensing_Results}
Using the methods described in Section~\ref{sec:Strong_Lens_Modeling}, we constructed mass models for each cluster that reproduced the observed image positions. We used separate lens elements for the baryonic and dark mass components of member galaxies, which is not a common approach in cluster lensing analysis. The strong lensing models were able to reproduce the image position data with good accuracy. We assumed measurement error of 0\farcs5 in each of the x and y coordinates of the data image positions. The models recovered the data image positions with root-mean-square position errors (combining the x and y components) ranging from 0\farcs32 to 1\farcs07, with a median of 0\farcs66, as shown in Table~\ref{tab:strong_lensing_results}. The reduced image plane $\chi^2$ (i.e., $\chi^2$ per degree of freedom) for the fits ranged from 0.36 to 5.79. The degrees of freedom are counted as follows: two for each image point (one each for the x and y components), less two for each source point as they are, in essence, free parameters, less one for each varied parameter in the MCMC model. For the 8 observed clusters, 5 have reduced chi square statistics < 2, indicating that the extent of the match between data and observation for those models is generally in accordance with the error variances. For the other 3 (Abell 2261, MACS 2129 and MACS 1720), this indicates that the assumed errors may have been underestimated. Systematic error is the likely dominant component, which is discussed in Section~\ref{sec:SL systematic errors}. 

All data images were reproduced except one. Occasionally, extra images were produced by the models, as noted in the relevant paragraphs of Section~\ref{sec:Models_of_Observed_Clusters}.  Posterior distributions for all the cluster models are shown in the supplemental online material.

\begin{figure*}
    \centering
    \caption{Histograms for 2-dimensional BCG offset, BCG mass-to-light ratio and DM halo core radius posteriors. The histograms are normalized to unity area, and have been smoothed slightly with a Gaussian kernel for display purposes.}
    \includegraphics{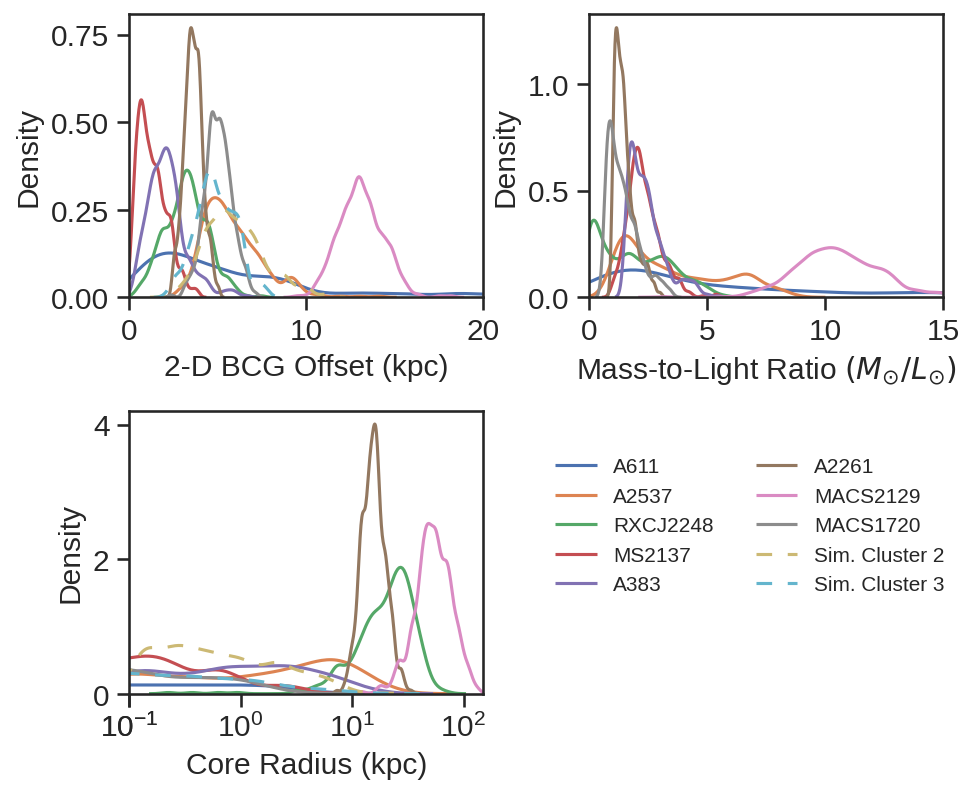}
    \label{fig:various_histograms}
\end{figure*}

\begin{figure}
    \centering
    \caption{Surface density inferred from strong lensing ("data", in color) and the corresponding median SIDM modeled surface density ("model", in black) versus radius for each real cluster. The area shown for each is the 68\% confidence interval. The clusters are offset by one decade each for display purposes.}
    \includegraphics[width=0.47\textwidth]{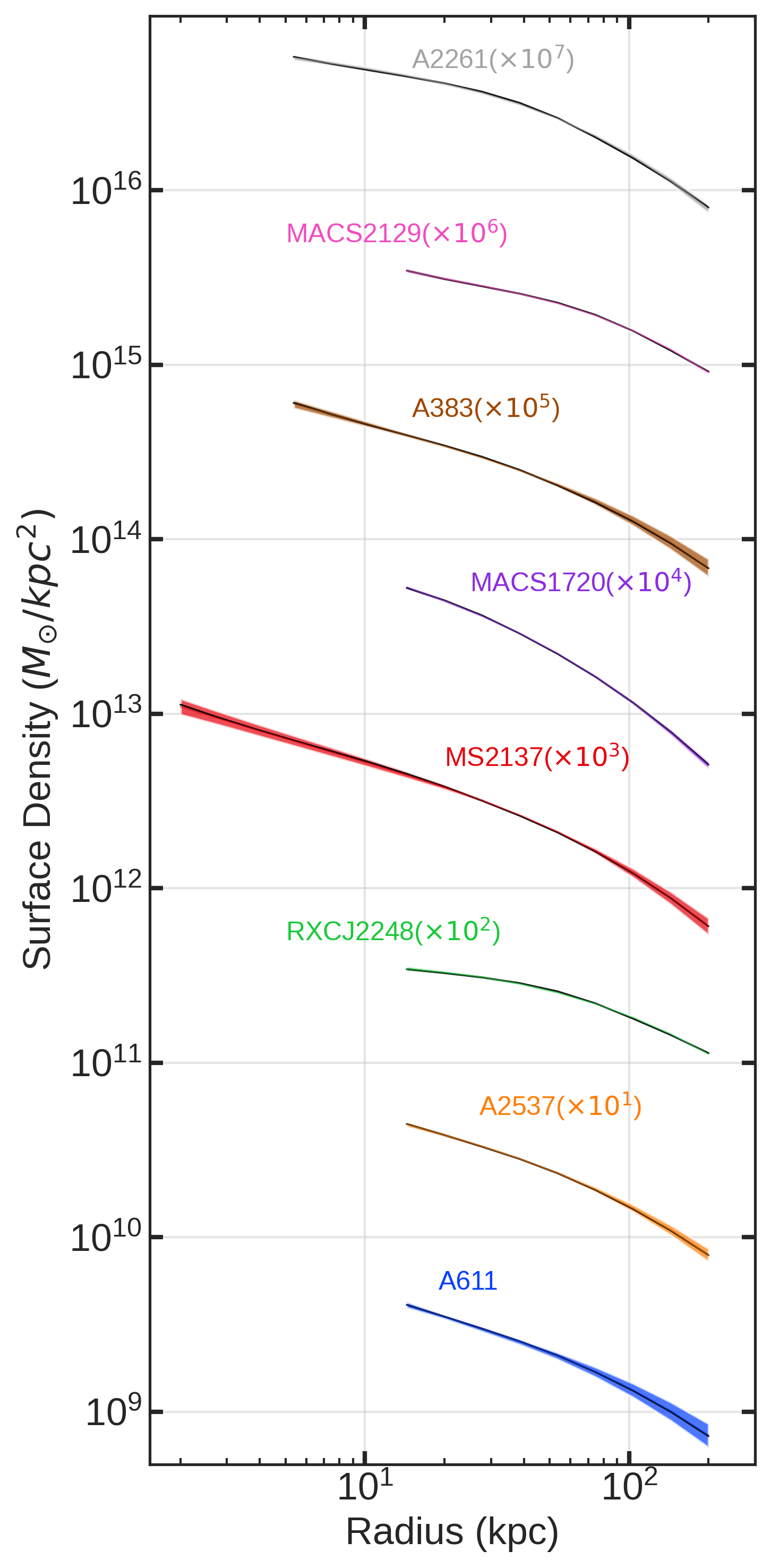}
    \label{fig:compare_SIDM_and_SL_kappa}
\end{figure}

\begin{table*}
	\centering
    \caption{Strong lensing results summary. Where confidence intervals are given, the 14th- and 86th-percentile values are indicated. The parameter $r_c$ is the cNFW core radius. The BCG luminosity is measured using the closest HST filter to the rest frame wavelength, so as to be approximately comparable to V-band. Measurement uncertainties in luminosity are <2\%. The halo offset from the BCG is the inferred projected offset in the plane of the sky.}
	\label{tab:strong_lensing_results}
	\renewcommand{\arraystretch}{1.3}
	\begin{tabular}{lccccccccc} 
		\hline
		& & RMS Pos. & & & Halo & & BCG & 2-D Halo  \\
		                      & Img. Plane  & Error & $r_c$  &  Halo Mass & Concen- & BCG Mass & Lum. & BCG M/L & Offset from  \\
        Cluster Name          & $\chi^2$[red.] & (arcsec) & (kpc)  & ($10^{14}M_{\sun}$) & tration & ($10^{12}M_{\sun}$) & ($10^{11}L_{\sun}$) & ($M_{\sun}/L_{\sun}$)& BCG (kpc) \\
        \hline
        Abell 611          & 20.7 [0.36] & 0.32  & $0.52_{-0.48}^{+6.48}$  & $8.77_{-2.30}^{+3.34}$ &  $6.20_{-1.05}^{+1.17}$ &$1.21_{-0.88}^{+2.74}$ & 3.25 & $3.72_{-2.71}^{+8.43}$ & $4.18_{-2.80}^{+5.75}$ \\
        Abell 2537          & 22.0 [1.57] & 0.59  & $1.18_{-1.12}^{+6.76}$ & $9.42_{-1.42}^{+1.77}$ & $6.77_{-0.76}^{+0.90}$ &$0.93_{-0.48}^{+1.09}$ & 3.32 & $2.79_{-1.43}^{+3.28}$ & $5.44_{-1.10}^{+1.18}$ \\
        RXC J2248.7-4431 & 116.8 [1.98] & 0.73 & $19.83_{-9.41}^{+13.03}$ & $18.39_{-1.30}^{+1.49}$  & $ 5.39_{-0.59}^{+0.71}$ & $0.82_{-0.74}^{+0.88}$ & 4.80 & $1.70_{-1.54}^{+1.84}$ & $3.11_{-1.17}^{+1.27}$ \\
        MS 2137.3-2353            & 39.0 [0.64] & 0.44 & $0.10_{-0.08}^{+0.52}$  & $5.85_{-0.96}^{+1.22}$  & $ 7.91_{-0.99}^{+1.13}$ & $0.72_{-0.19}^{+0.88}$ & 3.26 & $2.22_{-0.59}^{+0.75}$ &  $1.22_{-0.69}^{+0.96}$ \\
        Abell 383                & 24.9 [1.08] & 0.48 & $0.39_{-0.36}^{+2.61}$  & $7.79_{-1.38}^{+1.92}$ & $6.71_{-0.89}^{+1.01}$ & $1.05_{-0.33}^{+0.40}$ & 3.22 & $3.29_{-1.02}^{+1.25}$ & $2.53_{-1.13}^{+1.63}$ \\
        Abell 2261               & 83.5 [3.34] & 0.83 & $15.53_{-3.28}^{+4.43}$ & $8.23_{0.82}^{+0.98}$ &  $ 10.10_{-0.99}^{+1.08}$ & $0.77_{-0.16}^{+0.30}$ & $5.64$ & $1.36_{-0.28}^{+0.53}$ & $3.54_{-1.24}^{+4.22}$ \\
        MACS 2129.4-0741          & 75.6 [2.44]  & 0.79 & $54.80_{-17.57}^{+22.98}$ & $10.70_{-0.64}^{+0.74}$ & $ 7.23_{-0.99}^{+1.13}$ & $2.13_{-0.36}^{+0.38}$ & 2.03 & $10.49_{-1.75}^{+1.89}$ & $12.99_{-1.18}^{+1.30}$ \\
        MACS 1720.3+3536          & 86.8 [5.79] & 1.07  & $0.04_{-0.04}^{+0.24}$  & $4.28_{0.27}^{+0.31}$  & $ 10.83_{-0.66}^{+0.65}$ & $0.38_{-0.14}^{+0.25}$ & 3.01 & $1.27_{-0.46}^{+0.84}$ & $5.06_{-0.83}^{+0.82}$ \\
		\hline

	\end{tabular}
	\renewcommand{\arraystretch}{1.}
\end{table*}

Strong lensing parameter inferences are summarized in Table~\ref{tab:strong_lensing_results}. Figure~\ref{fig:various_histograms} shows histograms for the 2-dimensional BCG offset, mass-to-light ratio, and Core Radius. Below we discuss the parameter inferences for each cluster for these key parameters: DM halo mass (i.e., $m_{200}$), BCG mass, and core radius ($r_c$).  

\subsection{Abell 611}
Our inference of the mass of the central DM halo is $8.77_{-2.30}^{+3.34}\times10^{14}\;M_{\sun}$. In \citet{Newman2009} they infer a value of $6.2_{-0.5}^{+0.7}\times10^{14}\;M_{\sun}$ using a combination of weak lensing, strong lensing and kinematic data and employing a generalized NFW profile. In a subsequent work, \citet{Newman2013a} found $m_{200}$ of $8.31_{-1.23}^{+1.46} \times10^{14}\;M_{\sun}$ using weak and strong lensing. In \citet{Donnarumma2011} they use an NFW profile for the DM halo and infer strong lensing $m_{200}$ values of $4.68 \pm 0.31 \times10^{14}\;M_{\sun}$ using a dPIE profile for the BCG, and $6.32_{-0.23}^{+0.51}\times10^{14}\;M_{\sun}$ using a Sersic profile for the BCG. They arrive at an $m_{200}$ of $9.32 \pm 1.39 \times10^{14}\;M_{\sun}$ using X-ray gas temperature techniques. Our inference is consistent with all of these.

Our inferred BCG mass is $1.21_{-0.88}^{+2.74} \times10^{12}\;M_{\sun}$. In \citet{Newman2009} they infer a BCG mass of $1.01_{-0.29}^{+0.24} \times10^{12}\;M_{\sun}$, while in \citet{Donnarumma2009} they infer $6.17_{-1.79}^{+1.32} \times10^{12}\;M_{\sun}$ for their case 6, which is the most comparable to our configuration. Our inference falls in the middle, and is consistent with the \citet{Newman2009} result but in mild tension with the \citet{Donnarumma2009} result.

We infer a cNFW core radius of $0.52_{-0.48}^{+6.48}$ kpc, which is consistent with the findings of \citet{Newman2013b}, in which they infer a core size of $2.95_{-2.01}^{+4.29}$ kpc.

\subsection{Abell 2537}
For this cluster we infer a central DM halo mass of $9.42_{-1.42}^{+1.77} \times10^{14}\;M_{\sun}$. In \citet{Newman2013a} they infer a somewhat higher value of $13.1_{-1.2}^{+1.3} \times10^{14}\;M_{\sun}$ using a purely NFW profile and combining strong a weak lensing. In \citet{Cerny2018}, they employ a model with two dPIE DM halos with different center coordinates, and find a mass within 400 kpc of the BCG of $2.6 \pm 0.5 \times10^{14}\;M_{\sun}$. Our inferred value for mass within that radius is nearly the same, with a median value of $2.5 \times10^{14}\;M_{\sun}$.

Our inference for the BCG mass is $0.93_{-0.48}^{+1.09}\times10^{12}\;M_{\sun}$. Other authors do not specify their inferred BCG masses, but we note that in \citet{Newman2013a} they measure a luminosity of $5.86 \times10^{11}\;M_{\sun}$, and in  \citet{Cerny2018} they measure a luminosity of $5.27 \times10^{11}\;M_{\sun}$, each of which when combined with our inferred mass results in a reasonable mass-to-light ratio of approximately 1.6 to 1.8.

For Abell 2537, we infer a modest core radius of $1.18_{-1.12}^{+6.76}$ kpc. This is in tension with the inference of \citet{Newman2013b} of $46.8_{-19.2}^{+34.5}$ kpc. We are unsure of the reason for the discrepancy, but note that the measured half-light radius of the BCG is 15.7 kpc, which goes beyond the DM core radius, increasing the possibility for degenerate solutions between the two. Examination of the posterior plot (see supplemental online material) shows a mild bimodality in the BCG mass, possibly also admitting a larger core size. Note also that this cluster has the fewest images in the sample (16), and the fewest source points (4). The image position data is the same as that used in \citet{Newman2013a}. A small data set could potentially limit the ability of the model to accurately constrain the mass profile.

\subsection{RXC J2248.7-4431}
For this cluster we infer a central DM halo mass of $18.39_{-1.30}^{+1.49}\times10^{14}\;M_{\sun}$. In \citet{Caminha2016}, in their reference case 2 they find a mass of $2.90 \pm 0.02\times10^{14}\;M_{\sun}$ within an aperture of 250 kpc. The corresponding median value for our inference would be very similar; $2.88\times10^{14}\;M_{\sun}$. In \citet{Bonamigo2018a}, they infer an $m_{200}$ of $20.3 \pm 6.7 \times10^{14}\;M_{\sun}$, in good agreement with our inference. We infer a BCG mass for RXCJ2248 of $0.82_{-0.74}^{+0.88}\times10^{12}\;M_{\sun}$. We were unable to find comparable inferences from other authors. We infer a core radius of $19.83_{-9.41}^{+13.03}$ kpc, and we were not able to find comparable cNFW core size inferences from other authors.

\subsection{MS 2137.3-2353}
We infer a central DM halo mass of $5.85_{-0.96}^{+1.22}\times10^{14}\;M_{\sun}$ for this cluster. In \citet{Donnarumma2009}, the found an $m_{200}$ of $4.4_{-0.5}^{+0.6}\times10^{14}\;M_{\sun}$ using a strong lensing model, and \citet{Newman2013a} found an $m_{200}$ of $3.63_{-0.81}^{+1.26}\times10^{14}\;M_{\sun}$ using a combination of strong and weak lensing. Both of these figures are consistent with our inference.

While we infer a BCG mass of $0.72_{-0.19}^{+0.88}\times10^{12}\;M_{\sun}$, other authors do not report their BCG masses. However, we note that our meaured BCG luminosity of $3.26 \times10^{11}\;L_{\sun}$ and that measured in \citet{Newman2013a} of $3.20 \times10^{11}\;L_{\sun}$ are similar, and result reasonable mass-to-light ratios of approximately 2.2.

We infer a very small core radius of $0.10_{-0.08}^{+0.52}$ kpc, as does \citet{Newman2013b}, at $2.82_{-2.39}^{+3.01}$ kpc.

\subsection{Abell 383}

For this cluster we infer a central DM halo mass of $7.79^{+1.92}_{-1.38}\times10^{14}\;M_{\sun}$. In \citet{zitrin2011} they find a similar mass of $7.67\times10^{14}\;M_{\sun}$ from a generalized NFW profile. In addition, we find the 2D projected mass on the halo within 50 kpc, the distance to the large tangential arc of systems 1 and 2, is $2\times10^{13}\;M_{\sun}$. This value is consistent with $1.7\times10^{13}\;M_{\sun}$,~$2.2\times10^{13}\;M_{\sun}$, and $2\times10^{13}\;M_{\sun}$ from \citet{Monna2015}, \citet{Newman2011}, and \citet{zitrin2011} respectively.

We infer a core radius of $6.66^{+6.89}_{-4.20}$ kpc for the DM halo. We were not able to find comparable cNFW core size inferences from other authors. Our inferred BCG mass is $1.05^{+0.40}_{-0.33}\times10^{12}\;M_{\sun}$ which is lower than the results \citet{Monna2015} of $6.05\times10^{12}\;M_{\sun}$ or $6.13\times10^{12}\;M_{\sun}$, depending if they included or did not include velocity dispersion measurements. We note that our measured luminosity of $3.22\times 10^{11} L_{\sun}$ when coupled with our inferred BCG mass results in a reasonable mass-to-light ratio of $\sim 3.3$.

\subsection{Abell 2261}
For cluster Abell 2261, we infer a central DM halo mass of $8.23_{-0.82}^{+0.98} \times10^{14}\;M_{\sun}$. In \citet{Coe2012}, they found $m_{200}$ of $9.54_{-0.84}^{+0.84} \times10^{14}\;M_{\sun}$ by using a combination of weak lensing and strong lensing when assuming a spherical NFW profile halo, which is consistent with our inference.

We inferred a BCG mass of $0.77_{-0.16}^{+0.30}\times10^{12}\;M_{\sun}$, and a core radius of $15.53_{-3.28}^{+4.43}$ kpc. We were unable to find inferences from other authors for these parameters. 

\subsection{MACS 2129.4-0741}
For this cluster we infer a central DM halo mass of $10.70^{+0.74}_{-0.64}\times10^{14}\;M_{\sun}$. In \citet{2017MNRAS.466.4094M}, they specify the mass enclosed within the Einstein Parameter, $\Theta_{E}$, given as $29\pm4$". Within this radius, they find a mass of $8.6\pm0.6\times10^{13}\;M_{\sun}$. Using our model, we find a corresponding median value of $17.5\times10^{13}\;M_{\sun}$ within that same radius. 

Similarly, our model infers a median BCG mass within $\Theta_{E} = 29$" of $2.03\times10^{12}\;M_{\sun}$, whereas \citet{2017MNRAS.466.4094M} finds $8.4\pm2\times10^{12}\;M_{\sun}$.

This cluster has the largest inferred core size in our sample, at $54.8_{-17.6}^{+23.0}$ kpc. In \citet{2017MNRAS.466.4094M} they find an even larger core radius of $101^{+13}_{-11}$ kpc. However, as noted in \citet{2017MNRAS.466.4094M}, the DM halo core radius correlates with the BCG mass profile, thus they speculate that the core radius may be large because of an overestimate of the BCG mass.

\subsection{MACS 1720.3+3536}
For this cluster we infer a central DM halo mass of $4.28_{0.27}^{+0.31}\times10^{14}\;M_{\sun}$. In \citet{Zitrin2015} they use strong lensing to specify the mass enclosed within 136" of the center. They find a value of $3.35 \times10^{14}\;M_{\sun}$ for their NFW model. The corresponding median value for our model is a similar $3.57 \times10^{14}\;M_{\sun} $. Note that our model is very close to NFW, as the inferred core radius is quite small at $0.04_{-0.04}^{+0.24}$ kpc, the smallest in our sample. We can find no other works that present estimates of the core radius or BCG mass for this cluster.

\subsection{Halo and BCG masses}
The primary results of the strong lensing models are the surface density profiles, which are shown as the colored bands in Figure~\ref{fig:compare_SIDM_and_SL_kappa}. These profiles are used as data for the SIDM profile matching described in Section~\ref{sec:SIDM_Halo_Analysis}. The black lines in Figure~\ref{fig:compare_SIDM_and_SL_kappa} are the surface density profiles as modeled by the SIDM halo matching. The match is better than 0.1 dex for all clusters.

The posterior distributions of halo core radius, halo mass and BCG mass are shown in Table~\ref{tab:strong_lensing_results}. (Note that these inferences are not used for the SIDM halo analysis.) Our DM halo core inferences range from essentially zero (i.e., MS2137, MACS1720) to more than 50 kpc (MACS2129). Core size is an important characteristic in this analysis, as a small core rules out strong self-interactions.

We infer DM halo masses in the range of $\num{4e14}M_{\sun}$ to $\num{1.8e15}M_{\sun}$. These are broadly consistent with those found by other authors. We infer BCG masses in the range of $\num{7e11}M_{\sun}$ to $\num{2.1e12}M_{\sun}$. These are are generally consistent with those found for these objects by other authors, with the exceptions of A383 (where our inference is lower than that of \citet{Monna2015}) and MACS2129 (where our inference is lower than that of \citet{2017MNRAS.466.4094M}, however, those authors speculate that their BCG mass may be an overestimate).  

A potential degeneracy can occur in lens models between BCG mass and halo mass, especially if the BCG is closely co-centered with the DM halo, since the mass is a free parameter in both profiles and both deflect image positions similarly. In a similar way, degeneracy can occur between BCG mass and core size. As a sanity check for the inferred BCG masses, we calculate the mass-to-light ratio for each, using the closest HST filter to V-band in the rest frame. The luminosities were measured in Gaia/Source Extractor and corrected for galactic extinction. 

\citet{Newman2013a} used a stellar population synthesis model and assumed a Chabrier initial mass function (IMF) \citep{Chabrier2003} to arrive at a mass-to-light ratio $\Upsilon_V=M_*/L_V$ for 7 giant elliptical BCGs, finding a range of 1.80 to 2.32 in V-band, with low scatter. The assumption of a Salpeter IMF \citep{Salpeter1955} increases the ratio by a factor of 1.78, resulting in an upper value of 4.13. In \citet{Andrade2019}, they note the possibility of a super-Salpeter IMF in Abell 611, which would further increase the upper limit. Our results are consistent with those expected values, with the exception of MACS2129, for which we find $\Upsilon_V=10.49_{-1.75}^{+1.89}$. We do not know the reason for this outlier, but we speculate that it may be caused by difficulty in separating the BCG light from three other bright objects within the half-light radius of the BCG (approximately 17 kpc). As a test, we made a separate strong lens model for MACS2129 with $\Upsilon_V$ constrained to a value of 4. The resulting fit was inferior to the original, with $\chi^2$ increasing from 77 to 187, RMS position error increasing from 0\farcs79 to 1\farcs30, and only 28 out of 31 images matched. The inferred core size from that run was smaller than the original run, decreasing to a median posterior value of 39.4 kpc from the original 54.8 kpc, which would cause a decrease in the inferred SIDM cross section for that cluster. 

\subsection{Halo Concentrations}
\label{sec:halo concentrations}
We infer halo concentrations with median posterior values ranging from $5.39_{-0.59}^{+0.71}$ (RXCJ2248) to $10.83_{-0.66}^{+0.65}$ (MACS1720), with a median value of 7.0 (see Table~\ref{tab:strong_lensing_results}). For comparison, the concentration-mass relation of \citet{Diemer_2019} predicts median concentrations $\simeq4.0$ for halos at redshifts of 0.2 to 0.6 and in the mass range of $10^{14}M_{\sun}$ to $10^{15}M_{\sun}$. Also, \citet{Merten2015} observed concentrations of $3.7 \pm 0.65$ for 19 X-ray clusters in the CLASH sample. As we discuss in  Appendix~\ref{sec:biases_in_triaxial_halos}, this discrepancy can potentially be explained if the lines of sight to lensing clusters are preferentially oriented along the major axis of the clusters. The concentrations for prolate halos can be biased upward by up to 60\%. As described in Section~\ref{sec:SIDM_Halo_Analysis}, we vary the cluster's LOS axis ratio in our SIDM model to account for this effect. (The concentration parameter \textit{per se} is not used in the subsequent SIDM analysis; rather, the entire radial halo profile is used, as described in Section~\ref{sec:SIDM_Halo_Analysis}.) In addition, \citet{Fielder_2020} showed that concentration will be significantly higher for a DM halo when mass from its associated subhalos are excluded from the calculation. Our models do account for some subhalo mass separately (for the larger perturbers), so we would expect our primary halo to be more concentrated than that of the cluster as a whole.

\subsection{Strong Lensing Systematic Errors}
\label{sec:SL systematic errors}
Statistical errors in image positions in strong lensing studies are often quite small, on the order of 1 to 2 detector pixel widths. In our case, most images were from the Hubble Space Telescope ACS instrument, with a pixel with of 0\farcs05. Much more significant are systematic effects, which can include misidentified images, inaccurate image redshifts (especially for photometric redshift data; see \citet{Cerny2018}), unmodeled substructure and correlated mass along the line of sight. As in many other strong lensing studies, we account for the latter item via two external shear components in the model, which adds two degrees of freedom to the model. To account for other systematic errors, most studies increase the assumed position error well beyond that of the statistical uncertainty. Some assumptions for position error from other authors for the same clusters in our sample are as follows: 0\farcs2 for Abell 611 \citep{Donnarumma2011}, 0\farcs5 for MS2137, A383, A611 and A2537 \citep{Newman2013a}; 1\farcs4 for A2261 \citep{Coe2012}; 0\farcs5 to 1\farcs0 for MACS 2129 \citep{Monna2018a};  1\farcs0 for A383 \citep{Monna2015}; 0\farcs5 for RXCJ 2248 \citep{Caminha2016}. Note that some of those studies involved multiple data sets, combining strong lensing with weak lensing, X-ray analysis and/or stellar kinematic analysis. In that case, the position error assumption is more important because it scales the $\chi^2$ of the strong lensing component only, thus affecting the weighting of the strong lensing relative to the other data. In our case, we have only one component, so the relative weighting between data sources is not a concern. This was part of the motivation for using strong lensing alone in our study. We have therefore adopted a position error assumption of 0\farcs5, which seems broadly consistent with the residuals for all of the clusters in our our analysis. Note that this applies separately to the x and y components, so that the total uncertainty is 0\farcs71.

We note that the cluster MACS1720 has the highest reduced $\chi^2$ in our sample (5.79) and the highest RMS postion error (1\farcs07). In order verify that the assumed data position error of 0\farcs5 did not adversely impact the results, we made an additional run for that cluster with an assumed data position error of 1\farcs0. The resulting reduced $\chi^2$ fell to 1.63.  The inferred surface density profile declined slightly and its uncertainty approximately doubled; for example, at 20 kpc the inferred surface density changed from $4.42 \pm 0.09\times 10^9 M_\odot$ to $4.22 \pm 0.20 \times 10^9 M_\odot$.  The concentration fell from from $10.83 \pm 0.66$ to $9.71 \pm 0.84$. The resulting inference for $\log_{10}(\sigma/m)$ didn't change much, declining from $-1.35 \pm 0.18$ to $-1.40 \pm 0.20$.

As a check on the position error assumption, we discuss and plot the error residuals from strong lensing in Appendix~ \ref{sec:error_residuals} and Figure~ \ref{fig:error_residuals}. The residuals appear generally Gaussian in shape, and have standard deviation of approximately 0\farcs4 to 0\farcs8, roughly consistent with the assumed position uncertainty of 0\farcs5.

\subsection{BCG Offsets from DM Halo Centers}
An observable consequence of cored SIDM halos would be oscillations of BCG about the center of the halo after mergers, which would persist for several Gyr \citep{Kim2017}. In \citet{Harvey2019}, they used simulations to examine the effect of DM self-interaction on BCG offsets, and concluded that the distribution of such offsets from an ensemble of clusters would exhibit a median value of $3.8\pm 0.7$ kpc for a CDM scenario (i.e., $\sigma/m=0$ \cmsg), and $8.6 \pm 0.7$ kpc in a scenario where $\sigma/m=1$ \cmsg.  
The rightmost column of Table~\ref{tab:strong_lensing_results} and the first histogram in Figure~\ref{fig:various_histograms} show our inferred 2-dimensional offset of the DM halo centers for each cluster, which range from approximately 1 kpc to 10 kpc, with a median of approximately 4 kpc, except for MACS2129, which is an outlier at 10 kpc to 15 kpc. Since these are 2-dimensional offsets, we can estimate that the corresponding 3-dimensional offsets would be larger by a factor of approximately $\sqrt{3/2}=1.23$, although the precise value would depend on the distribution of the offsets. Using this factor results in a corresponding median value of approximately 5 kpc for our data, which would imply $\sigma/m \ll 1$ \cmsg in light of the findings of \citet{Harvey2019}. However we also note that \citet{Harvey2017} found that the error estimates from the posteriors of MCMC lens modeling could be understated by as much as a factor of 10. Applying this to the predicted offsets in Table 3 would result in uncertainties larger than the offsets themselves, making an inference discerning CMD from 1 \cmsg SIDM much weaker.

\section{SIDM Halo Analysis}
\label{sec:SIDM_Halo_Analysis}

It has been argued from observations of groups and clusters of galaxies, that the cross section must be velocity-dependent to have a significant effect for less massive galaxies \citep{Kaplinghat2016}. This velocity dependence can occur in several ways: resonant self-interaction of dark matter \citep{Chu2019}, light mediator models with either elastic or inelastic collisions \citep{Feng2010, Loeb2011, Tulin2013,Tulin2017,Alvarez2020,Boddy2016,Vogelsberger2019}, bound states \citep{Braaten2018}, and strongly interacting massive particles \citep{Hochberg2015,Choi2017}. See \citet{Chu2020} for a review of these models in addition to a model-independent approach for approximating velocity dependence using effective range theory. As the range of particle velocities in our cluster sample is relatively narrow, we opt for a constant velocity analysis here.

\subsection{SIDM Halo Model}
\label{sec:SIDM_halo_model}

For the SIDM halo model, we following the procedure in \citet{Kaplinghat2016} and write the full profile as an inner isothermal profile and an outer NFW profile, with the two profiles matched in mass and density at a radius $r_1$, which is determined by the cross section. The idea here is to get a smooth density profile that interpolates between an isothermal core and a collisionless outer envelope well enough to capture the halo profiles inferred from SIDM simulations. The characteristic radius $r_{1}$ dividing the inner and outer regions can be approximated by setting the average scattering rate per particle times the age of the halo to unity:
\begin{equation}
    \label{eq:rate_time}
    \text{rate x time}\approx \frac{<\sigma v>}{m}\rho(r_1)\;t_\text{age} = 1,
\end{equation}
where $\sigma$ is the scattering transfer cross section, v is the relative velocity between DM particles, $\rho(r_1)$ is the density of DM at the characteristic radius, $t_\text{age}$ is the age of the halo, and <...> denotes averaging over the velocity distribution. 
This is a simplification of the time-dependent process of halo assembly, but it compares well to numerical simulations \citep{Kaplinghat2016,Robertson:2020pxj} because of the approach to equilibrium. 

For the age of each halo, we adopt a value equal to the time at the redshift of the halo, as shown in Table~\ref{tab:cluster_list}.
The average relative velocity of SIDM particles, assuming a Maxwellian velocity distribution, can be shown to be $\frac{4}{\sqrt{\pi}}\sigma_0$, where $\sigma_0$ is the central velocity dispersion. Assuming a constant cross section $\sigma$ over the range of velocities accessible in the cluster, we have $\langle \sigma v \rangle = \sigma \langle v \rangle=\frac{4}{\sqrt{\pi}}\sigma\,\sigma_0$. For a model with a sharp velocity dependence, $\sigma$ should be interpreted as the velocity averaged transfer cross section \citep{Tulin2013a,Robertson2017a,Kahlhoefer2014,Boddy2016}. 

Considering the two regions, one which is thermalized by self-interactions and one which remains largely non-interacting, the complete profile can be written as
\begin{equation} \label{eq:r1-halo-model}
    \rho(r) = \begin{cases}
      \rho_{\text{iso}}(r), &   r<r_1 \\
      \rho_{\text{NFW}}(r), &   r\geq r_1\\
    \end{cases}
\end{equation}
where $\rho_{\text{iso}}(r)$ is a cored isothermal profile \citep{Kaplinghat2014}, and $\rho_{\text{NFW}}(r)$ is the NFW profile (see Equation~\ref{eq:cNFW}, but with $r_c \rightarrow 1$). For the region interior to $r_1$, interactions are common and we consider the DM particles to behave as an ideal gas, characterized by a pressure (p), density ($\rho)$ and (one-dimensional) velocity dispersion ($\sigma_0$) that obey an equation of state $p=\rho \sigma_0^2$. Assuming that the DM particles in the central region achieve hydrostatic equilibrium, we have
\begin{equation}
    \nabla p=-\rho \nabla \Phi_{\text{tot}}
\end{equation}
where $\Phi_{\text{tot}}$ is the total gravitational potential from both DM and baryons. In this analysis, we use the fitted BCG from strong lensing with a dPIE profile (Equation~\ref{eq:dPIE}) for the baryonic mass. The gravitational potential must satisfy the Poisson equation
\begin{equation}
    \nabla^2\phi_{\text{tot}}=4\pi G(\rho_{\text{DM}}+\rho_{\text{BCG}})
\end{equation}
where $G$ is Newton's gravitational constant, $\rho_{\text{DM}}$ is the mass density of DM and $\rho_{\text{BCG}}$ is the mass density of the BCG.
Thus the SIDM density profile explicitly depends on the stellar distribution \citep{Kaplinghat2014}. 
The solution to two equations above gives $\rho_{\text{iso}}(r)$ for the interior region. 

We assume that the isothermal solution is a cored profile at small radii with central density $\rho_0$ and central dispersion $\sigma_0$. For concreteness, we impose this boundary condition at 1\% of the stellar core radius ($r_\textrm{core}$ in the dPIE profile). We then evolve the isothermal Jeans equation outwards stopping at the radius $r_1$ when the SIDM density $\rho(r_1)$ satisfies Eq.~\ref{eq:rate_time}. By matching the mass and density of $\rho_{\text{iso}}$ to a NFW profile, we can find the NFW parameters $\rho_s$ and $r_s$ from $\rho_0$ and $\sigma_0$. Note that we do not search for a second solution at larger $r_1$ \citep{Kaplinghat2016,Robertson:2020pxj}, which could be just as good a fit. This second solution with the same $\rho_0$ is reminiscent of the core collapse phase \cite{Balberg:2002ue,Elbert2015,Essig:2018pzq,Nishikawa:2019lsc}, but it is not clear if this connection holds up in detail. 

As discussed in \ref{sec:halo concentrations} and Appendix \ref{sec:biases_in_triaxial_halos}, the apparent concentrations of halos that are elongated along the LOS are higher than the actual concentration. To account for this we transform our spherical halo profile $\rho(r)$ in Eq.~\ref{eq:r1-halo-model} into an ellipsoidal profile $\rho_e(r_e)=\rho(r_e)$ with $r_e^2=R^2+z^2 s^2$, and orient the $z$-direction along the line-of-sight. This implies that the ellipsoidal surface density $\kappa_e(R)=\kappa(R)/s$. Choosing the LOS axis ratio $s<1$ implies that for the same mass, concentration and cross section, the surface density is higher.  We allow $s$ to vary during the MCMC fitting procedure, which would allow for smaller $\rho_s$ and hence larger cross sections. Note that the central density does not vary linearly with $\rho_s$ as the cross section is changed.

\begin{figure}
    \centering
    \caption{Histogram plot for the line of sight axis ratio "s". The prior is shown as a dashed black line and is based on \citet{Vega-Ferrero2017} for s<0.5 and is flat for s>0.5.}
    \includegraphics[width=0.48\textwidth]{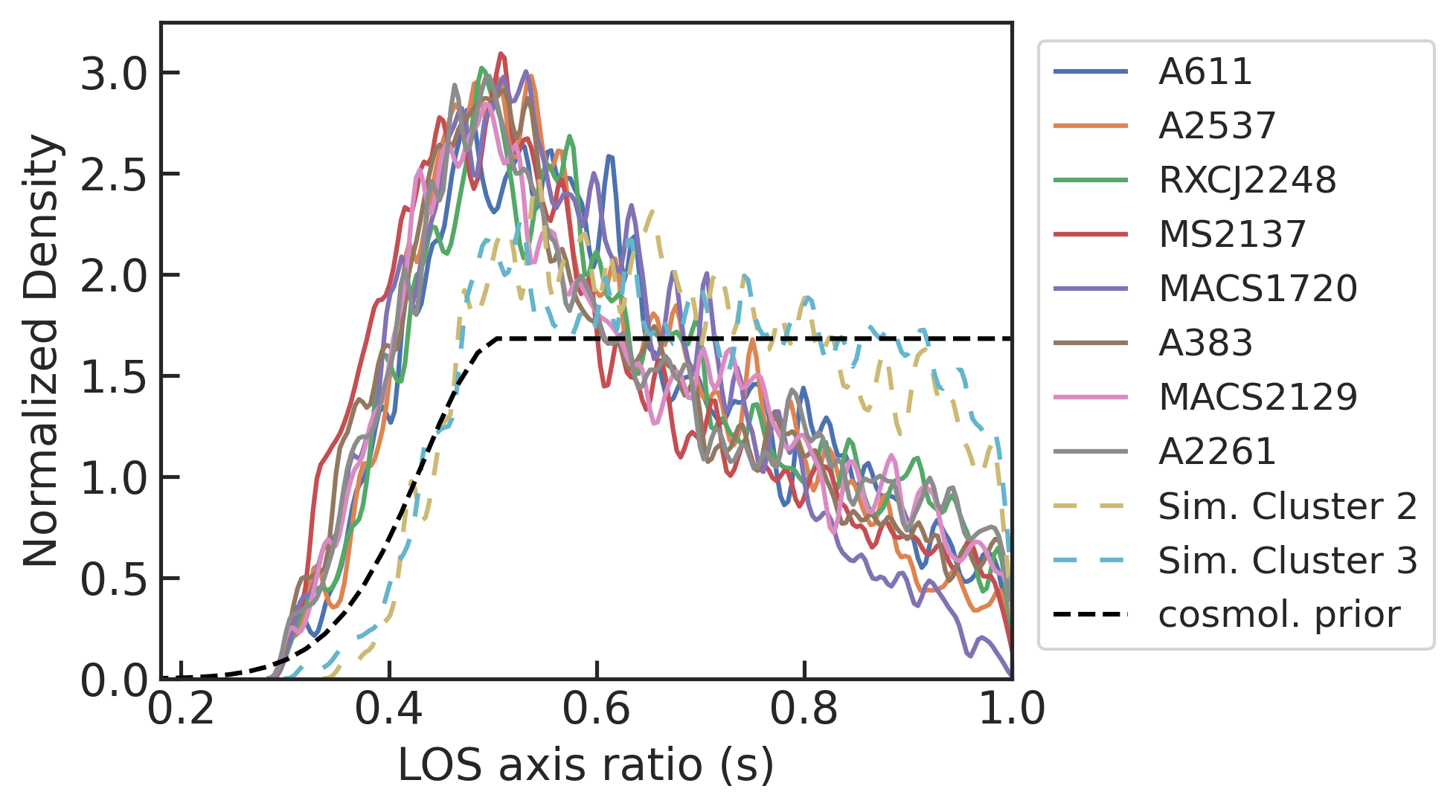}
    \label{fig:los_axis_ratio_hist}
\end{figure}

To arrive at a prior distribution for s, we use the results in \citet{Vega-Ferrero2017} and \citet{Bonamigo2015a} for relaxed clusters.  We utilize the PDF from \citet{Vega-Ferrero2017} for values of s < 0.5. The PDF reaches a maximum at $s\simeq 0.5$, and declines as s approaches unity. For $s > 0.5$, since our clusters may not always be fully oriented along the LOS, so we opt for a flat prior for $s>0.5$. The prior for $s$ is shown in Figure~\ref{fig:los_axis_ratio_hist}, along with the inferred posteriors for each cluster from the SIDM model. Our data prefers LOS axis ratios near 0.5 for all clusters.

There are two pitfalls that impede seamless parameter space exploration using the halo profile in Eq.~\ref{eq:r1-halo-model}. First is that when we scan over $\rho_0$ and $\sigma_0$, there is no guarantee that $r_1$ can be found. To insure against this failure, we change our variable from $\rho_0$ to $R_0 = \rho_0 (\sigma/m) (4/\sqrt{\pi}) \sigma_0 t_{\rm age}$ and put a prior on $R_0$ that is larger than unity \citep{Ren2019,Robertson:2020pxj}. Since the density falls with increasing radius, we are guaranteed to get a solution for $r_1$. For the present study we have adopted the following priors on the five parameters that are varied to find the SIDM solutions: $5<R_0<500$, $400 < \sigma_0 < 1400$, $0.01 < \sigma/m < 1.0$, $10^{11} M_\odot < M_{\rm BCG} < 10^{13} M_\odot$, $0.3 < s < 1$. With the exception of the lower limit of $\sigma/m$, our prior boundaries do not impact the inferred posteriors.

The second issue is that for a given $r_1$ and $\rho_{\text{iso}}(r)$, it may not be possible to find a matching NFW profile (see also,  \citet{Robertson:2020pxj}). The procedure to find the matching NFW profile proceeds by first matching $\gamma_M(r) = M(r)/(4\pi r^3 \rho(r))$ where $M(r)$ is the dark matter halo mass enclosed within radius $r$. Since for the NFW profile $\gamma_M(r)$ is only a function of $r/r_s$, we get $r_1/r_s$ and therefore $r_s$. Then we can match the isothermal and NFW $M(r)$ at $r_1$ to infer $\rho_s$. However, the first step can fail if  $\gamma_M(r_1) < 0.5$ for the isothermal profile, because $\gamma_M(r) \geq 0.5$ for the NFW profile. 
However, these cases are rare and they are not physically interesting, as it requires a core unlike that seen in simulations with $\rho(r_1) \simeq \rho_0$. With these checks in place, it is possible to find unique NFW matches for all $\rho_{\text{iso}}(r)$ profiles generated by the likelihood sampler. Once we have the full density profile, then we can compute the 2D density profiles to compare to the data. 

Before we discuss the results from the SIDM halo matching process discussed above, it is worth noting this analytic model (described in \citet{Kaplinghat2016}) has been remarkably successful in capturing the density profiles in simulations from dwarf galaxies to clusters of galaxies \citep{Kaplinghat2016,Robertson2017,Sokolenko:2018noz,2019MNRAS.490.2117R,Ren2019,Robertson:2020pxj}. The study in \citet{Sokolenko:2018noz} specifically focused on clusters and investigated different ways of matching the isothermal and NFW solutions. They found that the model of \citet{Kaplinghat2016} can reproduce the core density of the halos, as shown in the left panel of Figure 22 of their paper. The inferred core is recovered at the 10-20\% level for cross sections below 1 \cmsg, which are relevant for our study. Remarkably, the analytic model provides an unbiased description over almost two orders of magnitude in cross section from 0.1 to 10 \cmsg. There seems to be a bias creeping in at $\sigma/m=0.1$ \cmsg; if this trend continues to lower cross sections, this would imply that our inferred cross sections are biased somewhat high. 

The right panel of the same figure from \citet{Sokolenko:2018noz} shows that the kinetic energy inferred from the central dispersion is systematically lower than the average kinetic energy within $r_1$.  This bias is important for us since we need both the cross section and the average relative velocity of DM particles to extract constraints on SIDM models. The results from  \citet{Sokolenko:2018noz} indicate that the average relative velocity can be biased low by about 20\% if it is inferred from the central dispersion $\sigma_0$ using the halo profile in Eq.~\ref{eq:r1-halo-model}. While this seems like a small effect, it can be an important systematic if we are constraining a cross section model with a sharp velocity dependence. Since most models predict cross sections that fall with increasing relative velocity at these high velocities, a conservative way of constraining SIDM models is to use the largest possible average relative velocity for each cluster. We can estimate this maximum average relative velocity as $1.4 v_\textrm{max}$ using the fact that the maximum RMS velocity of DM particles is very close to $v_\textrm{max}$ \citep{Rocha2012}. We provide both $\sigma_0$ and $V_{\rm max}$ posteriors for the clusters. 

\subsubsection{The Importance of the Characteristic Radius $r_1$}
The impacts of DM self-interaction on galaxy cluster properties was examined in \citet{Robertson2019} using the Bahamas simulations. In reviewing the SIDM halo-matching model, those authors conclude ``we find the model to provide a good description of simulated SIDM density profiles, and (importantly) find that the isothermal Jeans model can be used to infer the cross-section from a simulated halo`s density profile...."  In the upper right panel of Figure 2 from that work, they show that the median density of the simulated cluster of mass $\sim 10^{15} M_{\sun}$ at a radius of 10 kpc is approximately 70\% higher in CDM as compared to SIDM with $\sigma/m$  of 0.1 \cmsg. The divergence in the median densities persists out to radii $\gtrsim$ 40 kpc. The scatter about the mean in the inner parts is not random -- it is mostly driven by the scatter in the outer profiles due to the concentration--mass relation and the impact of the baryons on the inner profile. Both of these effects are captured by the isothermal halo profile used in \citet{Robertson2019} and our work. 

\begin{figure*}
    \centering
    \caption{Comparison of the characteristic radii $r_1$ to the BCG radii and the radii of the innermost strong lensing data points. The dotted lines indicate 1:1 equality. Note that $r_1>r_{\mathrm{BCG}}$ and $r_1>r_{\mathrm{inner}}$ in all cases except A611, which allows a solution mode with low cross section and an $r_1$ value extending down to $\sim$ 25 kpc at the 1$\sigma$ level.}
    \includegraphics[width=0.48\textwidth]{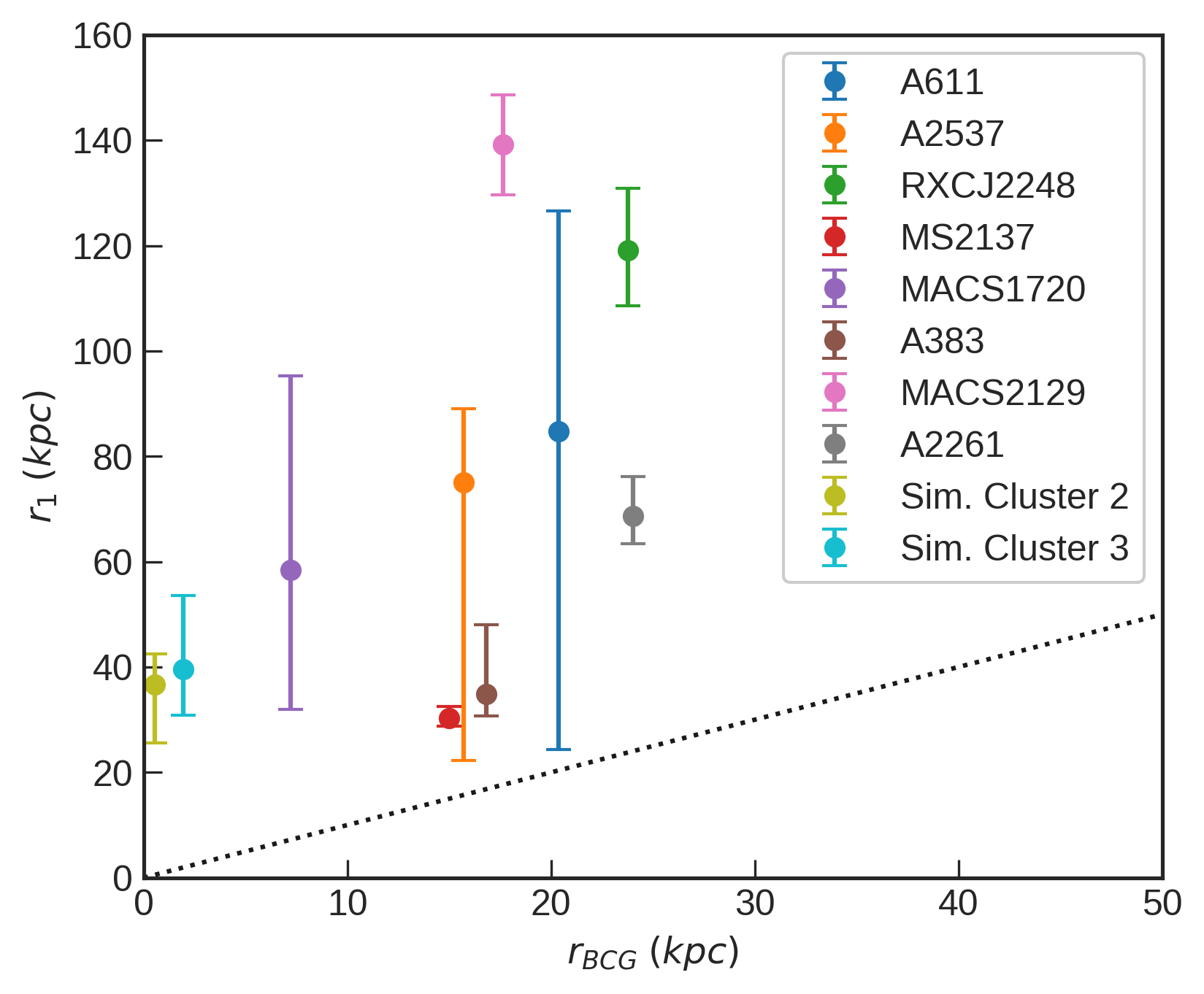}
    \includegraphics[width=0.48\textwidth]{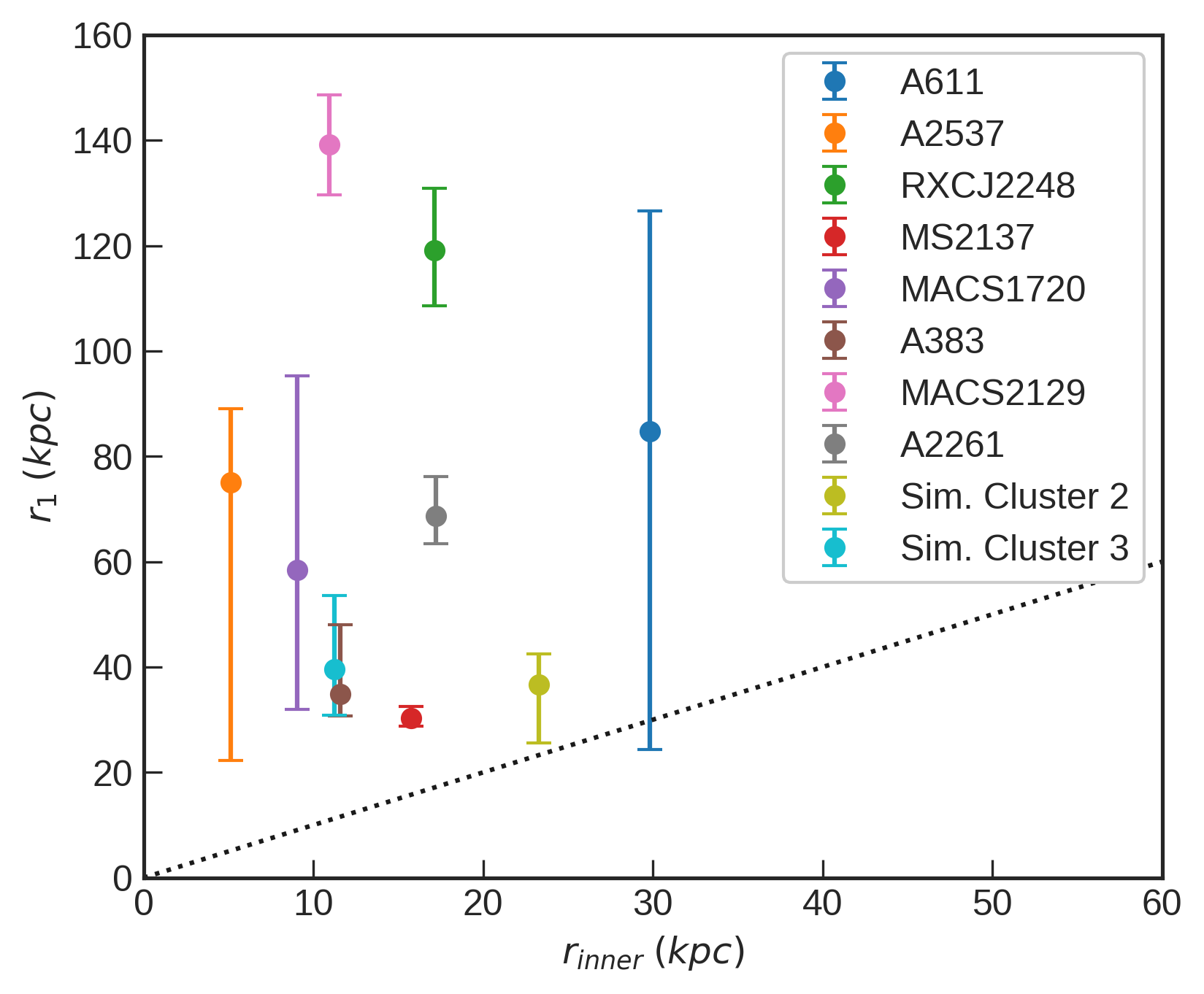}
    \label{fig:r1_plots}
\end{figure*}

The model parameter $r_1$ indicates the matching point between the isothermal and NFW halo profiles. If $r_1$ occurs well outside the BCG, then we can be reasonably confident that adiabatic contraction does not impact the outer NFW profile significantly. If the value of $r_1$ is within the range of strong lensing data, then we can be reasonably confident that the isothermal profile is being constrained by the data. Figure~\ref{fig:r1_plots} shows $r_1$ for the cluster sample in relation to the BCG radius (left panel) and the innermost strong lensing data point (right panel). The $r_1$ radii are larger than both, with the exception of A611 cluster. A611 has a bi-modal solution that extends down to low cross sections and values of $r_1$ that are comparable to those of the BCG radius and innermost data points. 

Writing equation~\ref{eq:rate_time} in terms of the density at $r_1$, we see that $\rho(r_1)=1/(\frac{<\sigma v>}{m} \; t_\text{age})$. Since $t_\text{age}$ varies only modestly over our data set, constraints on $\rho(r_1)$ are the drivers of the constraints on cross section. For example, taking a cluster at redshift 0.3, $<v>=$1500 km/s, and $\sigma/m \geq $0.1 \cmsg, we arrive at $\rho(r_1) \leq \num{3.12e6}  M_{\sun} \mathrm{kpc}^{-3}$. Halos that are more dense than this at their characteristic radius would disfavor SIDM cross sections at or above 0.1 \cmsg. Note that in \citet{Robertson2019}, Figure 2, the densities of simulated CDM and SIDM halos (with SIDM cross section of 0.1 \cmsg) do indeed begin to depart from one another at approximately the density calculated here.

\subsection{SIDM Halo Matching Inference Results}
\label{sec:SIDM Halo Matching Results}

We employ an MCMC fitting code to implement the second stage of our analysis, using as input the strong lensing posterior for the DM surface density profile for each cluster. By using the entire surface density profile, the model can appropriately match the halo density profile, and accommodate multiple possible modes in the strong lensing solution set. We used the Dynesty sampler \citep{Speagle2020, Higson2019} to generate the MCMC posterior chain for this stage. The key posterior parameters are: cross section, $r_1$ (the characteristic radius), average particle speed, and BCG mass, as noted in Figure~\ref{fig:analysis_pipeline}.  
From this we infer the central density ($\rho_0$) and the matched NFW halo parameters ($M_\mathrm{vir}$ and $c_\mathrm{vir}$). 
We fit cross sections at constant velocity, noting that the median maximum circular velocities (a rough proxy for relative particle speed) for the clusters in the sample are in the range of 1,000 km/s to 1,800 km/s, with most around 1,200 km/s. 

As shown schematically in Figure~\ref{fig:analysis_pipeline}, surface density posteriors at radii ranging from 2 kpc to 200 kpc are generated by the strong lensing model and were first decomposed into principal components using the Principal Component Analysis (PCA) in the Scikit-Learn software package \citep{Pedregosa2011a}, which typically yielded 3 or 4 components for any given cluster. We found that supplying 10 logarithmically spaced radial data points ranging from 10 kpc to 200 kpc enabled good surface density profile reproduction for most clusters, but three clusters (A383, A2261 and MS2137) required that an additional 5 bins extend farther inward, to 2 kpc, to ensure that the model made a good match to the shape of the surface density profile inferred from strong lensing. 

A Gaussian Mixture Model (GMM) was used to model the likelihood using the principal components of the surface density profiles. We calculated the Bayesian Information Criteria (BIC) for each cluster while varying the number of GMM components, and found that having more than four components in the GMM did not improve the BIC significantly. We therefore used four components in the GMM.

In addition to the surface density profiles, the SIDM model also requires the parameters of the BCG density profile as described above. We use the BCG scale radius as measured from photometry for this purpose. However, we do not use the BCG mass posterior from strong lensing. By varying the BCG mass independently, we are able to explore the degeneracy between the BCG mass and cross section; see Figure~\ref{fig:mbcg_cvir_compare} for a comparison of the BCG mass inferred from the SIDM and cNFW fits, and the associated discussion in Section~\ref{sec:SIDM Halo Matching Results}. Inference of the BCG mass is sometimes multimodal, with larger BCG masses being associated with larger cross sections (lower DM core densities). Posterior distributions for the SIDM halo model are shown in the supplemental online material.

\begin{figure}
    \centering
    \caption{Posterior histograms for $\log_{10}(\sigma/m)$. The simulated clusters are shown as dashed lines.}
    \includegraphics[width=0.48\textwidth]{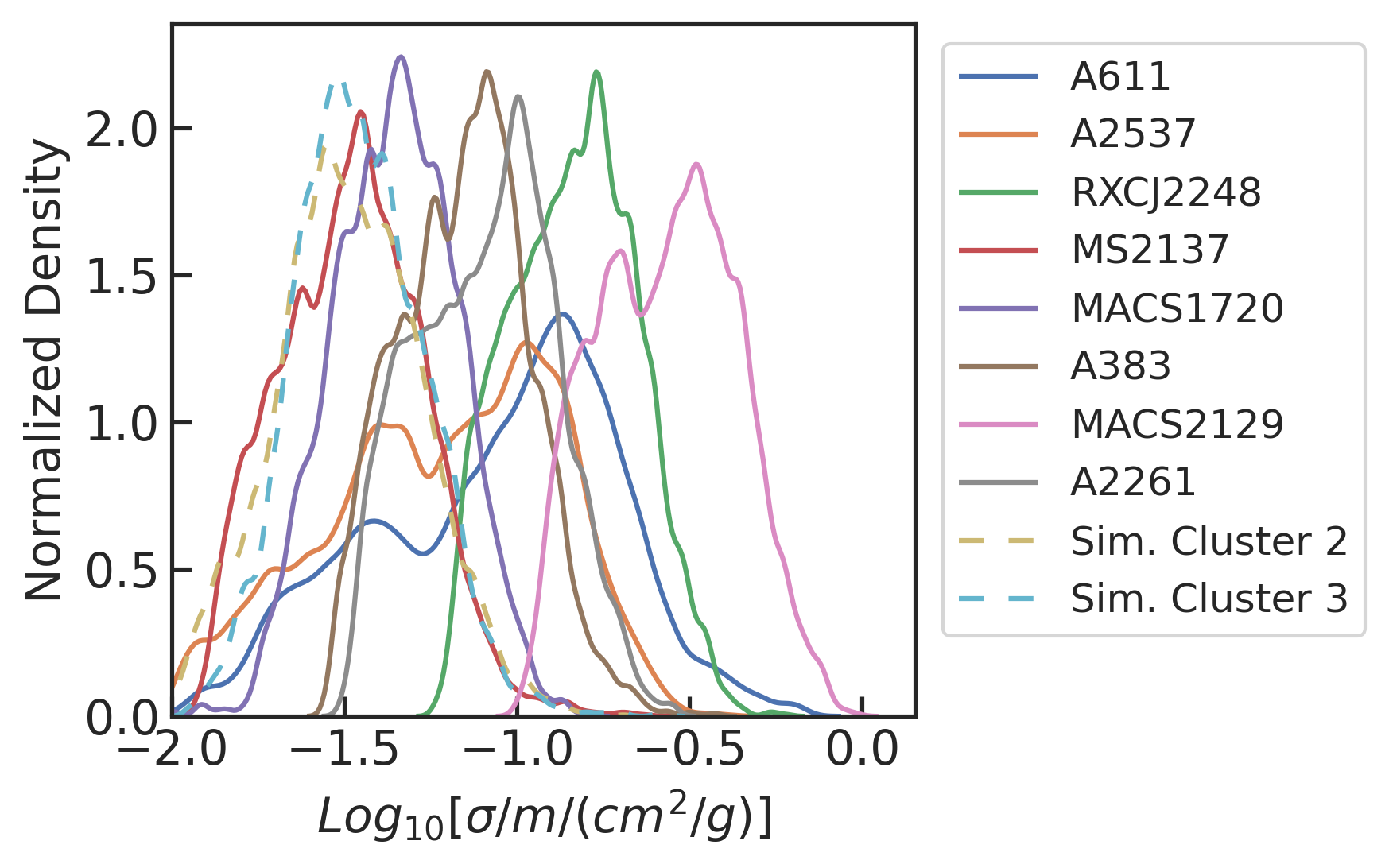}
    \label{fig:logcross_hist}
\end{figure}

\begin{figure*}
    \centering
    \caption{Posteriors for relative velocity $V_{\mathrm{rel}} = 4 \sigma_0 / \sqrt{\pi}$ (horizontal axes, in km/s) versus SIDM cross section (vertical axes, in $cm^2/g$) from the SIDM halo matching model. The smaller boxes are for the clusters individually, and the larger box at the bottom is the composite of all 10 clusters. The $2\sigma$ regions are shaded, and the $1\sigma$ regions are shown in a darker shade. }
    \includegraphics[width=0.97\textwidth,height=0.94\textheight]{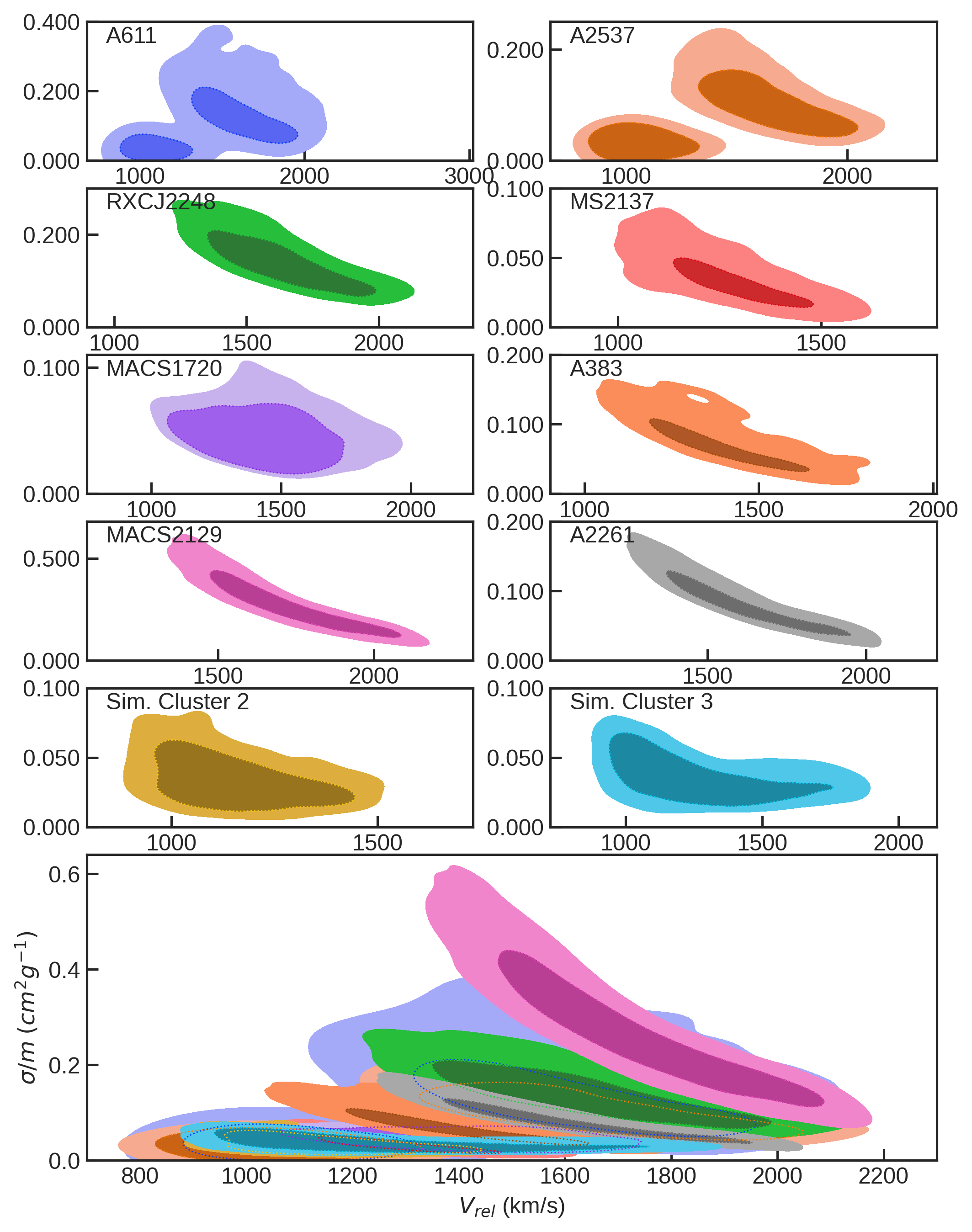}
    \label{fig:cross sec vs v combo}
\end{figure*}

\begin{figure*}
    \centering
    \caption{Posteriors from the SIDM model for $r \, \rho(r)$ (vertical axes, in $M_{\sun}$kpc$^{-2}$) versus radius (horizontal axes, in kpc). The median posterior and $2\sigma$ band are shown in orange. The dashed blue line shows the median concentration-mass relation found from \citet{Diemer_2019} using the redshift and the average virial mass of the specific cluster. For the simulated clusters, the actual $r \, \rho(r)$ is shown, but note that this includes subhalo density and is not exactly comparable to the other lines.}
    \includegraphics[width=0.96\textwidth,height=0.92\textheight]{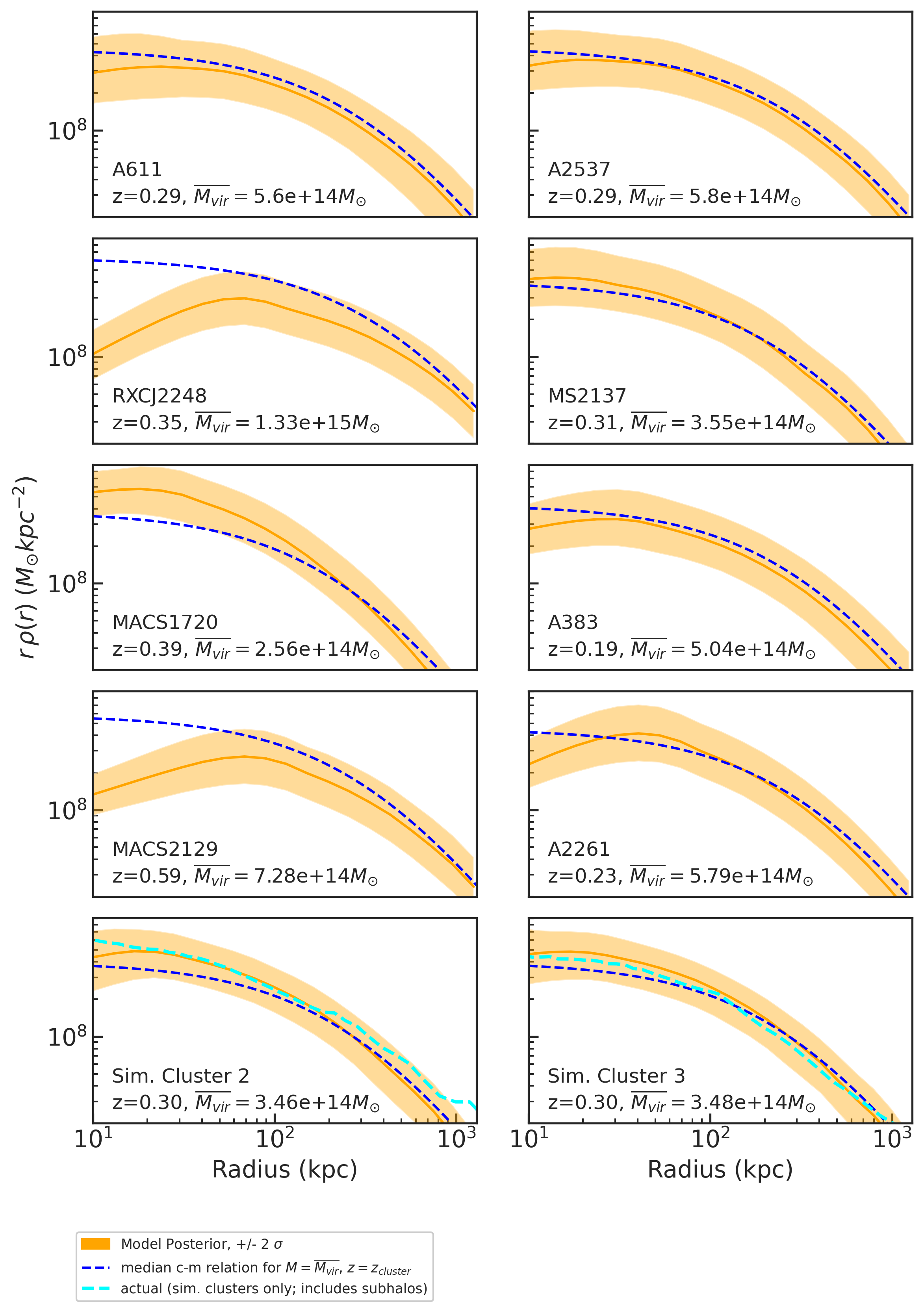}
    \label{fig:Robertson_comparison_grid}
\end{figure*}

\begin{table*}
	\centering
    \caption{SIDM parameter posterior summary. The columns are: (1) cluster name, (2) $\log_{10}$ cross section per unit mass, median and 68\% C.L., (3) $\log_{10}$ cross section per unit mass, mean $\pm$ standard deviation, (4) characteristic radius, (5) central density, (6) central velocity dispersion, (7) halo maximum circular velocity, (8) BCG mass, (9) halo virial mass, and (10) concentration. The indicated ranges are the 68\% confidence intervals.}
	\label{tab:SIDM_params}
	\renewcommand{\arraystretch}{1.5}
	\resizebox{\textwidth}{!}
	{
	\begin{tabular}{l|cccccccccc} 
		\hline
		 & $\log_{10}\Big(\frac{\sigma/m}{cm^2g^{-1}}\Big)$ & $\log_{10}\Big(\frac{\sigma/m}{cm^2g^{-1}}\Big)$ & $r_1$ & $\rho_0$ & $\sigma_0$ & $v_{\text{max}}$  & $m_{\text{BCG}}$ & $m_{\text{vir}}$ & $c_{\text{vir}}$ \\
        Cluster Name          &  (median)  & (mean)  & (kpc) & ($10^6\,M_{\sun}\,kpc^{-3}$) & ($km\;s^{-1}$) & ($km\;s^{-1}$)  & ($10^{12}\,M_{\sun}$) & ($10^{14}\,M_{\sun}$) & \\
        \hline
A611           & ${-1.00}_{-0.47}^{+0.27}$ & ${-1.06} \pm {0.35}$ & $ 94.5_{-64.8}^{+15.4}$ & $165.1_{-60.5}^{+97.8}$ & $  649_{-168}^{+138}$ & $ 1205_{-169}^{+206}$ & $ 2.32_{-1.72}^{+0.90}$ & $ 5.05_{-1.79}^{+2.97}$ & $ 5.86_{-1.11}^{+1.22}$ \\
A2537          & ${-1.18}_{-0.40}^{+0.30}$ & ${-1.21} \pm {0.33}$ & $ 83.6_{-56.3}^{+18.3}$ & $205.4_{-108.1}^{+58.5}$ & $  629_{-179}^{+151}$ & $ 1235_{-150}^{+209}$ & $ 1.88_{-1.57}^{+0.62}$ & $ 5.35_{-1.65}^{+2.65}$ & $ 6.20_{-0.82}^{+0.94}$ \\
RXCJ2248       & ${-0.83}_{-0.21}^{+0.18}$ & ${-0.83} \pm {0.18}$ & $117.3_{-13.1}^{+16.9}$ & $ 17.5_{-5.2}^{+7.2}$ & $  706_{ -92}^{+121}$ & $ 1554_{-194}^{+277}$ & $ 0.67_{-0.28}^{+0.72}$ & $12.24_{-3.78}^{+6.94}$ & $ 3.78_{-0.45}^{+0.60}$ \\
MS2137         & ${-1.47}_{-0.23}^{+0.20}$ & ${-1.48} \pm {0.21}$ & $ 32.8_{-2.5}^{+4.6}$ & $256.6_{-43.9}^{+46.3}$ & $  560_{ -55}^{ +70}$ & $ 1085_{-136}^{+205}$ & $ 1.04_{-0.16}^{+0.18}$ & $ 3.25_{-1.04}^{+1.74}$ & $ 7.46_{-1.14}^{+1.34}$ \\
MACS1720       & ${-1.35}_{-0.19}^{+0.18}$ & ${-1.35} \pm {0.18}$ & $ 47.2_{-6.2}^{+17.6}$ & $453.7_{-337.4}^{+930.9}$ & $  628_{ -87}^{ +99}$ & $ 1072_{-129}^{+159}$ & $ 0.78_{-0.56}^{+0.66}$ & $ 2.41_{-0.68}^{+1.03}$ & $10.10_{-1.15}^{+1.08}$ \\
A383           & ${-1.14}_{-0.22}^{+0.18}$ & ${-1.15} \pm {0.19}$ & $ 61.0_{-3.1}^{+9.3}$ & $142.6_{-33.0}^{+32.8}$ & $  606_{ -63}^{ +89}$ & $ 1146_{-152}^{+215}$ & $ 1.33_{-0.18}^{+0.20}$ & $ 4.56_{-1.54}^{+2.60}$ & $ 6.47_{-1.05}^{+1.21}$ \\
MACS2129       & ${-0.56}_{-0.24}^{+0.20}$ & ${-0.57} \pm {0.20}$ & $153.0_{-11.1}^{+12.7}$ & $ 54.0_{-9.5}^{+11.9}$ & $  741_{ -84}^{+117}$ & $ 1349_{-177}^{+245}$ & $ 1.90_{-0.28}^{+0.30}$ & $ 6.63_{-2.14}^{+3.81}$ & $ 3.71_{-0.47}^{+0.60}$ \\
A2261          & ${-1.07}_{-0.24}^{+0.19}$ & ${-1.09} \pm {0.20}$ & $ 89.6_{-5.4}^{+7.4}$ & $ 52.2_{-7.0}^{+8.3}$ & $  696_{ -77}^{+117}$ & $ 1253_{-159}^{+242}$ & $ 1.20_{-0.13}^{+0.19}$ & $ 5.27_{-1.59}^{+3.00}$ & $ 7.59_{-0.99}^{+1.15}$ \\
Sim. Cluster 2 & ${-1.49}_{-0.20}^{+0.22}$ & ${-1.48} \pm {0.21}$ & $ 35.3_{-10.7}^{+6.9}$ & $1414.5_{-910.6}^{+2223.6}$ & $  509_{ -62}^{ +78}$ & $ 1115_{-143}^{+156}$ & $ 0.16_{-0.04}^{+0.07}$ & $ 3.28_{-0.99}^{+1.39}$ & $ 8.24_{-1.10}^{+1.22}$ \\
Sim. Cluster 3 & ${-1.46}_{-0.17}^{+0.20}$ & ${-1.45} \pm {0.19}$ & $ 37.3_{-7.6}^{+10.1}$ & $222.9_{-123.1}^{+787.1}$ & $  549_{ -90}^{+138}$ & $ 1124_{-162}^{+153}$ & $ 0.35_{-0.20}^{+0.77}$ & $ 3.37_{-1.11}^{+1.37}$ & $ 8.17_{-1.11}^{+1.19}$ \\
		\hline
	\end{tabular}
	}
	\renewcommand{\arraystretch}{1.}
\end{table*}

The surface density models produced by the SIDM halo matching model reproduced the data to within 0.1 dex or better. Plots of the matches for each cluster are shown in the Appendix in Figure~\ref{fig:kappa_match}. Note that the model relies on surface density profile rather than direct inferences of parameters such as BCG mass, halo mass, concentration, etc. This allows multimodal solutions in those parameters to be accurately incorporated into the model.

Figure~\ref{fig:Robertson_comparison_grid} plots the inferred $r \, \rho(r)$ of the central DM halo versus radius for each cluster (orange line and bands), and allows comparison to several other curves. We plot $r \, \rho(r)$ rather than density to allow better data visualization. The dashed blue line represents the median concentration-mass relation from \citet{Diemer_2019}, accounting for the redshift and using the median virial mass of each specific cluster. Sim. clusters 2 and 3 are somewhat more dense than that of the median concentration-mass relation of \citet{Diemer_2019}. The cyan dashed line shows the actual $r \, \rho(r)$ for the simulated clusters, but note that this data includes subhalo DM, and so is not exacly comparable to the other lines, especially in the outer parts of the halo. Our results show notable cores (i.e., a positive slope to the $r \, \rho(r)$ curve in the inner radii) in clusters RXCJ2248, MACS2129, and A2261.

Histograms for the inferred SIDM cross section per unit mass for each cluster are shown in Figure~\ref{fig:logcross_hist}, and the tabulated results are summarized in Table~\ref{tab:SIDM_params}.  Cross sections for the simulated clusters are inferred to be $0.037 \pm 0.019$ \cmsg and $0.039 \pm 0.020$ \cmsg for Sim. Clusters 2 and 3, respectively, at the 68\% confidence level. 

The median inferred cross section for the observed clusters ranges from 0.04 \cmsg (MACS1720) to 0.28 \cmsg (MACS2129). The inferred values show $\sigma/m<0.23$ \cmsg (68\% C.L.) for all clusters except MACS2129, which shows $\sigma/m<0.44$ \cmsg at the 68\% C.L. 

Our inferred cross sections per unit mass versus mean relative particle speed are shown in Figure~\ref{fig:cross sec vs v combo}. We calculate the central relative velocity between particles as $V_{\mathrm{rel}} = 4 \sigma_0 / \sqrt{\pi}$, assuming a Maxwellian velocity distribution. Using the same method as that described above in Section~\ref{sec:Constraints on the Cross Section} to account for systematic error in the sample, but in this case for $V_{\mathrm{rel}}$ rather than cross section, we find the relative velocities in the sample are $1458_{-81}^{+80}$ km/s. However we note that this could be biased low by up to 20\%, as discussed in Section~\ref{sec:SIDM_halo_model}. There is an anti-correlation between inferred cross section and $V_{\mathrm{rel}}$, as can be seen in Figure~\ref{fig:cross sec vs v combo} and Table~\ref{tab:SIDM_params}. Note that the trend when the results from all the clusters are put together is that higher cross section inferences go with higher $V_{\mathrm{rel}}$ and higher $\sigma_0$. This is likely because a higher cross-section results in a larger interaction rate and, in turn, a larger $r_1$ (characteristic radius) and greater average particle speed within $r_1$.  

The cross section posteriors for the observed clusters are shown in Fig.~\ref{fig:logcross_hist}. It seems clear by eye that there is some support for a common cross section around $0.1\rm cm^2/g$. We will not simply average the results, but allow for a systematic error in inferring the underlying cross section in Section~\ref{sec:Constraints on the Cross Section}.

\begin{figure}
    \centering
    \caption{Comparison of the BCG mass inferred (top panel) and $c_{\mathrm{vir}}$ (bottom panel) from strong lensing with that from the SIDM model. The error bars indicate the 68\% confidence interval. The diagonal lines represent equality.}
    \label{fig:mbcg_cvir_compare}
    \includegraphics[width=0.47\textwidth]{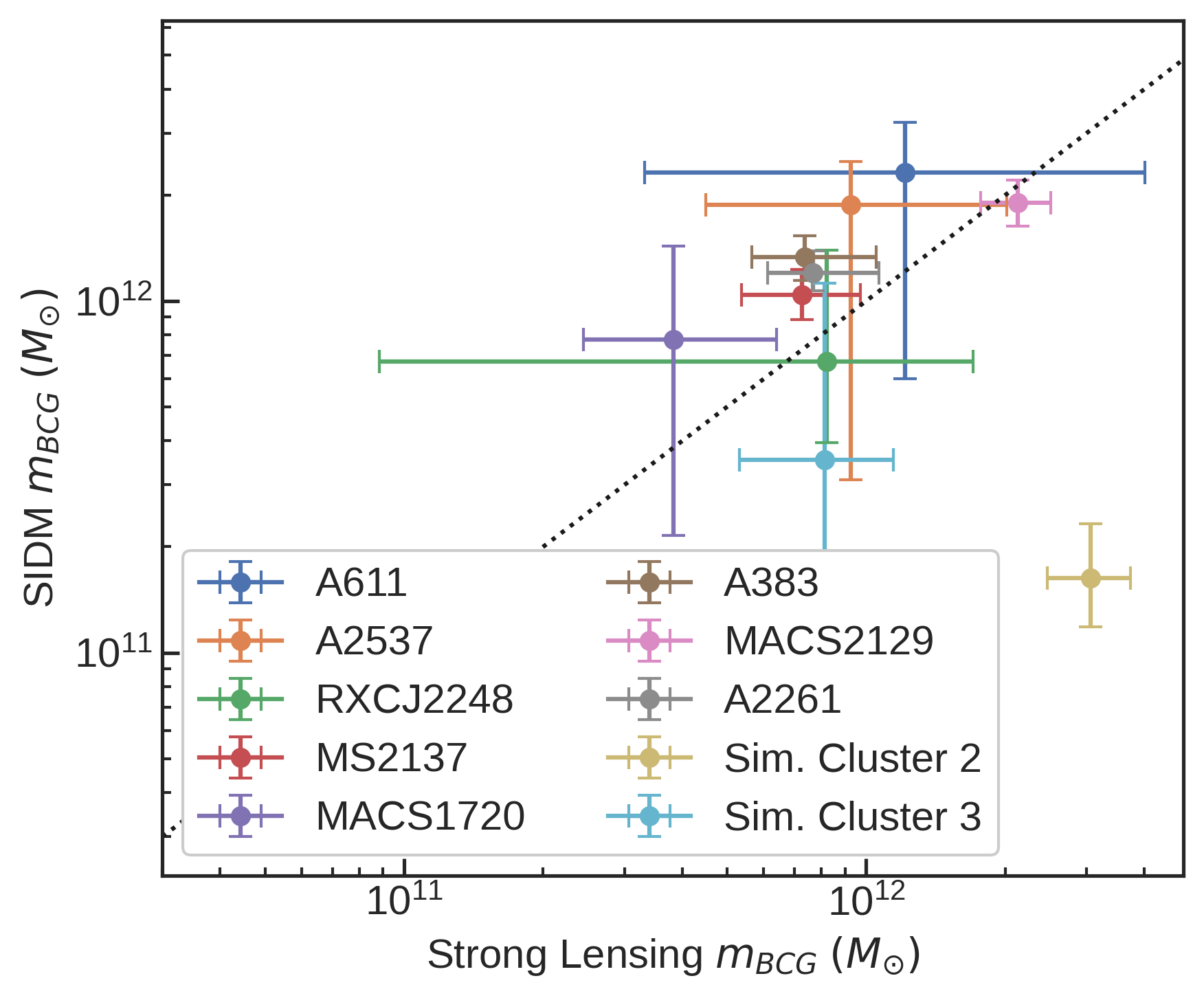}
    \includegraphics[width=0.47\textwidth]{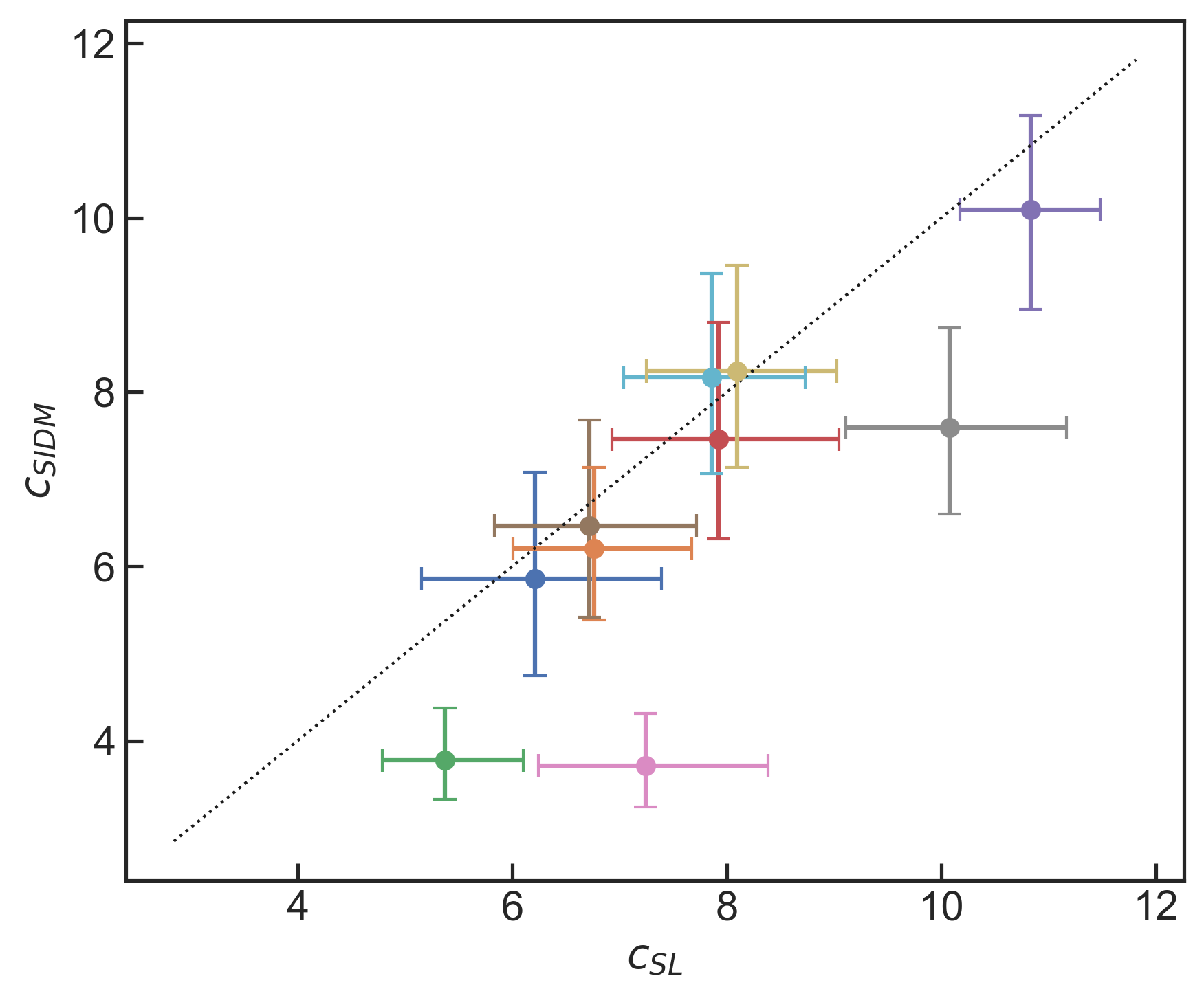}
    \end{figure}

\begin{figure}
   \caption{\textit{Top:} Concentration versus inferred cross section. The error bars indicate the 68\% confidence interval. The square markers are the simulated clusters. \textit{Bottom:} DM halo surface density at a radius of 28 kpc versus inferred cross section.}
    \centering
    \includegraphics[width=0.47\textwidth]{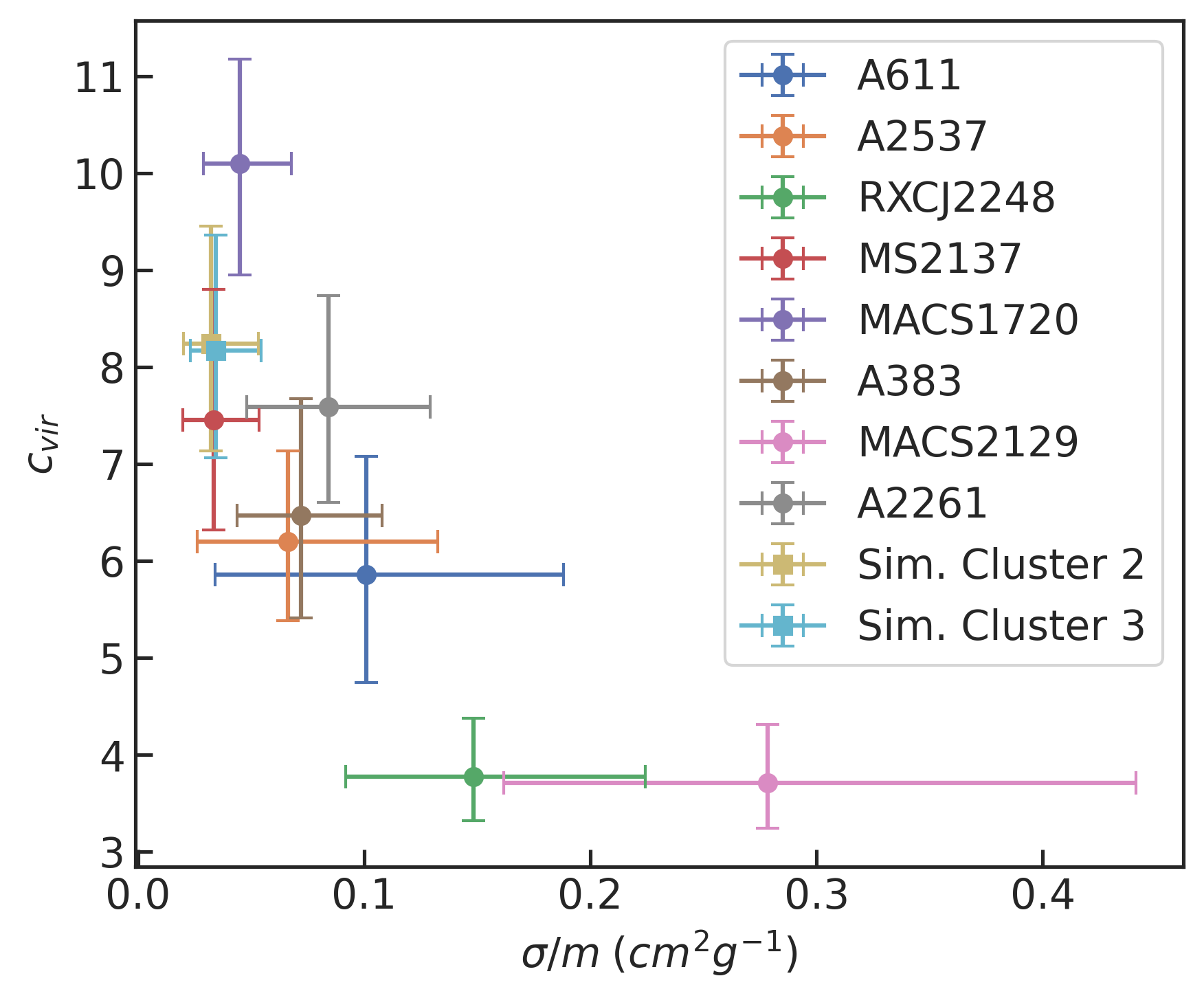}
    \includegraphics[width=0.47\textwidth]{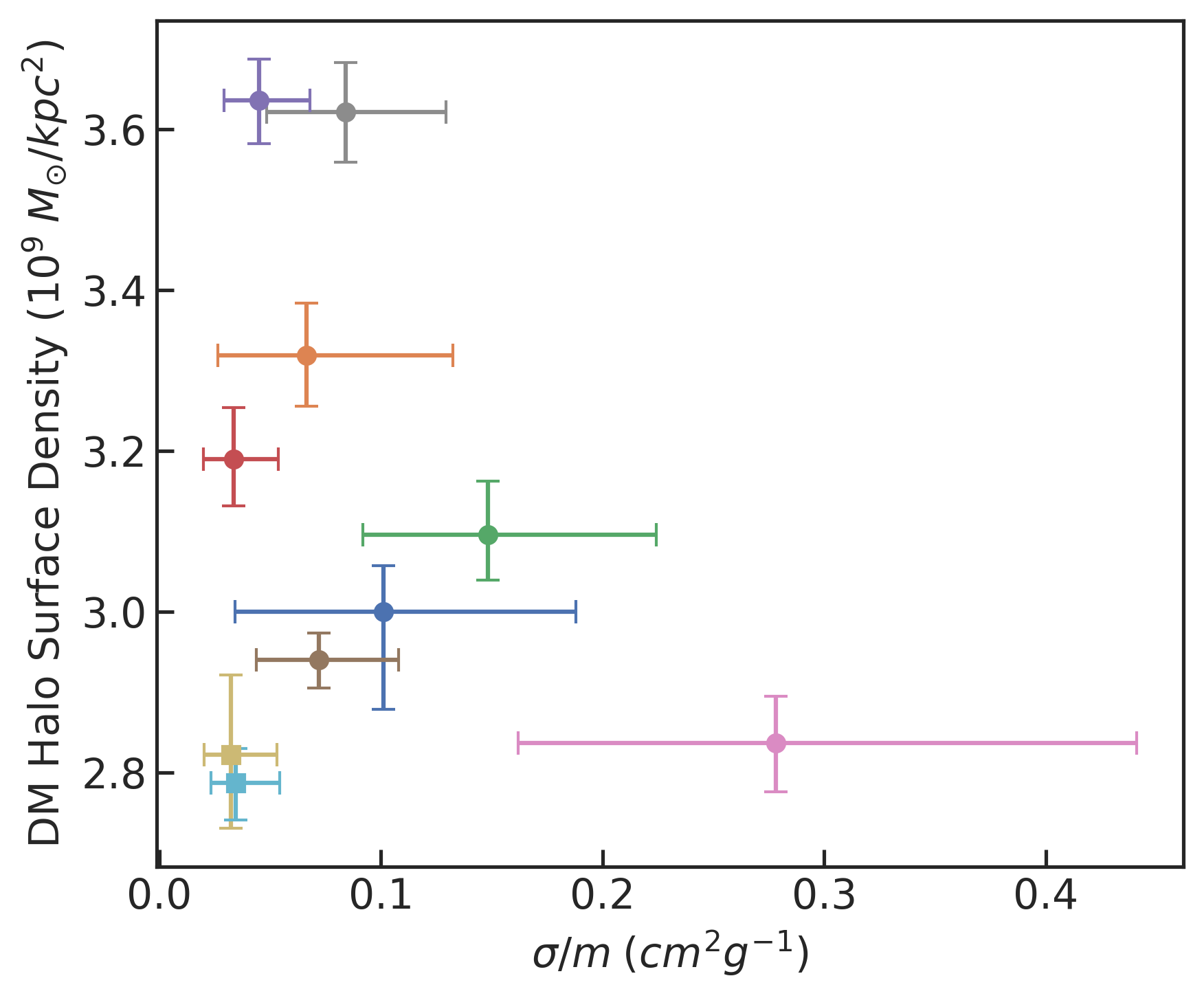}
    \label{fig:cvir_and_surf_density_vs_cs}
\end{figure}

The top panel of Figure~\ref{fig:mbcg_cvir_compare} shows the BCG masses for each cluster inferred from strong lensing compared to the inference results from the SIDM model. The correlation between the the two models is good, with the notable exception of Sim. Cluster 2. We reran the SIDM analysis for that cluster with a strong Gaussian prior on BCG mass: $\log{(m_{\text{BCG}}/M_{\sun})}=12.54\pm0.1$ (i.e., the value inferred from the strong lensing analysis of that cluster). Our results were robust under that test, showing a only very small increase in cross section. 

The bottom panel of Figure~\ref{fig:mbcg_cvir_compare} shows the concentration for each cluster inferred from strong lensing compared to the inference results from the SIDM model. For the strong lensing cNFW fits, we defined $c_\mathrm{vir}=r_s/r_\mathrm{vir}$, which should be comparable to the NFW $c_\mathrm{vir}$ when the cNFW core radius $r_c \ll r_s$. The SIDM model allows for elongation along the LOS by allowing the axis ratio s to vary, resulting in generally lower concentrations than that inferred in the SL analysis. This in turn results in lower inferred central densities and higher SIDM cross sections. 

In Figure~\ref{fig:cvir_and_surf_density_vs_cs} we show the concentration inferred in the SIDM model for each cluster versus the SIDM cross section in the top panel. There is a large scatter in the concentrations from about 3 to 10, while the cross sections are scattered around $0.1~\rm cm^2/g$. Note that MACS2129 has a high median cross section and a low concentraion. This suggests that carefully combining strong and weak lensing data with a tailored analysis such as what we have done could be fruitful. 
The bottom panel shows DM halo surface density at 28 kpc versus inferred SIDM cross section. The observed clusters are systematically higher in surface density than the simulated clusters. Among the observed clusters there does not seem to be a correlation between surface density at a particular radius and cross section, although cluster MACS2129 does appear to have the highest inferred cross section while its surface density is among the lowest in the group.

Besides having the highest inferred SIDM cross section, Cluster MACS2129 is unusual in several regards. It has the highest redshift in our sample (0.589), and 
is the only cluster for which our lens model was unable to reproduce one of the data images. MACS2129 also has the largest DM halo core inferred from strong lensing, at $54.8_{-17.6}^{+23.0}$ kpc, which is directly relevant for setting the cross section. In addition, this cluster's mass-to-light ratio is an outlier, at $\Upsilon_V=10.49_{-1.75}^{+1.89}$, while the inference for the other 7 clusters is for $1<\Upsilon_V<4$. As discussed in Section~\ref{sec:Strong_Lensing_Results}, we investigated the impact of constraining $\Upsilon_V$ to a value of 4 for MACS2129, and it would indeed result in a smaller core size (a median of 39.4 kpc compared to the original 54.8 kpc) but the fit of the model was significantly inferior to the original. A smaller core would lead to a lower inferred SIDM cross section for that cluster. It is possible that a better strong lensing model is required for MACS2129 but it is not clear what form that would take. 

We also compared the inferences for the virial mass of the DM halo, $m_{\mathrm{vir}}$, from the strong lensing fits and the SIDM halo matching results.  Our results quoted in Tables \ref{tab:strong_lensing_results} and \ref{tab:SIDM_params} show that the estimates are largely consistent with one another at the 10\% level. The exceptions include the two highest mass systems, RXCJ2248 and MACS2129, that had virial masses inferred from the SIDM halo matching procedure that were about 30\% higher than those from the cNFW fit to the strong lensing data. There is a systematic trend for the SIDM halo matching procedure to prefer slightly higher halo virial masses, but mostly within individual error bars.

\subsection{Constraints on the Cross Section}
\label{sec:Constraints on the Cross Section}

After having obtained posterior chains that include cross section per unit mass ($sm=\sigma/m$) for the simulated and observed clusters, we turn to estimating the true underlying cross section. One option is to average the cross sections obtained for the observed clusters. However, this would not be appropriate without checking if the inferred cross sections are statistically consistent with each other. Therefore, we estimate the average cross section from the sample by allowing for a systematic error. 

This systematic error can also allow us to (partially) account for effects not included in our analysis, including variations of the radial profile and shape away from our cored elliptical NFW profile model, and variations of subhalo mass profile away from the scaling relations used to model them. We also recall that the work of \citet{Sokolenko:2018noz} showed a spread of about 10-20\% in the inferred core sizes for small cross sections ($\sigma/m<1$ \cmsg). We expect that this will translate into roughly constant error in $\log(\sigma/m)$. We also explored one systematic effect related to the SIDM halo profile -- the elongation of halos along the line of sight by assuming an ellipsoidal halo profile that is oriented along the LOS. A more sophisticated non-spherical SIDM halo profile may be able to bring the cross section inferences for the different clusters into better agreement with each other.

For the observed clusters, we denote the means and standard deviations for the posteriors of $\log_{10}[(\sigma/m)/(\rm cm^2/g)]$ from the SIDM analysis as $\mu_i$ and $\epsilon_i$, with $i={1,...,8}$. These are listed in Table~\ref{tab:SIDM_params}. We used $\log_{10}(\sigma/m)$ because its posterior is more Gaussian for all the clusters compared to the posteriors for $\sigma/m$. We then add a common systematic error $\epsilon_{\rm model}$ in quadrature to the inferred errors of all the clusters. The combined likelihood can be written as, 
\begin{align}
    L = \prod_{i} N \Big( \mu_i | \mu_{\mathrm{true}}, \sqrt{\epsilon_i^2 + \epsilon_\mathrm{model}^2} \Big)
\end{align}
where $N$(x | mean, standard deviation) denotes the PDF of a normal distribution with specified mean and standard deviation. This allows us to estimate the true underlying value of $\log_{10}[(\sigma/m)/(\rm cm^2/g)]$ denoted by $ \mu_{\mathrm{true}}$ simultaneously with $\epsilon_\mathrm{model}$. We adopt uniform priors between -2 and 0 for $\mu_\mathrm{true}$ (which means the true cross section is between 0.01 and 1 \cmsg) and a uniform prior between 0 and 2 for $\epsilon_\mathrm{model}$. 

We inferred $\mu_{\rm true}=-1.086_{-0.125}^{+0.125}$, which implies that the average SIDM cross section ($=10^{\mu_\mathrm{true}}$) is $0.082_{-0.021}^{+0.027}$ \cmsg (68\% C.L.), and it is less than 0.13 \cmsg at the 95\% confidence level. We inferred a model (systematic) error ($\epsilon_{\mathrm{model}}$) of 0.27 dex. This common error is larger than the individual errors from the fits in Table~\ref{tab:SIDM_params}. Thus, we can approximate each cluster as having an error of about 0.27 dex. Within this approximation, the error on the mean is $0.27/\sqrt{8} \simeq 0.1$ dex, and hence the $2\sigma$ upper bound on $\sigma/m$ should be close to $0.082\times 10^{0.2} \rm cm^2/g = 0.13 \rm cm^2/g$, which it is. 
We repeated our analysis using means and standard deviations for $\sigma/m$ rather than $\log_{10}(\sigma/m)$, and an analysis with a normal distribution for $\sigma/m$ truncated below zero--- the results of the three methods were consistent with one another. 

In comparison, a constraint on SIDM $\sigma/m$ of $0.10_{-0.02}^{+0.03}$ \cmsg for galaxy clusters was reported in \citet{Kaplinghat2016}. This was based on data from 6 clusters from \citet{Newman2013a,Newman2013b}, which employed strong lensing, weak lensing and stellar kinematic approaches to determine the cluster profiles. In  \citet{Sagunski2021}, they use a Jeans analysis technique and find $\sigma/m<0.35$ \cmsg (95\% C.L.) for galaxy clusters. Cluster mergers can also put constraints on SIDM cross section, as collisionless DM would not be slowed in a cluster merger. In \citet{Randall2008} they analyze the collision in the Bullet Cluster (galaxy cluster 1E 0657-56) and find $\sigma/m<1.25$ \cmsg at the 68\% confidence level. In \citet{Robertson2017}, they use simulations on the same cluster and find a more relaxed constraint of 2 \cmsg in their fiducial model. \citet{Harvey2015} analyzed an ensemble of 72 clusters to find $\sigma/m<0.47$ \cmsg at the 95\% confidence level, but see \citet{Wittman:2017gxn}. Analyses of other cluster collisions that put an upper constraint on SIDM cross section include  $\sigma/m<4$ \cmsg in MACS J0025.4-1222 \citep{Bradac2008}, $\sigma/m<3$ \cmsg in Abell 2744 \citep{Merten2011} and $\sigma/m<7$ \cmsg in cluster DLSCL J0916.2+2951 \citep{Dawson2012}. \citet{Harvey2019} found $\sigma/m< 0.39$ \cmsg at the 95\% confidence level based on BCG oscillations in otherwise relaxed galaxy clusters.

The limits we have obtained are very stringent, as the comparison above shows. Unlike previous work, our dedicated SIDM analysis used the full radial surface density profile, allowing for stronger constraints. The fact that we selected clusters with masses close to $10^{15}M_\odot$ and with images over a wide range of radii also likely played a part in this. We can test this hypothesis with more heterogeneous cluster data sets in the future. One may worry that our reliance on the analytic model down to such low cross sections is leading to overly stringent constraints. As we discussed previously, the model has only been validated for cross sections of $0.1$ \cmsg and higher by \citet{Sokolenko:2018noz}. However, the cNFW fits to the strong lensing data recovered uniformly high surface densities at 30 kpc for the clusters, close to or larger than that of the simulated clusters.
These high surface densities directly lead to the strong constraints on the cross section. A related issue is that the analytic model may not be capturing the effects of BCG stellar distribution on the SIDM density profile well enough at low cross sections because the inner halo is not close to isothermal. However, it is not clear why there would be a bias to lower cross sections in this case. Note that the inferred $r_1$, despite the small cross sections, is in the range of 30 to 140 kpc, where the cluster is dark matter dominated. This is again related to the high dark matter densities required to model the strong lensing images. It is also useful to keep in mind that the impact of the baryons on the SIDM density profile of these $\sim 10^{15} M_\odot$ clusters is much smaller than lower mass clusters (which would have almost similar stellar masses). Nevertheless, we cannot rule out that systematic errors of order 0.1 \cmsg are introduced by the method, and this needs to be investigated using simulations. 

\section{Conclusions}
\label{sec:Conclusions}
With the aim of constraining the self-interaction cross section of DM particles, we have constructed strong lensing models of 8 observed clusters (see Figure~\ref{fig:image_plane_plots}). We included DM subhalos for the perturbing galaxies to more accurately model the perturber mass distribution. The inferred surface density from strong lensing for the main halo was fit with a SIDM profile to infer the  self-interaction cross section. For the SIDM profile, we used a well-tested analytic model, in which the outer region follows a NFW profile and the inner region is isothermal due to DM self-interactions, with the transition radius being set by the self-interaction cross section. We allowed the SIDM halo profile to be elongated along the line-of-sight with a cosmological prior to regulate axis ratios larger than 2.  We used the posteriors for the cross sections inferred from all the clusters to infer the true underlying SIDM cross section, allowing for a common systematic error due to mismodeling. We have tested our inference pipeline on two mock data sets obtained from the Illustris-TNG simulation. Our key findings are summarized below.

Using strong lensing alone we were able to reproduce the image positions in our cluster sample with RMS image position errors ranging from 0\farcs32 to 1\farcs07. Our methodology includes separate lens model components for the member galaxy's baryonic and DM components, allowing for more flexible characterization of each galaxy's matter distribution. The models reproduced all data images, with the exception of one image in the outlier MACS2129. 

We find that the strong lensing inference for concentration was biased high for the two simulated relaxed clusters (see Table~\ref{tab:simulated_cluster_list}) compared to the cosmological expectation for spherically-averaged halos. For the observed clusters we find a median concentration of 7.0 from strong lensing, which is also higher than would be expected from the concentration-mass relation, and higher than that found in X-ray observations of CLASH clusters. As discussed in Appendix~\ref{sec:biases_in_triaxial_halos}, this may be expected in situations where the line of sight is preferentially oriented along the major axis of halos, which boosts strong lensing probability. 
In the common case of prolate halos with their major axis oriented along the line of sight, halo concentration can be biased by up to $\sim 60\%$. 
Previous works \citep[e.g.,][]{Newman2013a} have also noted that observed concentrations for a sample of clusters similar to ours are indeed higher than what is predicted from mass-concentration relations in CDM simulations. In our SIDM halo model, we allow the halos to be elongated along the line of sight, which reduces the inferred concentrations when fitting the SIDM halo model to the inferred surface density profiles.

We found that the BCGs centers in our models are either coincident with the DM halo centers or offset by few to 10 kpc (Figure~\ref{fig:various_histograms}). The median offset is about 4 kpc, as measured in the plane of the sky, which would correspond to approximately 5 kpc in 3D distance assuming isotropy. \citet{Harvey2019} used simulations to conclude that the distribution of offsets from an ensemble of clusters would exhibit a median value of $3.8\pm 0.7$ kpc for a CDM scenario (i.e., $\sigma/m=0$), and $8.6 \pm 0.7$ kpc in a scenario where $\sigma/m=1$ \cmsg. Comparing to  \citet{Harvey2019}, the median value we obtained is consistent with $\sigma/m \ll 1$ \cmsg. However we also note that \citet{Harvey2017} found that MCMC lensing code inferences of offset are susceptible to substantially understating the error in position offset, thus tempering the ability to make strong inferences from the offsets.

Using the SIDM halo model, the cross section per unit mass for all observed clusters had median values in the range of 0.034 \cmsg to 0.15 \cmsg (Figure~\ref{fig:logcross_hist} and Table~\ref{tab:SIDM_params}), with the exception of outlier MACS2129 at 0.28 \cmsg. We combined these individual measurements allowing for an unknown common modeling error, and inferred an SIDM cross section of $0.082_{-0.021}^{+0.027}$ \cmsg (68\% C.L.), with an upper limit of 0.13 \cmsg at the 95\% C.L. (see Sections~\ref{sec:Constraints on the Cross Section}). We infer a systematic (modeling) error of 0.27 dex. In comparison, a constraint on SIDM $\sigma/m$ of $0.10_{-0.02}^{+0.03}$ \cmsg for galaxy clusters was reported in \citet{Kaplinghat2016} using a simpler analysis, and \citet{Sagunski2021} found $\sigma/m<0.35$ \cmsg (95\% C.L.) for galaxy clusters using Jeans analysis. 
The mean relative velocity of dark matter particles for the eight clusters we have analyzed is $1458_{-81}^{+80}$ km/s. Since relative particle speeds are much lower in galaxies, the cross sections for dark matter interactions can be larger in galaxies, as many concrete particle physics models predict \cite{Tulin2017}.

\section*{Acknowledgements}

We gratefully acknowledge a grant of computer time from XSEDE allocation TG-AST130007. QM was supported by NSF grant AST-1615306 and MK was supported by NSF grant PHY-1620638.

\section*{Data Availability}
The observational raw image data underlying this article are
available the Mikulski Archive for Space Telescopes (MAST) at \url{https://archive.stsci.edu/access-mast-data}. Additional data was used as noted in the references in Table~\ref{tab:cluster_list} and is available in those references.
	
%%%%%%%%%%%%%%%%%%%%%%%%%%%%%%%%%%%%%%%%%%%%%%%%%%

%%%%%%%%%%%%%%%%%%%% REFERENCES %%%%%%%%%%%%%%%%%%

% The best way to enter references is to use BibTeX:

\bibliographystyle{mnras}
\bibliography{MyCollection.bib} 

%%%%%%%%%%%%%%%%%%%%%%%%%%%%%%%%%%%%%%%%%%%%%%%%%%

%%%%%%%%%%%%%%%%% APPENDICES %%%%%%%%%%%%%%%%%%%%%

\appendix

\section{Relevant lensing formulas for the cNFW halo model}\label{sec:appendix_lensing_formulae}

The cNFW (cNFW) model is defined by modifying the NFW profile as follows:

\begin{equation}
\rho = \frac{\rho_s r_s^3}{\left(r_c + r\right)\left(r_s + r\right)^2}.
\end{equation}

Defining $x = r/r_s$ and $\beta = r_c/r_s$, by integrating the density profile along the line of sight we find an analytic expression for the projected density profile,

\begin{equation}
\kappa(x) = \frac{2\kappa_s}{(\beta-1)^2}\left\{\frac{1}{x^2-1}\left[1-\beta-(1-x^2\beta)\mathcal{F}(x)\right] - \mathcal{F}\left(\frac{x}{\beta}\right)\right\}
\label{eq:cNFW_kappa}
\end{equation},
where we have defined $\kappa_s = \rho_s r_s/\Sigma_{cr}$, and

\begin{equation}
{\mathcal F}(x) =
\begin{cases}
  \frac{1}{\sqrt{x^2-1}}\,\mbox{tan}^{-1} \sqrt{ x^2-1 } & (x>1) \\
  \frac{1}{\sqrt{1-x^2}}\,\mbox{tanh}^{-1}\sqrt{ 1-x^2 } & (x<1) \\
  1                                                      & (x=1)
\end{cases}
\end{equation}

When using the pseudo-elliptical approximation, it is useful to have an analytic formula for the deflection angle generated by a spherical cNFW lens. By integrating Equation~\ref{eq:cNFW_kappa}, we obtain

\begin{eqnarray}
\alpha(x) & = & \frac{2\kappa_s r_s}{(1-\beta)^2 x}\biggl\{(1-\beta)^2\ln\left(\frac{x^2}{4}\right) - \beta^2\ln\beta^2  ~ ~ + ~ ~ ~ ~ ~ ~ ~  \\
& & ~ ~ ~ 2(\beta^2-x^2)\mathcal{F}\left(\frac{x}{\beta}\right) + 2[1+\beta(x^2-2)]\mathcal{F}(x)\biggl\}.
\end{eqnarray}

It can be easily verified that in the limit $\beta \rightarrow 0$, these formulae reduce to the usual analytic formulas for an NFW profile \citep{Golse2002a}. Numerical convergence of these formulae becomes difficult in the neighborhood of either $x \approx \beta$, $x \approx 1$ or $\beta \approx 1$; in each of these cases, series expansions can be used for greater accuracy, all of which have been implemented and tested in the \qlens code.

\FloatBarrier
\section{Stellar Mass--Halo Mass  Relation}
\label{sec:power_law}
\begin{figure}
    \centering
    \caption{Stellar mass-halo mass data from \citet{Niemiec_2017}, and the best fit power law which was used to determine the mass for DM subhalos in the lens models.}
    \includegraphics[width=0.47\textwidth]{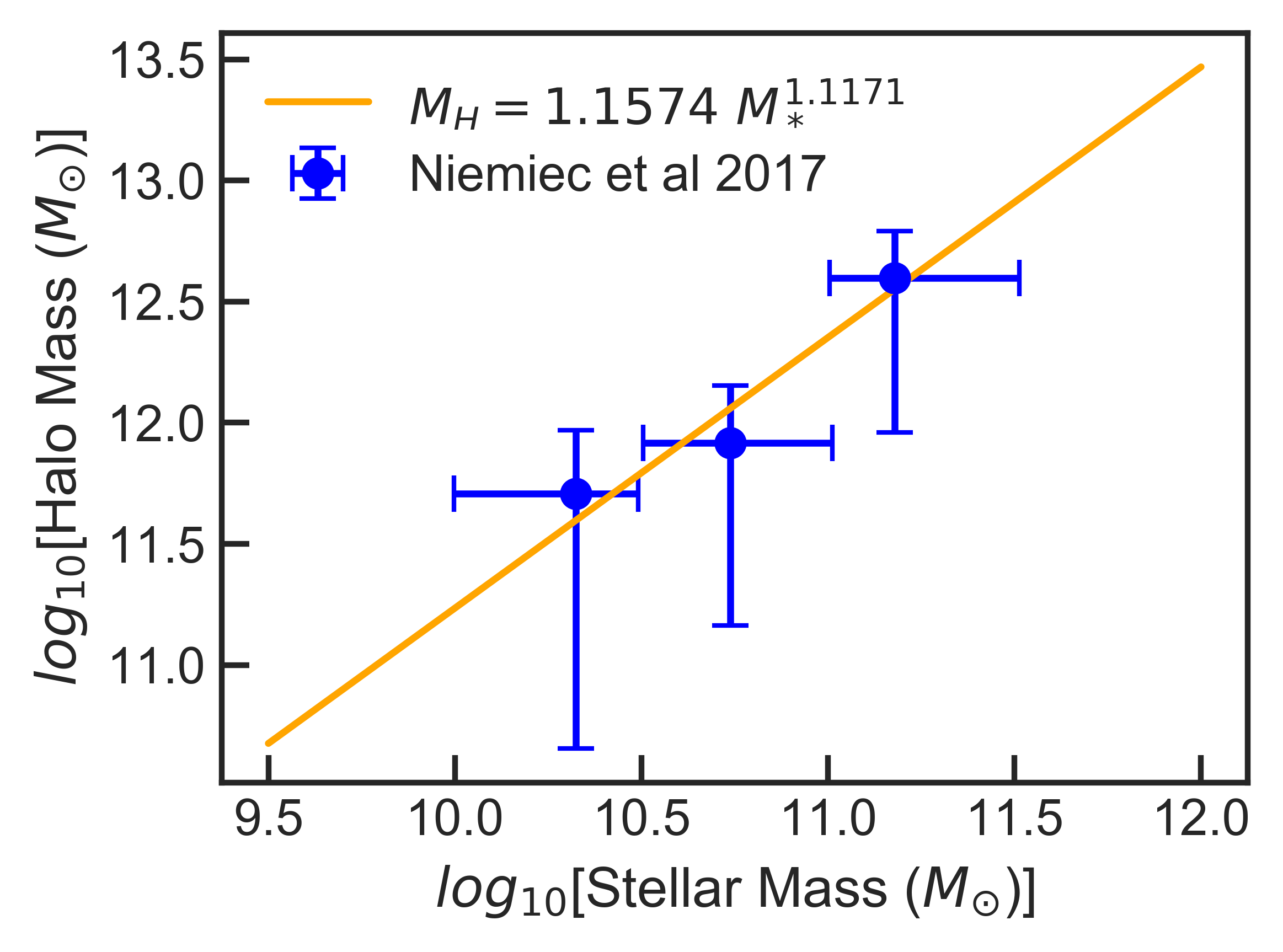}
    \label{fig:SMHM_Niemiec_fit}
\end{figure}

\citet{Niemiec_2017} examined the stellar mass-halo mass relation for cluster galaxies. A power law fit to their data is shown in Figure~\ref{fig:SMHM_Niemiec_fit}. The resulting power law is $M_{\textrm{halo}} = 1.1574M_*^{1.1171}$ .

\FloatBarrier
\section{Concentration Bias in Triaxial Halos}
\label{sec:biases_in_triaxial_halos}
Since strong lensing analysis is based on 2-D projections of 3-D bodies, effectively hiding one dimension from direct measurement, biases can arise in model parameters such as concentration and scale radius. Halo elongation parallel (perpendicular) to the line of sight can increase (decrease) the strength of the lensing. Moreover, alignment of the major axis with the line of sight has shown to increase as lensing cross section increases, and leads to an upward bias in concentration \citep{Meneghetti2010}.

Galaxies and galaxy clusters are thought to form along filaments of the cosmic web. The major axis of elliptical galaxies tend to be parallel to the direction of the host filament \citep{Tempel2013}. In cosmological N-body simulations, prolate halos are more common than oblate ones, especially so for massive halos \citep{Bonamigo2015}. We therefore expect the preponderance of our sample to be prolate halos with their major axis oriented preferentially along the line of sight, creating significant upward bias in concentration.
\begin{figure}
    \centering
    \caption{Bias in concentration when fitting 2-D projections of triaxial halos. The bias is $\sim+60\%$ for prolate halos oriented along the LOS.}
    \includegraphics[width=0.40\textwidth]{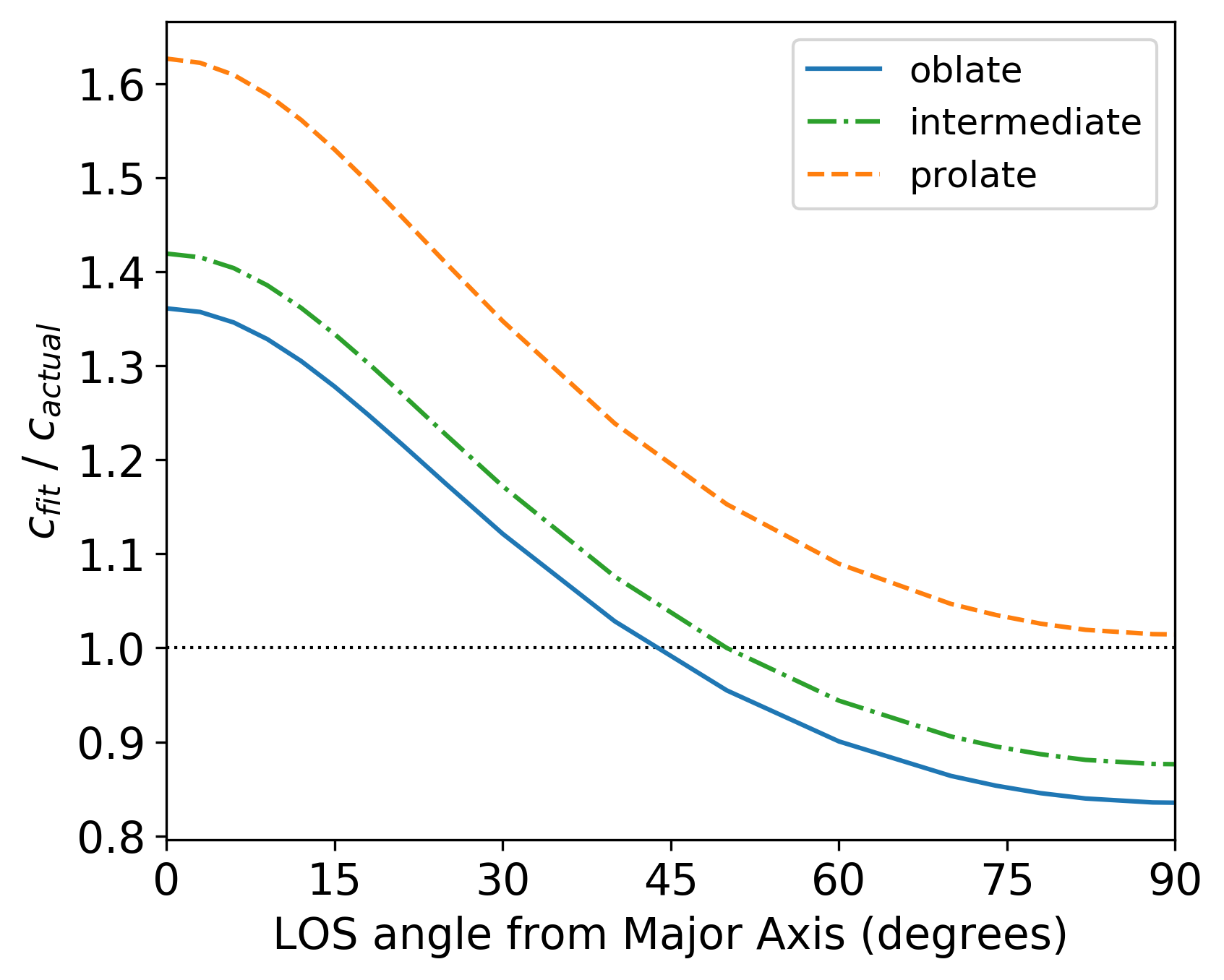}
    \label{fig:concentration_bias}
\end{figure}
To quantify the effect of concentration bias in our modeling pipeline, we created simulated triaxial halos, projected them in various orientations, and fitted the resulting 2D shapes with our fitting code. The simulated halos were NFW ellipsoids, with triaxiality introduced by scaling the x, y and z axes by factors a, b and c. The triaxiality is thus defined by the ellipsoid with form
\begin{equation}
    \frac{x^2}{a^2}+\frac{y^2}{b^2}+\frac{z^2}{c^2}=1.
\end{equation}
For the oblate case, we chose $a=0.5,\;b=c=1.0$. For the intermediate case, $a=0.5,\;b=0.75,\;c=1.0$. The prolate case used $a=b=0.5,\;c=1.0$. Thus the z-axis is the major axis in all cases. Figure \ref{fig:concentration_bias} shows the results of that analysis. Biases in concentration range from approximately +60\% to -20\%. The strongest bias of $\sim+60\%$ occurs in the common case of prolate halos with major axis orientation nearly along the line of sight. We therefore expect that our modeled concentrations are often biased high by as much as 60\%.

\FloatBarrier
\section{Strong Lensing Error Residuals}
\label{sec:error_residuals}
\begin{figure}
    \centering
    \caption{Residual position errors in image locations from strong lensing. Both the x and y error components are included; positive residuals mean errors in the westerly direction for x-components and the northerly direction for y-components.}
    \includegraphics[width=0.47\textwidth]{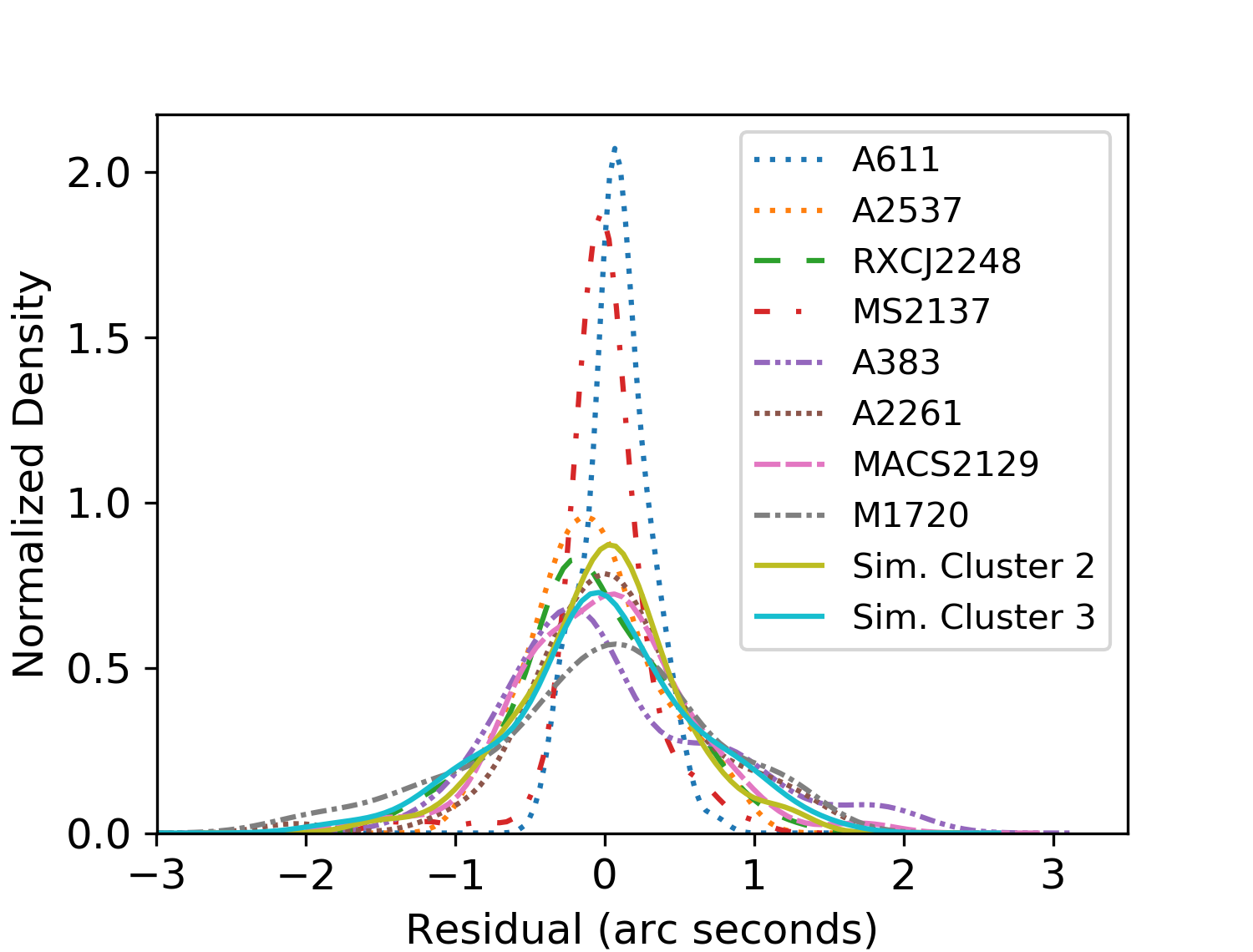}
    \label{fig:error_residuals}
\end{figure}

In the Figure~\ref{fig:error_residuals} we show the distribution of residual errors in image locations from the strong lensing analysis. For the simulated clusters, random Gaussian error of 0\farcs50 was added to each of the x and y components, resulting in a total position error of 0\farcs71. In the strong lensing $\chi^2$, we have again assumed image position errors of 0\farcs50 for each of the x and y components. The figure demonstrates that the image position errors are distributed in a Gaussian shape, as would be expected for measurement errors.

\FloatBarrier
\section{Surface Density Match between Strong Lensing and SIDM Models}
\label{sec:SIDM_kappa_match}
\begin{figure*}
    \centering
    \caption{Plots of the 68\% confidence intervals for strong lensing ("data") and SIDM ("model") surface densities for each cluster. The model is shown in blue and the data in red, and the regions appear purple where they overlap.}
    \includegraphics[height=0.94\textheight]{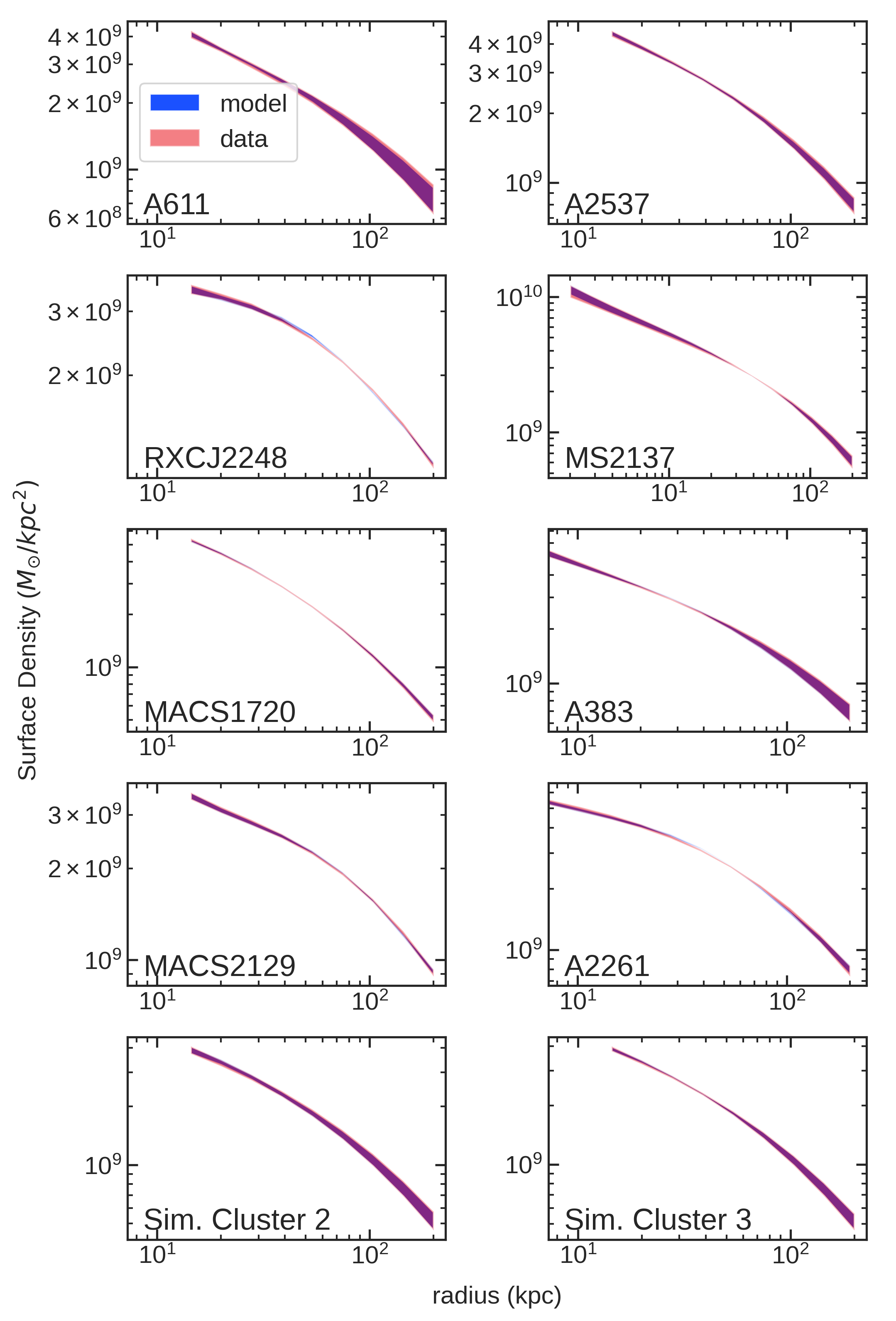}
    \label{fig:kappa_match}
\end{figure*}

\label{lastpage}
\end{document}